\documentclass[dvips]{acta}
\usepackage{supertabular,lscape,epsfig}
\usepackage{amssymb}
\usepackage{amsmath}

\SetPages{0}{0}

\SetVol{70}{2019}

\begin{document}

\begin{Titlepage}
\Title{Mount Suhora high cadence photometric survey of T~Tauri-type stars}

\Author{Micha{\l} Siwak, Marek Dr{\'o}{\.z}d{\.z}, Karol Gut, Maciej Winiarski, Waldemar Og{\l}oza, Grzegorz Stachowski}
{Mount Suhora Astronomical Observatory, Cracow Pedagogical University,\\ 
ul. Podchora{\.z}ych 2, 30-084 Krak{\'o}w\\
e-mail: michal.siwak@gmail.com}

\Received{Month Day, Year}
\end{Titlepage}

\Abstract{
Results of high-cadence multi-colour observations of 121 pre-main sequence stars available 
from the northern hemisphere are presented. 
The aim of this survey was to detect transit-like signatures caused by occultation 
of these young stars and their accretion-induced hot spots by close-in planets and/or dusty clumps. 
Although none planetary transits were detected, our data allow to determine rotational periods 
for some T Tauri stars, characterise accretion processes operating in classical T~Tauri-type stars 
in time scales ranging from a few minutes to days, as well as the large-scale dips caused by dusty 
warped discs.}
{T Tauri stars, Herbig Ae/Be stars, accretion}

\section{Introduction}

Followed by Class~0 and Class~I protostar phases lasting about $10^5$ yr, classical T~Tauri-type stars
(CTTS) and their more massive ($\sim 2-10$~M$_{\odot}$) counterparts - Herbig Ae/Be (HAEBE) stars - represent 
the first observable 
in visual wavelengths phases of young stars, still being surrounded by accretion discs. 
Currently it is well observationally established that accretion onto the central star occurs 
through magnetically controlled accretion, in a way originally proposed for neutron stars 
by Ghost \& Lamb (1977) and further suggested to apply to CTTS by K{\"o}nigl (1991).
According to this mechanism, accretion disc is threaded by stellar magnetic field lines. 
The star-disc locking through magnetic field prevents the star to accelerate its rotation 
while it contracts well before 
reaching the breakup speed; instead its rotational period is equalized to the Keplerian period 
at the disc co-rotation radius.
In the narrow transition zone, extending for a few stellar radii from an inner disc radius, 
the disc plasma is controlled in its motion by the magnetic field and transferred along its lines 
to the high and moderate stellar latitudes, where it strikes the photosphere producing hot spots. 
It is normally assumed that the hot spots, together with solar-type, but huge
dark spots produce complex, large amplitude (1-2 mag) photometric variability of the stars (Herbst
et al., 1994, 2002). 
The hot accretion spots are responsible for many peculiarities observed in CTTS, particularly
the ultraviolet and visual excess of radiation (veiling) and strong and variable emission lines. 
While accretion is the dominant mechanism responsible for variability of stars observed well above 
the disc plane, in stars observed closer to its plane persistent dusty disc warps often
cause semi-periodic or variable obscuration of central stars, in combination with the aforementioned accretion-induced
variability (e.g. Bouvier et al., 2003; Stauffer et al., 2015; McGinnis et al., 2015; Ansdell et al., 2016). 

Hot and cold spots on CTTS should last sufficiently long on stellar photosphere to produce periodic 
light curves dominated by stellar rotation. 
Instead, the light curves frequently show quasi-periodic variations or no periodicity at all 
(Herbst et al., 1994, 2002; Grankin et al., 2007; Rucinski et al., 2008; Alencar et al., 2010; 
Siwak et al., 2011a; Stauffer et al., 2014). 
The subject of plasma flows in magnetized stars was numerically investigated with three dimensional 
magneto hydro dynamical (3D~MHD) simulations by Romanowa et al. (2004, 2008), 
Kulkarni \& Romanowa (2008, 2009) and Kurosawa \& Romanowa (2013).
According to the authors  of these publications, for magnetospheres a few times the stellar radius in size 
(but no less than two stellar radii), the accretion from the surrounding disk can occur in either {\it stable}, 
{\it moderately stable} or {\it unstable regime}. 
The regime of accretion which is prevalent at a given time is primarily controlled by the mass accretion rate 
and it may alternate between moderately stable and unstable depending 
on the actual value of the mass accretion rate (Romanowa \& Kulkarni, 2008). 
Our {\it MOST} satellite observations have qualitatively confirmed these theoretical findings: 
for low mass accretion rates, plasma is transported toward the star in two stable funnel flows originating 
near the disc co-rotation radius, and it strikes the star near its magnetic poles. 
Two antipodal fairly stable hot spots formed during this process produce fairly regular light curves, 
revealing rotational period of a star (e.g. in IM~Lup, see in Siwak et al., 2016). 
Once the mass accretion rate becomes higher, the inner disc radius decreases.
Instabilities developing in the ''inner disc -- magnetosphere'' boundary produce 
a few chaotic equatorial tongues, which penetrate through the magnetosphere and produce short-term chaotic 
hot spots leading to irregular brightness changes, as observed in RU~Lup (Siwak et al., 2016) 
and TW~Hya (Rucinski et al., 2008; Siwak et al., 2011a, 2014, 2018). 
Except of the above, it was recently observationally established that in CTTS accreting in stable 
regime either ''disc-planet'' or ''disc-low mass companion'' interactions may lead to temporary 
inner disc instabilities and related episodes of inhomogenous accretion, as seen in CI~Tau (Biddle et al., 2018), 
DQ~Tau (Tofflemire et al., 2017a, Muzerolle et al., 2019), and TWA~3A (Tofflemire et al., 2017b).
 
Planet formation processes appear to start when protostars are still deeply hidden behind dust 
(e.g. Drazkowska \& Dullemond, 2018) and continues in protoplanetary discs during the CTTS and HAEBE phase. 
It is widely accepted that planets form beyond the ''snow
line'', where is cold enough for volatile compounds to condense into solid ice grains either through
core accretion (Pollack et al. 1996), or through instabilities in the disc (Boss, 2001).
The disc instability mechanism is very efficient in forming giant 
planets and brown dwarfs due to gravitational fragmentation in outer parts of massive
protoplanetary discs after several thousand -- 1 Myr. 
The core accretion is efficient in forming giant gas planets 
in less massive protoplanetary discs, with typical time scales of about 3-5~Myr. 
The most recent discoveries of large populations of pebbles in protoplanetary discs initiated works
on planet formation via pebble accretion (e.g. Johansen \& Lambrechts, 2017).
The state-of-art instruments utilizing mid-, far-infrared, sub-mm and mm wavelengths provide numerous 
observational evidences for ongoing planets formation at distances of 10-100~AU from central stars 
(e.g. Flaherty et al., 2011; Benisty et al., 2015, Perez et al., 2016;, Stolker et al., 2016). 
As the ''snow line'' is located on a few astronomical units for Solar-type stars, the new born planets 
have to interact with the disc material and gradually migrate to new orbits (Nelson \& Papaloizou 2003) 
to explain the orbits of Earth-like planets in Solar System, and tight 0.03-0.1~AU orbits with 3-10 days 
orbital periods of ''Hot Jupiters'', commonly discovered around main-sequence stars during the 90s. 
Orbital parameters of these giant planets were theoretically explained by Romanova \& Lovelace (2006), 
who obtained that for typical misalignment angle between magnetic and rotational axes of the star, 
the magnetospheric gap could stop inward migration of all planets at 0.05~AU. 
Discovery of the first Neptun-like planet K2-33b transiting weak-lined T Tauri star (WTTS) with 
a period of 5.4 days by {\it Kepler-K2} mission (David et al. 2016) fits well in the above image. 
After a decade of unsuccessful application of the radial velocity technique 
(e.g. Crockett et al., 2012; Lagrange et al., 2013) 
only recently sophisticated methods of removal of features induced by cold spots and plagues 
in stellar line profiles, allowed to detect possible long-term periodic radial velocity variations 
in V830~Tau and TAP~26, which are WTTS (Donati et al., 2015, 2016; Yu et al., 2017), and also around 
CI~Tau (Johns-Krull et al., 2016a), which is CTTS. 
However, the planetary status of the very-first short-periodic planet candidate orbiting CVSO~30 
every 0.44~d on inclined orbit (van Eyken et al., 2012), is still uncertain and the subject 
of debate (Yu et al., 2015). 
Time-series H-alpha spectroscopy strongly suggests that this planet may be subject to tidal disruption, 
as it appears that it is losing its mass through the Roche lobe overflow 
mechanism (Johns-Krull et al., 2016b). 

Encouraged by the discovery of the putative planet CVSO~30b, as well as our discovery 
of semi-periodic/random dips in 2011 {\it MOST} satellite observations of TW~Hya, 
likely the results of hot spots occultations by dusty clumps (Siwak et al., 2014), we decided to conduct 
high-cadence (30-90~sec) multi-colour survey of pre-main sequence stars available from the northern 
hemisphere. 
The main goal was to deliver evidences for more similar phenomenons for further sophisticated studies. 
Additional benefits of such high-cadence data sampling, uncommon for the major space-based photometric telescopes, 
is the ability for study of rapid accretion-induced 
effects occurring in the time scales as short as a few minutes. 
Photometric data obtained during irregularly distributed nights can also be used for rough 
characterisation of large-scale dips, caused by dusty disc warps and/or massive disc winds. 
These data are also well suited for determination 
of rotational periods -- a basic ''clock'' in a star-disc system -- which has still very high 
importance for a proper description 
of the overall spin-age relation for the pre-main sequence stars and is essential for our 
understanding of the rotation-braking process (Bouvier et al., 2014; Karim et al., 2016). 
Its later stages are now better understood: After roughly 5-10 Myr, when an accretion disc material 
is depleted and the star-disc locking mechanism can no longer operate, the former CTTS and now 
WTTS spins up due to conservation of its angular momentum during 
its contraction and falls into a region of 1-2 day periodicities. 
At this stage, hot accretion-induced spots no longer form and light curves are mainly modulated 
by stable colds spots, permitting even studies of differential surface rotation (e.g. Grankin et al., 2008; Siwak et al., 2011b). 

Details of our Mount Suhora Observatory ({\it MSO}) observations are described in Section~2. 
Variety of results obtained during this survey are described and immediately discussed in Section~3. 
We summarize our main findings in Section~4.

\section{Target selection, observations and data reduction}

We compiled initial list of targets based on Herbig et al. (1994) and Grankin et al. (2007). 
As the time progressed, the list was slightly modified to increase efficiency of this survey: 
observations of fields containing a handful instead of only single or two young stars were preferred. 
Poor weather conditions during all but first Winter (2013/2014) of our survey nominal duration time 
forced us to skip observations of a few preselected fields from Taurus, Auriga and Orion constellations. 
Fortunately, most of these missed members of ''Taurus-Auriga'' Star Forming Region 
(SFR) was later observed during {\it Kepler-K2} mission (Borucki et al., 2010) and by {\it TESS} mission 
(Ricker et al., 2015) -- results obtained by means of these space-telescopes 
are incomparably richer and will be the subject of separate publications. 
The list of all young stars (compiled from {\sc Simbad} database) contained in 20 fields 
(column 1)\footnote{These fields are numbered according 
to their observation start dates.  
Field \#1 is centered on ''R1~Mon'' association, fields \#4, 8, 10, 13 and 14 contain members of ''Taurus-Auriga'' 
SFR, field \#2 includes most members of ''Gulf-Mexico'', i.e. the small part of ''North America'' SFR, 
field \#7 includes a fraction of ''Pelican Nebulae'' (IC~5070) SFR members, while field \#9 members of NGC~6914 
association.
Targets from fields \#5, 6, 11, 15, 16 and 19 were selected among members of ''Cepheus-flare'' SFR; 
more specifically fields \#5 and 11 contain young stars forming in (and nearby) the famous NGC~7023 and NGC~7129 nebulaes, 
respectively, while field \#15 nearby L1261 cloud. 
BM~And (field \#3) is the only CTTS observed in ''GAL 110-13'' association, while V594~Cas is the only HAEBE 
from ''RfN VDB 4'' cloud. 
Field \#12 show stars emerging from dusty ''Serpens Molecular Cloud''. 
None membership is known for the two HAEBEs from fields \#17 and 18.
} 
and bright enough for {\it MSO} telescopes is presented in column 2 in Tables~1-4. 
Many of these stars were previously observed by other authors, who focused on investigation 
of long-term variability, determination of accretion diagnostics to estimate major stellar 
parameters, and also disc-related infrared fluxes. 
We do not present detailed time-ranked historical results for every individual target. 
Instead, in the next section we will shortly discuss our major achievements in the historical 
context only for a couple of stars, which show puzzling variability and were therefore 
more widely investigated in the past by other authors.

{\MakeTable{cccccc}{14cm}{Stars (column 2) observed in consecutive fields (column 1) 
and their coordinates in ICRS 2000 system (column 3 and 4). 
Classification given either in {\sc Simbad} database, by other authors and/or firmly deduced from 
our light curves is given in column 5. 
In column 6 we indicate mechanism causing observed variability if firmly established, 
otherwise the question mark is present. 
Stars showing constant light during our monitoring (both in short and long time scales) are also indicated. 
Periods (in days) with errors in parentheses are given for eight periodic stars. 
Two FU~Ori-type stars are marked in bold -- data obtained during this survey 
will be the subject of separate study.}
{\hline\hline
Field  & Stars  & RA [hh:mm:ss]   & DEC [$^{\circ}:':''$]  & Classification & Variability  \\ \hline \hline
{\#1}  &         VY Mon                  & 06:31:06.92 & +10:26:04.98   & HAEBE          & dipper \\
       &         V481 Mon                & 06:31:09.14 & +10:26:10.73   & CTTS           & dipper \\
       &         V540 Mon                & 06:31:07.66 & +10:26:19.85   & CTTS           & uncertain\\
       &         V541 Mon                & 06:31:07.96 & +10:28:36.92   & CTTS           & dipper \\
       &         V542 Mon                & 06:31:15.74 & +10:28:13.30   & CTTS?          & uncertain \\
       &         V687 Mon                & 06:31:10.13 & +10:26:04.47   & CTTS           & dipper \\  
       &         V698 Mon                & 06:30:50.17 & +10:33:09.81   & HAEBE          & dipper \\ 
       &         LkHa 342                & 06:31:30.12 & +10:32:33.55   & CTTS           & dipper \\ 
       &         2MASS J06310484+1027317 & 06:31:04.84 & +10:27:31.81   & YSO            & uncertain\\
       &         2MASS J06311294+1027468 & 06:31:12.94 & +10:27:46.92   & YSO            & dipper \\ 
       &         2MASS J06311457+1027305 & 06:31:14.57 & +10:27:30.59   & YSO            & uncertain \\
       &         2MASS J06312223+1020134 & 06:31:22.24 & +10:20:13.55   & YSO            & const. \\
       &         2MASS J06312967+1023230 & 06:31:29.67 & +10:23:23.07   & YSO            & const. \\
       &         KHA 17                  & 06:31:13.56 & +10:26:59.99   & YSO            & uncertain \\ 
       &         KHA 22                  & 06:31:37.49 & +10:26:58.05   & YSO?           & uncertain \\ \hline
{\#2}  &         V521 Cyg                & 20:58:23.81 & +43:53:11.39   & CTTS           & dipper  \\    
       &         V1538 Cyg               & 20:57:57.51 & +43:50:08.91   & WTTS?          & const.,flare\\  
       &         V1539 Cyg               & 20:57:59.88 & +43:53:26.00   & CTTS           & accretor\\ 
       &         V1716 Cyg               & 20:58:06.12 & +43:53:01.12   & WTTS           & 4.16(9)~d \\
       &         V1929 Cyg               & 20:57:22.25 & +43:57:53.43   & WTTS?          & 0.4263(15)~d \\ 
       &         V1957 Cyg               & 20:57:56.52 & +43:52:36.21   & WTTS?          & 5.2875(80)~d \\
       &         V2051 Cyg               & 20:57:48.81 & +43:50:23.55   & WTTS?          & const.,flare \\ 
       &         {\bf V2493 Cyg}         & 20:58:17.03 & +43:53:43.34   & FUor           & --     \\
       &         LkHa 187                & 20:58:21.54 & +43:53:44.89   & CTTS?          & dipper \\ 
       &         LkHa 189                & 20:58:24.01 & +43:53:54.56   & CTTS?          & 2.42(5)~d \\  
       &         LkHa 191                & 20:59:05.83 & +43:57:03.14   & CTTS           & dipper \\   
       &         2MASS J20573452+4359547 & 20:57:34.54 & +43:59:54.72   & YSO            & uncertain \\
       &         [RGS2011]J205745.44+434845.1 & 20:57:45.44 & +43:48:45.14 & WTTS?       & 0.3692(15)~d \\
       &         2MASS J20580604+4349328 & 20:58:06.06 & +43:49:32.90   & WTTS?          & 1.333(10)~d \\
       &         2MASS J20580885+4346598 & 20:58:08.86 & +43:46:59.84   & CTTS?          & 5.85(1)~d \\   
       &         2MASS J20581082+4353082 & 20:58:10.82 & +43:53:08.25   & CTTS?          & uncertain \\
       &         TYC 3179-925-1          & 20:57:17.48 & +43:49:48.43   & YSO            & uncertain \\ \hline
{\#3}  &         BM And                  & 23:37:38.48 & +48:24:11.84   & CTTS           & dipper  \\ \hline
{\#4}  &         GH Tau                  & 04:33:06.22 & +24:09:33.63   & CTTS           & dip/acc \\  
       &         GI Tau                  & 04:33:34.06 & +24:21:17.07   & CTTS           & dip/acc \\
       &         GK Tau                  & 04:33:34.56 & +24:21:05.85   & CTTS           & dip/acc \\            
       &         V807 Tau                & 04:33:06.63 & +24:09:55.04   & CTTS           & accretor\\ \hline
}

\MakeTable{cccccc}{14cm}{Table 1 - continuation}
{\hline\hline 
Field     & Stars         & RA [hh:mm:ss]   & DEC [$^{\circ}:':''$] & Classification & Variability  \\ \hline \hline  
{\#5}     & EH Cep                  & 21:03:24.39 &	+67:59:06.52 & CTTS           & dipper  \\
          & FU Cep                  & 21:01:46.75 &	+68:08:45.24 & CTTS           & accretor\\
          & FW Cep                  & 21:02:33.01 &	+68:07:29.10 & CTTS           & dipper  \\
          & HZ Cep                  & 21:01:36.2  &	+68:08:22.00 & CTTS           & accretor\\
          & SX Cep                  & 21:01:37.57 &	+68:11:30.95 & WTTS?          & uncertain\\ 
          & LkHa 428 N+S            & 21:02:28.30 &	+68:03:28.52 & CTTS?          & uncertain\\ 
          & 2MASS J21005550+6811273 & 21:00:55.56 &	+68:11:27.21 & CTTS?          & uncertain \\ 
          & 2MASS J21005808+6809382 & 21:00:58.13 &	+68:09:38.21 & CTTS?          & uncertain \\
          & 2MASS J21010367+6813092 & 21:01:03.72 &	+68:13:09.26 & CTTS?          & dipper? \\
          & 2MASS J21011252+6810195 & 21:01:12.53 &	+68:10:19.52 & CTTS?          & uncertain \\ 
          & 2MASS J21012706+6810381 & 21:01:27.06 &	+68:10:38.11 & CTTS?          & uncertain \\ 
          & 2MASS J21014250+6812572 & 21:01:42.51 &	+68:12:57.25 & CTTS?          & uncertain \\
          & 2MASS J21014358+6809361 & 21:01:43.59 &	+68:09:36.18 & WTTS?          & uncertain \\
          & 2MASS J21025943+6806322 & 21:02:59.44 &	+68:06:32.24 & CTTS?          & uncertain \\
          & unnamed            & 21:01:05.19 & +68:04:21.79 & CTTS?      & dipper \\ \hline
{\#6}     & DI Cep             & 22:56:11.54 & +58:40:01.77 & CTTS       & accretor     \\ \hline
{\#7}     & LkHa 156           & 20:51:26.99 & +44:13:15.45 & CTTS?      & uncertain    \\ 
          & LkHa 160           & 20:51:41.42 & +44:15:07.10 & CTTS?      & accretor?    \\ 
          & LkHa 161           & 20:51:41.91 & +44:16:08.20 & CTTS?      & uncertain    \\  
          & LkHa 163           & 20:51:58.65 & +44:14:56.75 & CTTS?      & uncertain    \\ 
          & LkHa 168           & 20:52:06.05 & +44:17:16.04 & HAEBE      & uncertain    \\ 
          & LkHa 172           & 20:52:26.76 & +44:17:06.63 & CTTS       & const.?      \\ 
          & LkHa 174           & 20:52:30.89 & +44:20:11.62 & CTTS?      & uncertain    \\ 
          & LkHa 175           & 20:52:34.37 & +44:17:40.26 & CTTS?      & uncertain    \\ 
          & LkHa 176           & 20:52:58.83 & +44:15:03.79 & HAEBE      & uncertain    \\    
          & 2MASS J20523726+4418145 & 20:52:37.27 & +44:18:14.52 & YSO?       & const.?      \\ \hline
{\#8}     & DD Tau             & 04:18:31.13 & +28:16:29.16 & CTTS       & accretor     \\
          & CY Tau             & 04:17:33.73 & +28:20:46.81 & CTTS       & accretor     \\ 
          & CZ Tau             & 04:18:31.59 & +28:16:58.16 & CTTS       & uncertain    \\
          & V892 Tau           & 04:18:40.61 & +28:19:15.64 & CTTS       & const.       \\
          & V1023 Tau          & 04:18:47.03 & +28:20:07.49 & CTTS       & accretor     \\ \hline
{\#9}     & LkHa 228           & 20:24:31.02 & +42:16:05.07 & CTTS       & uncertain    \\
          & HBC 694            & 20:24:29.54 & +42:14:02.00 & HAEBE      & dipper?      \\
          & {\bf V1515 Cyg}    & 20:23:48.02 & +42:12:25.78 & FUor       & --           \\  
          & HBHA 4202-32       & 20:24:23.05 & +42:20:02.82 & HAEBE      & dipper?      \\
          & [D75b] Em*20-080   & 20:24:22.35 & +42:16:53.80 & HAEBE      & const.       \\
          & V1383 Cyg          & 20:24:11.20 & +42:17:09.10 & CTTS?      & accretor     \\
          & V1385 Cyg          & 20:24:39.59 & +42:21:09.02 & CTTS?      & uncertain    \\ \hline \hline
}

\MakeTable{cccccc}{14cm}{Table 1 - continuation}
{\hline\hline 
Field     & Stars    & RA [hh:mm:ss]   & DEC [$^{\circ}:':''$]  & Classification & Variability  \\ \hline \hline
{\#10}    & HI Ori             & 05:31:23.59 & +12:09:43.83 & CTTS           & dipper       \\
          & HK Ori             & 05:31:28.05 & +12:09:10.15 & CTTS?          & dipper       \\ 
          & HM Ori             & 05:31:47.80 & +12:18:08.20 & CTTS?          & dipper       \\
          & V448 Ori           & 05:30:51.70 & +12:08:36.73 & CTTS           & accretor     \\
          & V450 Ori           & 05:31:24.55 & +12:12:11.01 & CTTS           & dipper       \\ 
          & 2MASS J05315128+1216208 & 05:31:51.29 &	+12:16:20.75 & CTTS      & dipper       \\ 
          & [DM99] 135         & 05:31:08.02 & +12:06:06.37  & CTTS?          & const.       \\
          & [DM99] 136         & 05:31:15.50 & +12:11:23.70 & CTTS           & dipper       \\
          & [DM99] 142         & 05:31:21.66 & +12:05:47.69 & CTTS?          & uncertain    \\
          & [DM99] 152         & 05:31:40.48 & +12:10:46.88 & WTTS?          & uncertain    \\
          & [DM99] 161         & 05:31:58.48 & +12:22:47.36 & WTTS?          & const.       \\
          & [DM99] 164         & 05:31:59.94 & +12:08:06.80 & WTTS?          & uncertain    \\ \hline    
{\#11}    & V350 Cep           & 21:42:59.99 & +66:11:27.90 & CTTS           & uncertain    \\ 
          & V361 Cep           & 21:42:50.18 & +66:06:35.18 & HAEBE          & uncertain    \\
          & V373 Cep           & 21:43:06.82 & +66:06:54.24  & HAEBE          & uncertain    \\ 
          & BD+65 1636         & 21:42:46.04 & +66:05:13.80 & HAEBE          & const.       \\  
          & BD+65 1638         & 21:42:58.56 &  +66:06:10.56 & HAEBE          & uncertain    \\  
          & 2MASS J21424031+6610069 & 21:42:40.31 &	+66:10:07.00 & YSO       & const.       \\
          & 2MASS J21424705+6604578 & 21:42:47.06 &	+66:04:57.90 & YSO       & dipper?      \\
          & 2MASS J21424707+6610512 & 21:42:47.07 &	+66:10:51.30 & YSO?      & const.       \\ 
          & 2MASS J21425349+6608053 & 21:42:53.48 &	+66:08:05.29 & YSO?      & uncertain    \\
          & 2MASS J21430168+6607089 & 21:43:01.68 &	+66:07:08.96 & YSO       & const.       \\
          & 2MASS J21431161+6609114 & 21:43:11.59 &	+66:09:11.46 & YSO?      & uncertain    \\
          & 2MASS J21432695+6609365 & 21:43:26.95 &	+66:09:36.45 & YSO       & const.       \\
          & 2MASS J21432932+6603319 & 21:43:29.34 &	+66:03:31.96 & YSO       & const.       \\
          & 2MASS J21433625+6611329 & 21:43:36.25 &	+66:11:32.89 & YSO       & uncertain    \\
          & 2MASS J21435035+6608477 & 21:43:50.36 &	+66:08:47.65 & YSO?      & uncertain    \\
          & TYC 4261-842-1          & 21:42:26.93 & +66:07:42.60 & YSO?      & const.       \\ \hline
{\#12}    & VV Ser                  & 18:28:47.86 &	+00:08:39.92 & HAEBE     & dipper       \\                           
          & 2MASS J18282143+0010409 & 18:28:21.45 &	+00:10:41.14 & CTTS?     & const.       \\
          & 2MASS J18285808+0017243 & 18:28:58.09 &	+00:17:24.38 & CTTS?     & const.       \\
          & 2MASS J18290025+0016578 & 18:29:00.25 &	+00:16:57.85 & CTTS?     & const.       \\
          & 2MASS J18290394+0020212 & 18:29:03.94 &	+00:20:21.26 & CTTS?     & accretor     \\ 
          & 2MASS J18290576+0022324 & 18:29:05.77 &	+00:22:32.41 & YSO       & const.       \\ \hline
{\#13}    & LkHa 208                & 06:07:49.53 &	+18:39:26.49 & HAEBE     & uncertain    \\
          & LkHa 209                & 06:08:14.39 &	+18:37:25.00 & CTTS?     & uncertain    \\ \hline
{\#14}    & DI Tau                  & 04:29:42.47 &	+26:32:49.12 & CTTS      & const.       \\ 
          & DH Tau                  & 04:29:41.56 & +26:32:58.27 & CTTS      & accretor     \\        
          & JH 507                  & 04:29:20.70 &	+26:33:40.44 & CTTS      & accretor     \\ 
          & 2MASS J04293623+2634238 & 04:29:36.24 &	+26:34:23.49 & CTTS      & const.       \\ \hline      
{\#15}    & V395 Cep                & 23:20:52.12 &	+74:14:07.08 & CTTS      & 3.42(1)~d    \\ \hline\hline
}

\MakeTable{cccccc}{14cm}{Table 1 - continuation. Stars observed with 20-cm telescope.}
{\hline\hline 
Field  & Stars              & RA [hh:mm:ss]   & DEC [$^{\circ}:':''$]  & Classification & Variability  \\ \hline \hline
{\#16} & BH~Cep             & 22:01:42.87 & +69:44:36.42 & HAEBE          & uncertain    \\
{\#17} & HD~174571          & 18:50:47.17 & +08:42:10.09 & HAEBE          & uncertain    \\ 
{\#18} & HD~190073          & 20:03:02.51 &	+05:44:16.66 & HAEBE          & uncertain    \\
{\#19} & HD~203024          & 21:16:03.05 &	+68:54:52.10 & HAEBE          & uncertain    \\
{\#20} & V594~Cas           & 00:43:18.26 &	+61:54:40.14 & HAEBE          & uncertain    \\ \hline\hline    
}

The vast majority of the {\it MSO} data was collected with the 60~cm Carl-Zeiss telescope. 
Originally in Cassegrain configuration, the telescope was later modified to operate in the primary focus 
for the purposes of M-dwarfs survey (Baran et al., 2011). 
Three CCD cameras and four photometric systems were utilized. 
Between March, 2012 -- September, 2013, we used the SBIG ST10XME CCD camera equipped with the Johnson-Morgan 
filters provided by SBIG.
In October, 2013, the Apogee~Alta U42 CCD camera, equipped with a set of Johnson-Bessel filters 
manufactured by Custom Scientific was installed on the telescope. 
In August, 2014, we started to collect our data through Sloan filters, which have several advantages over 
the former filter sets: the pass bands of Sloan system are practically independent of each other, 
and transmission efficiency of every single filter is higher. 
At the time the Apogee~ASPEN~CG47 camera replaced the above one. 
Only for V395~Cep, which is about 10 magnitude bright star, the data were collected 
through Str{\"o}mgren filters. 
Observations of five HAEBE stars, which are too bright for the 60-cm telescope, were obtained by means 
of the remotely operated 20-cm Ritchey-Chretien {\it MSO} telescope, equipped with SBIG ST10- XME CCD 
camera and the Johnson-Morgan filters provided by SBIG. 
We present detailed log of our observations in Tables 5-7. 

In order to make a balance between information about colour indices, which is necessary to characterise 
mechanisms leading to specific light variations, and high cadence (30-90~sec) required 
for sufficient sampling of planetary transits and rapid accretion-induced events (Figure 1), 
most of our targets was observed in {\it BVI}, {\it VI}, {\it gi} or {\it gri} filters only. 
In order to find transits similar to CVSO~30, we made attempts to get full phase coverage for periods 
up to about 0.8~day. 
In practice, this requirement was met for most fields after 4-11 hours of continuous observations 
during 5-20 different nights (depending on the season), non-uniformly distributed over the time 
span of a few weeks. 
Fields containing particularly interesting targets were observed for one or two more seasons. 

All frames were {\it bias}, {\it dark} and {\it flat-field} calibrated in a standard way within 
the {\sc MIDAS} package (Warmels, 1991). 
Aperture photometry was obtained by means of {\sc C-Munipack} software (Motl, 2011) which utilizes 
{\sc DAOPHOT} package (Stetson, 1987). 
As the weather conditions which are prevailing in Poland do not allow for frequent and accurate 
calibrations to the standard photometric systems, we decided to left all data in instrumental systems. 
Light curves of all stars showing any variability occurring 
either in short, long, or in both time scales are shown in Figures 6-13\footnote{Only data obtained 
by means of the 60-cm telescope are shown -- we decided to skip data gathered by means of the 20-cm 
telescope due to unstable {\it flat-field}. For the same reason, we omit the respective upper values 
of covered periods in Table~7}. 
These light curves were also left uncorrected for the first and the second-order atmospheric 
extinction terms. 
However, using approximate colour extinction coefficients determined for {\it MSO}, we corrected 
the data used for calculation of colour-magnitude diagrams (Figs.~2,3,5) and phase-ordered 
observations in periodic stars (Fig.~4).
We estimate that the uncertainties associated with inaccurate (up to 30\%) determination 
of the color exinction values should not exceed 0.04~mag for stars having colour index redder/bluer 
than comparison star(s) by as much as 2~mag. 
The uncertainty should not exceed 0.01~mag for typical colour index difference of 0.5~mag. 
We note that these estimates relate to upper error values: they are calculated for 
$B-V$ and $g-r$ colour indices, which are most sensitive on innacurate colour extinction correction, and 
airmass 2 -- the value which was rather exception than the rule during our observations.

Based on detailed visual inspection of the light curves shown in Figures~6-13, we have assigned 
our targets to one of the three major groups of variability, as indicated in column 3 in Tables~1-4. 
The group of ''dippers'' consists of stars observed close to the disc plane and their 
variability is owing to dusty disc warps, acting as occulting screens as they rotate around the star. 
Light curves of ''accretors'' are dominated, or at least occasionally affected, by chaotic hot spots 
created during inhomogeneous accretion. 
Finally, light curves of ''periodic'' stars are produced either by rotational modulation in visibiliy 
of persistent photospheric cold spots (in WTTS), or relatively stable hot spots (in CTTS), produced 
at the footprints of stable accretion funnel(s). 
We stress that classification given in this paper for stars previously not intensely observed by other 
authors do base on our sparse data only and may be the subject to changes, e.g. thanks 
to the ongoing {\it TESS} mission and long-term monitoring programs like {\it ROTOR} (Grankin et al., 2008) 
and that conducted at the Rozhen National Astronomical Observatory of Bulgaria (Ibryamov et al., 2014). 

\section{Results and discussion}

\subsection{The search for planetary-like transits}

In the first step, we examined all light curves collected continuously for at least several hours 
for signatures of any brightness dips caused by transiting Jupiter-like planets or low-mass stellar 
companions, but with a negative result. 
Each light curve was examined several times: first immediately after 
the night, using preliminary set of comparison stars and later, during more detailed analyses 
using light curves received with final, more appropriate comparison stars (see in column 2 in Tables 5-7). 
In these searches we based on estimates of Neuhauser et al. (2011) that young ''inflated'' 
planets having 1-10 Jupiter masses and orbiting typical T~Tauri-type stars should produce 
0.02-0.07~mag dips, which would be directly visible in our data. 
Even though it is a promising opportunity, we did not apply any automatic procedure for search 
for even more shallow transits. 
This is owing to numerous little ($\lesssim 0.005$~mag), accretion- and/or colour extinction-related 
variations (the latter caused by passing thin cirrus clouds), both leading to false-positive detections.\\ 
In the first column in Tables 5-7, just below the field numbers, we list the appropriate upper values of periods 
almost completely covered by our observations.
We stress that single-site observations disabled us to get good phase coverage for periods 
equal to $0.5\pm0.01$~d. 
In addition, only limited amount of stars (see in the second column in Tables~1-4), for which 
photometric quality of 1\% of single observational points was achieved, is suitable for detection 
of 1-5\% transit-like dips. 
As the vast majority of our targets are M-K type dwarfs, which are often additionally heavily 
reddened by dusty environments, in many cases search for the transits was only possible 
in light curves obtained in $i$-bands. 
This band is also least "polluted" by hot spots, which decrease amplitude of potential 
transits. 
Finally, the light curves gathered for targets 1.5-2~mag fainter than mean (artificial) comparison 
star were usually suitable only for studies of large-amplitude light variations.

\begin{figure}[htb]
\centerline{%
\begin{tabular}{l@{\hspace{0.1pc}}}
\includegraphics[width=.5\linewidth]{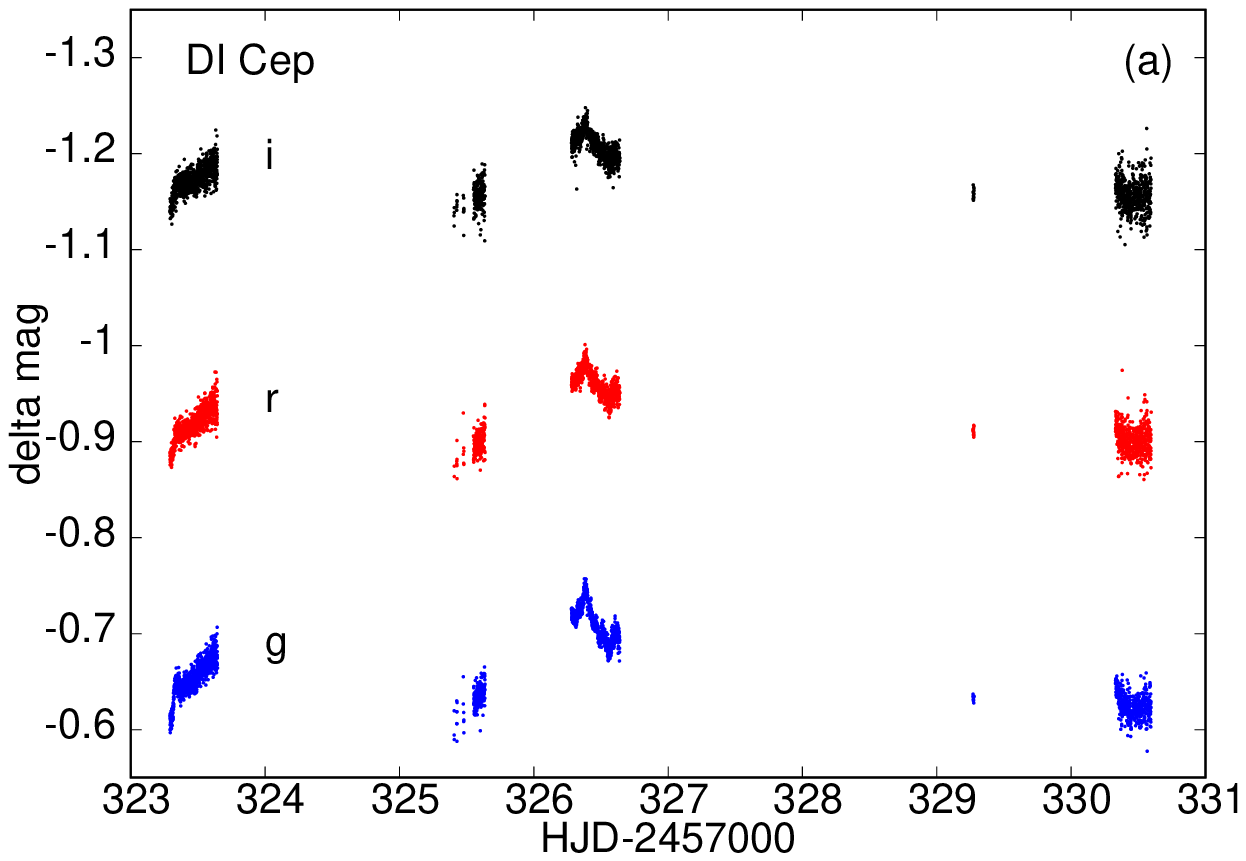} 
\includegraphics[width=.5\linewidth]{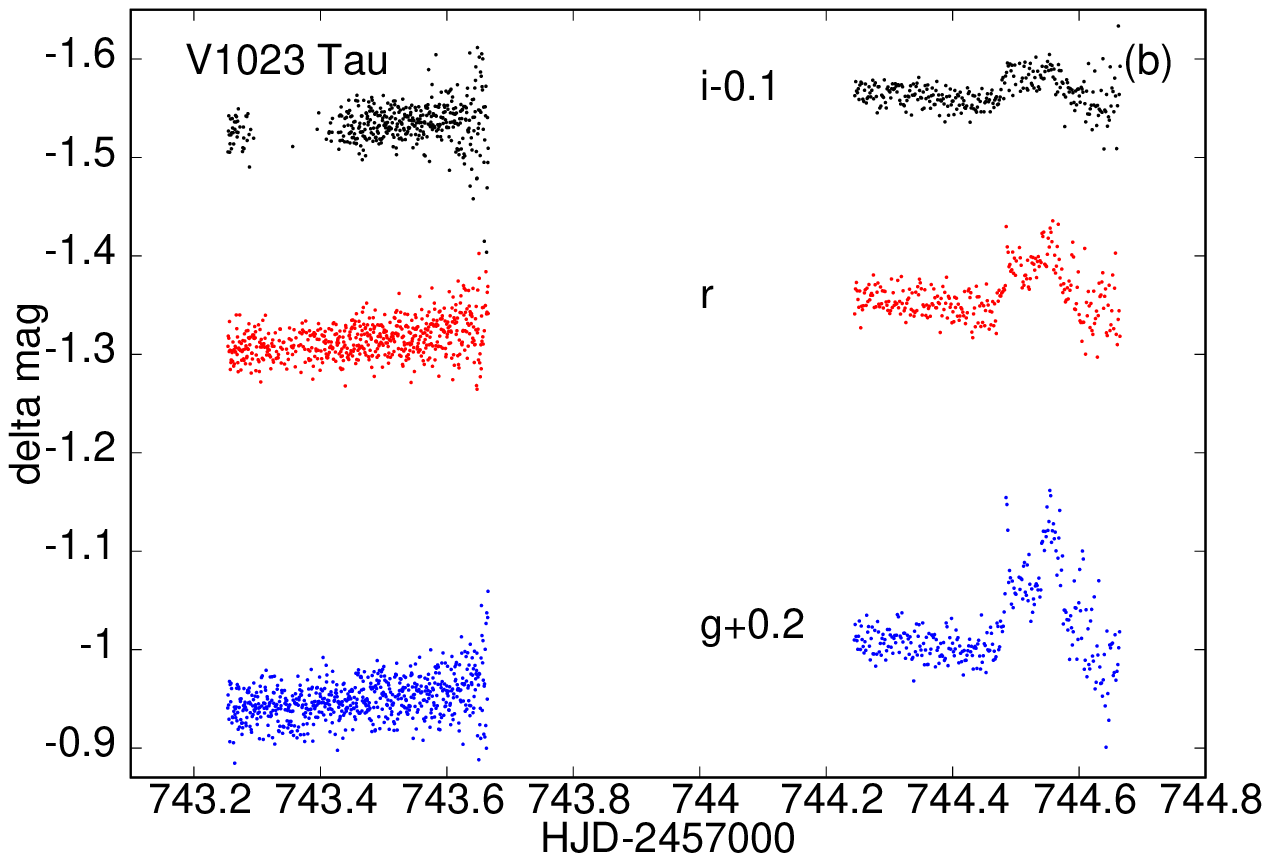}\\
\includegraphics[width=.5\linewidth]{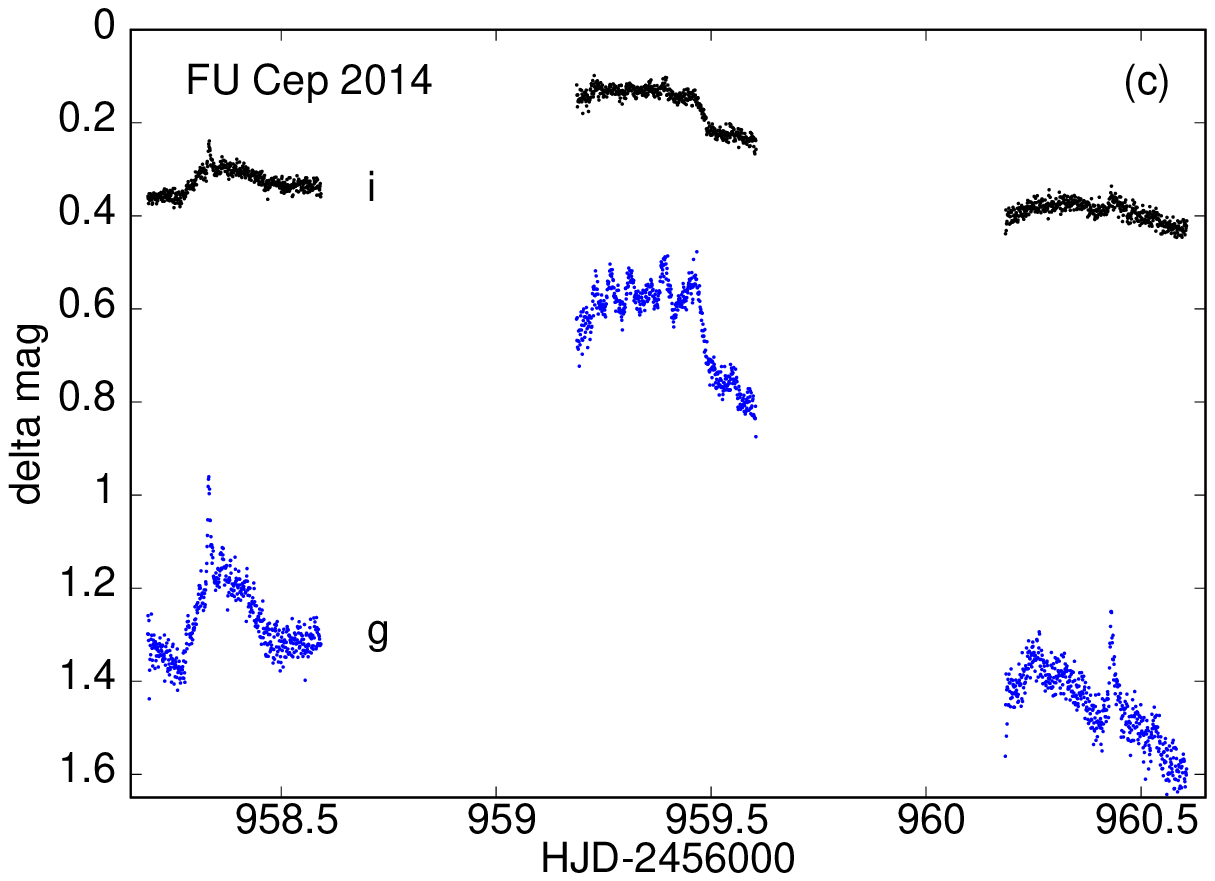}  
\includegraphics[width=.5\linewidth]{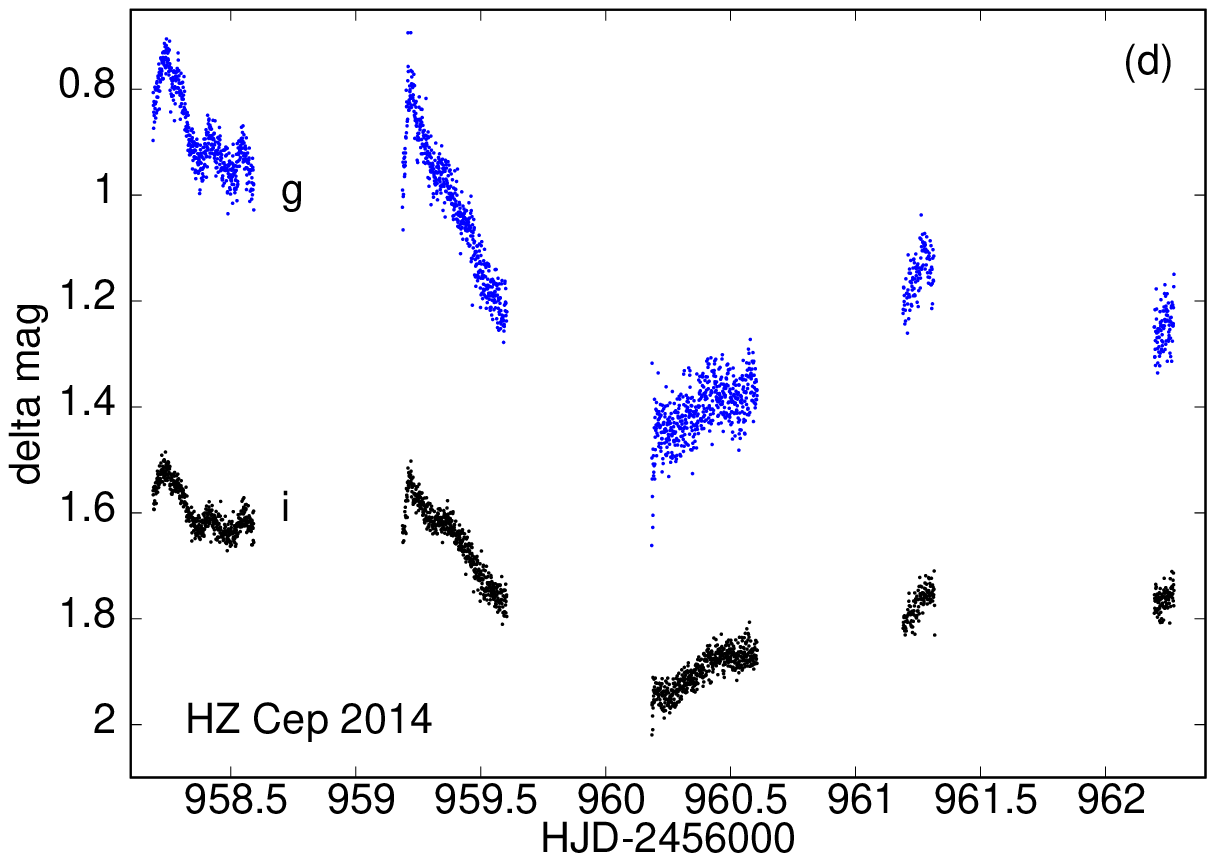}\\
\includegraphics[width=.5\linewidth]{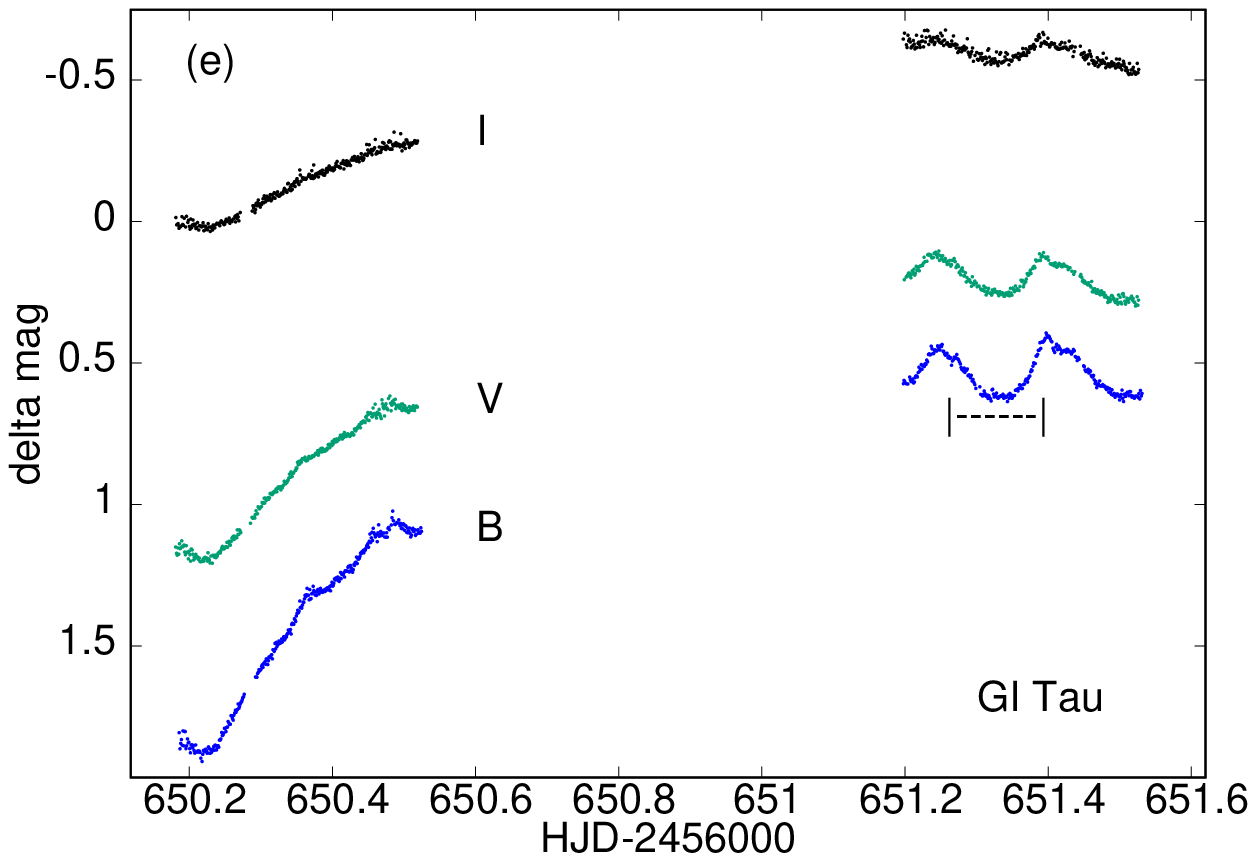}  
\includegraphics[width=.5\linewidth]{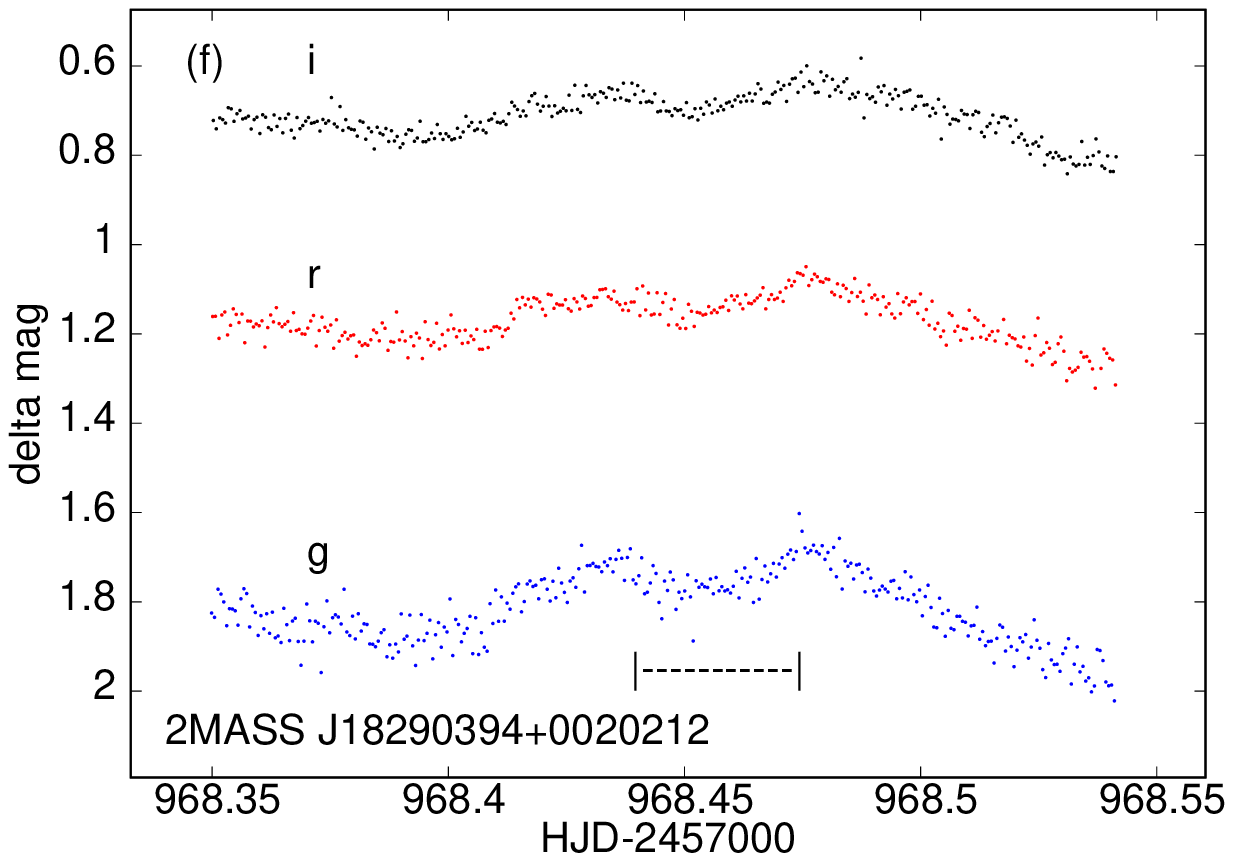}
\end{tabular}}
\FigCap{Light curves of six stars classified as accretors. The best defined oscillatory behavior, 
with the characteristic period of 56.4~min, is visible in the local maximum of FU Cep, but single 
flare-like accretion bursts are also present during other nights. 
The two bottom panels show possible occultations of hot spots during the local brightness maxima 
of two stars, as indicated by marks.}
\end{figure}

\subsection{Accretors}

Most of our targets belong to CTTS, which are constantly transferring disc plasma 
in unstable regime or just show episodes of inhomogeneous accretion. 
Although space-based uninterrupted light curves are best suited for investigation of these phenomena, 
time-coherent variability caused by accretion can sometimes be as short as 10-30~min 
and can be still successfully investigated from the ground (Gullbring et al., 1996; Siwak et al., 2018).\\ 
Stars in which accretion effects dominate in overall brightness variations are indicated 
in the third column in Tables~1-4 as ''accretors''.  
We also include into this group several stars showing smooth brightness variations for most of the time, 
but interrupted by several well-defined ''accretion bursts'' (Fig.~1 a,b,c,d). 
These bursts are of triangular shape and can be well distinguished from other rapid brightness 
variations, e.g. stellar flares. 
As mentioned above, due to the temperature contrast between typical hot spot ($\sim 5000-10000$~K) 
and photosphere ($\sim 4000$~K), the largest amplitudes of these light variations are observed in the blue 
part of the spectrum and decrease towards longer wavelengths. 
This dependence is true both for large-scale and small-scale accretion-induced variability. 

Siwak et al. (2018) showed that time-coherent variability in TW~Hya mimicking oscillatory behavior 
can be as short as 10-30~min. 
Only one star from our sample, namely FU~Cep, is in many aspects similar to TW~Hya: 
its 2014 observations (Fig.1c) showed two single accretion bursts of the duration times 
33~min (at $HJD=2456958.331$) and 56~min (at $HJD=2456960.430$), respectively. 
But the most intriguing finding is the oscillatory behavior during the local brightness 
maximum at $HJD\approx2456959.35$, lasting for the entire maximum plateau. 
To calculate their period first we removed small trend in the maximum 
using low-order polynomial and then we transformed the data to flux units. 
We obtained frequency spectrum which shows single well-defined peak at 56.4~min. 
Interestingly, it is similar in terms of duration to the second aforementioned accretion burst (at $HJD=2456960.430$). 
We have preliminarily proposed to explain these oscillations in TW~Hya as caused 
by post-shock plasma oscillations. 
This conclusion was based on the theoretical result of Matsakos et al. (2013), who presented a range of models, 
among them one with an oblique impacting surface leading to QPO of 15~min. 
However, the period found in FU~Cep is almost four times longer. 
Similarity of the second single accretion burst duration time and these oscillations period 
(56~min) gives hint that perhaps other hypothetical scenario 
(already also considered by us for TW~Hya) can more plausibly explain observed phenomenon: 
one can not exclude the possibility that for some reason small plasma clumps hit the star 
at regular intervals, what mimics oscillatory behavior. 
Theoretical models of this phenomenon are strongly desirable.

\begin{figure}[htb]
\centerline{%
\begin{tabular}{l@{\hspace{0.1pc}}}
\includegraphics[width=.5\linewidth]{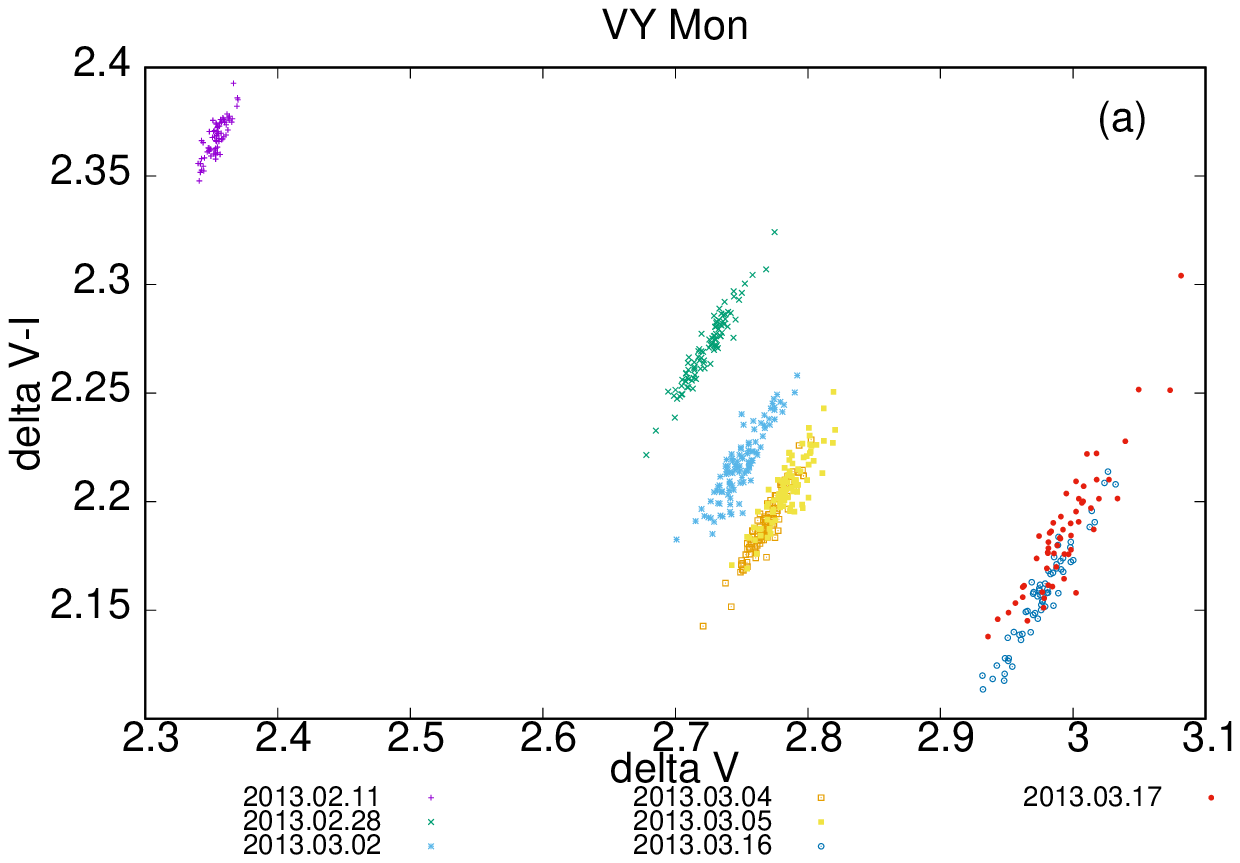} 
\includegraphics[width=.5\linewidth]{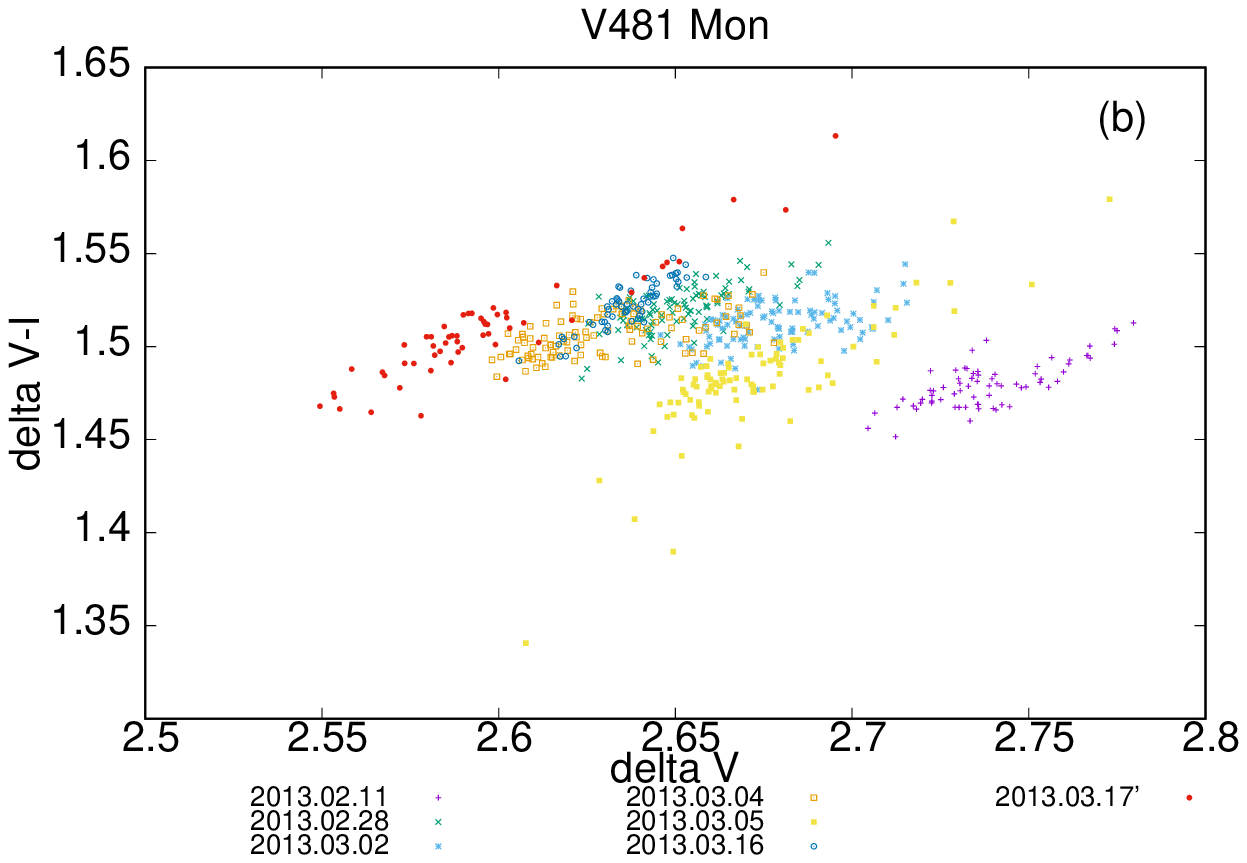}\\
\includegraphics[width=.5\linewidth]{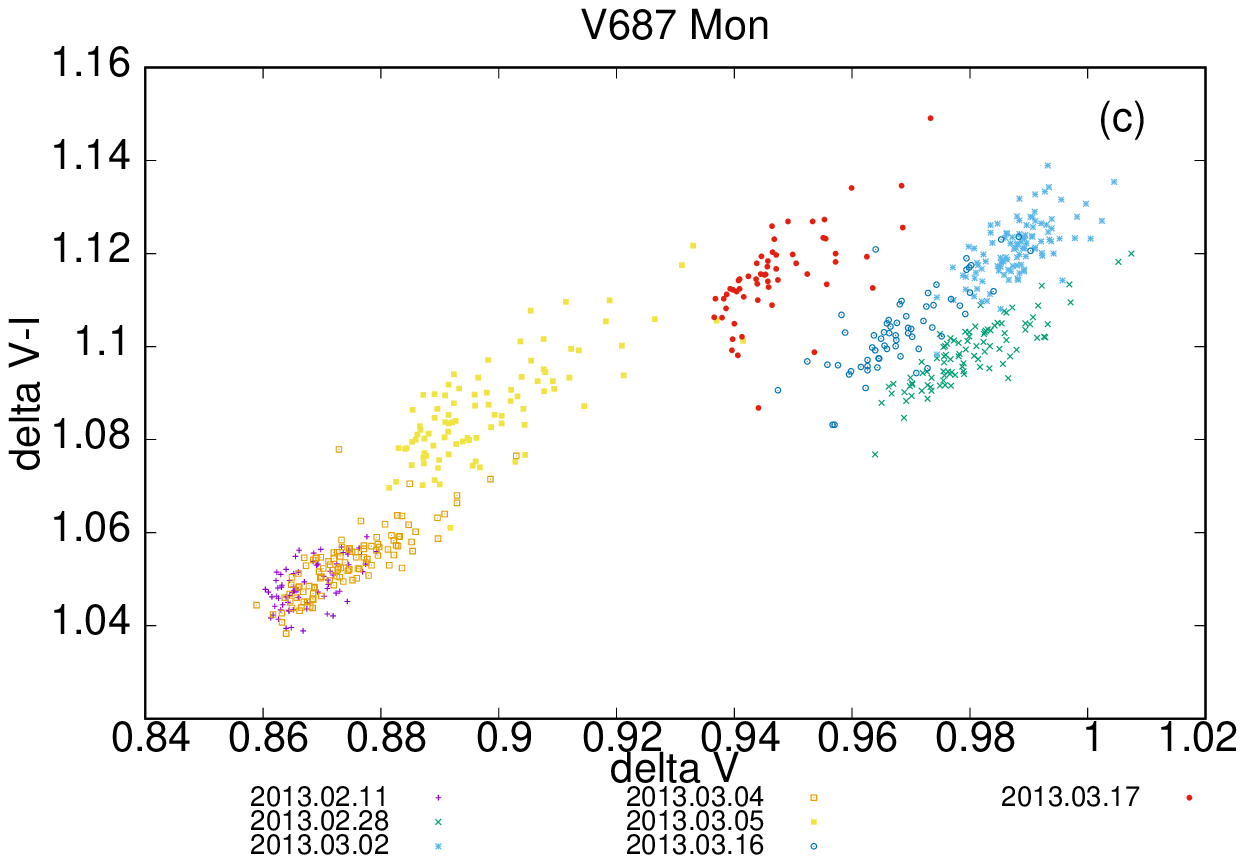}  
\includegraphics[width=.5\linewidth]{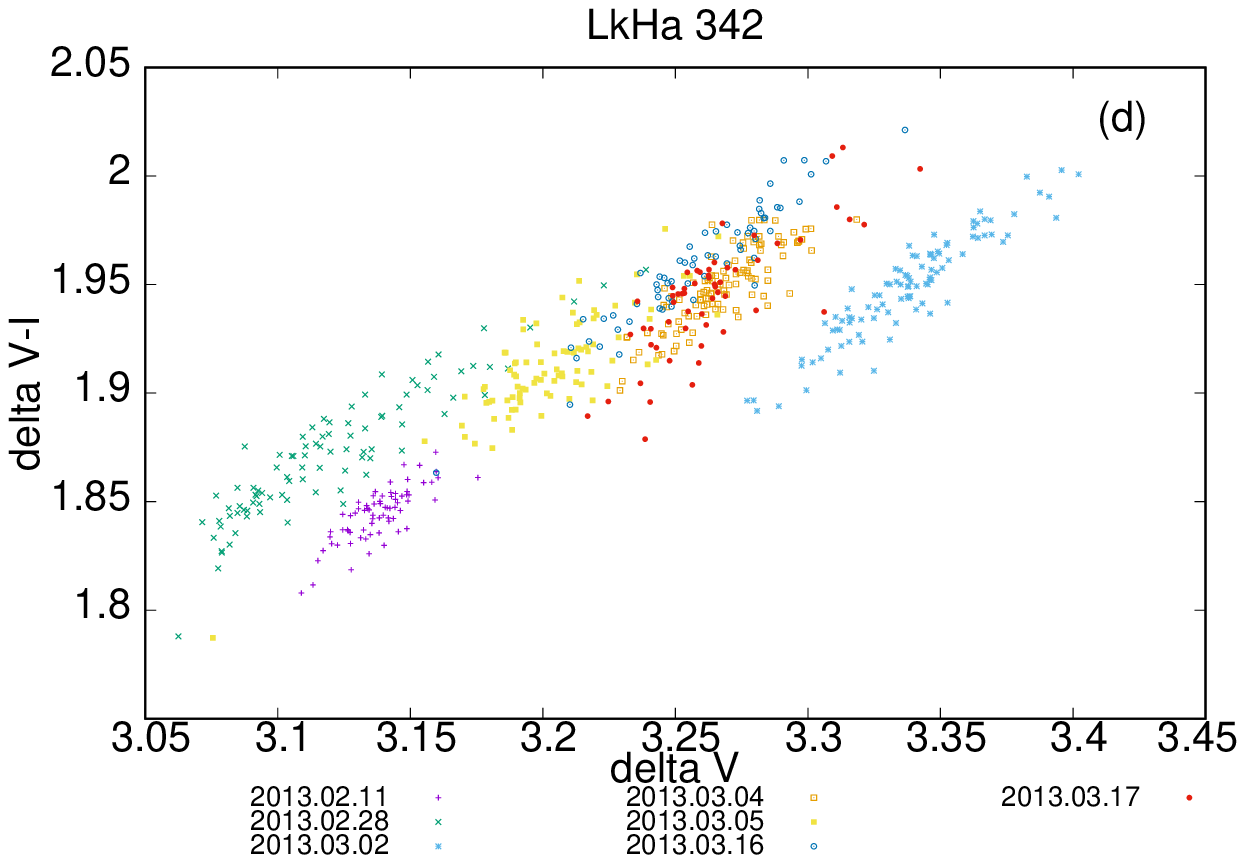}\\
\includegraphics[width=.5\linewidth]{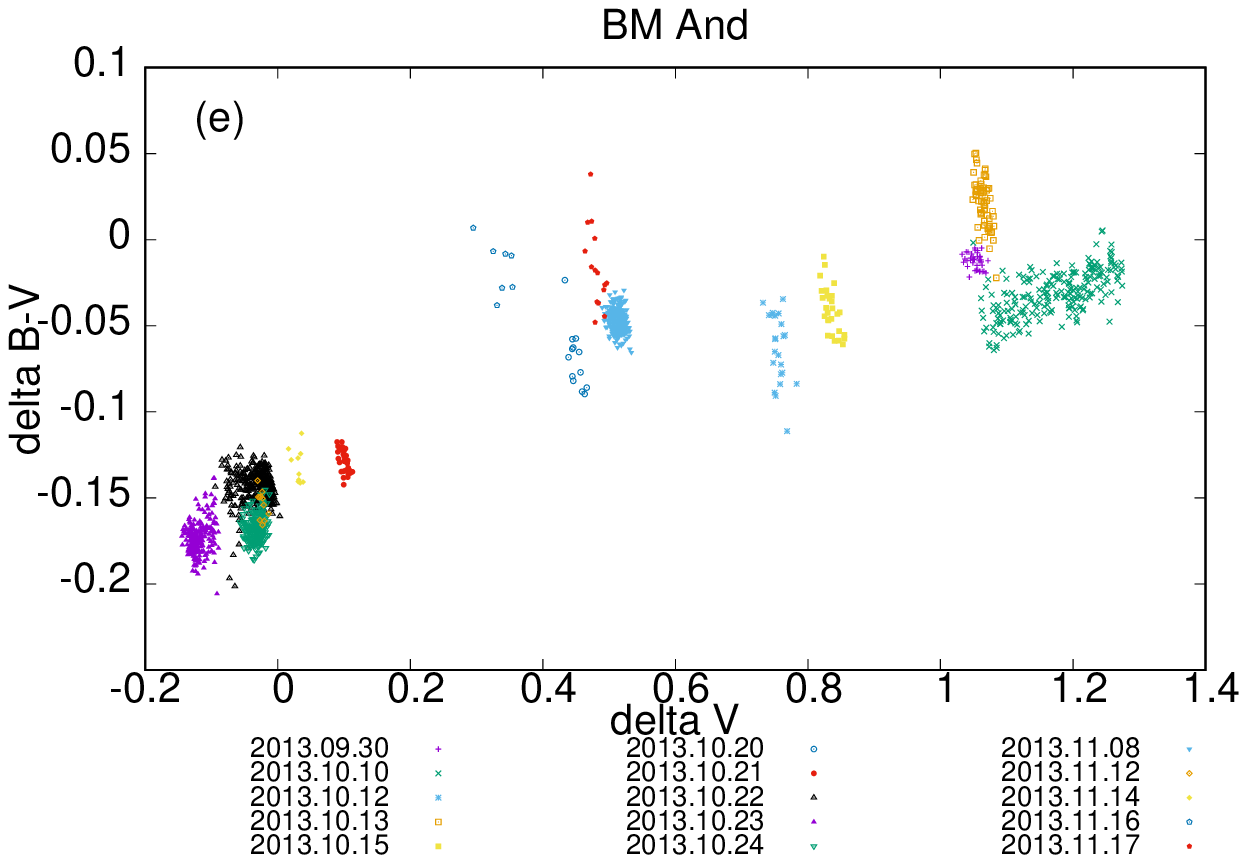}  
\includegraphics[width=.5\linewidth]{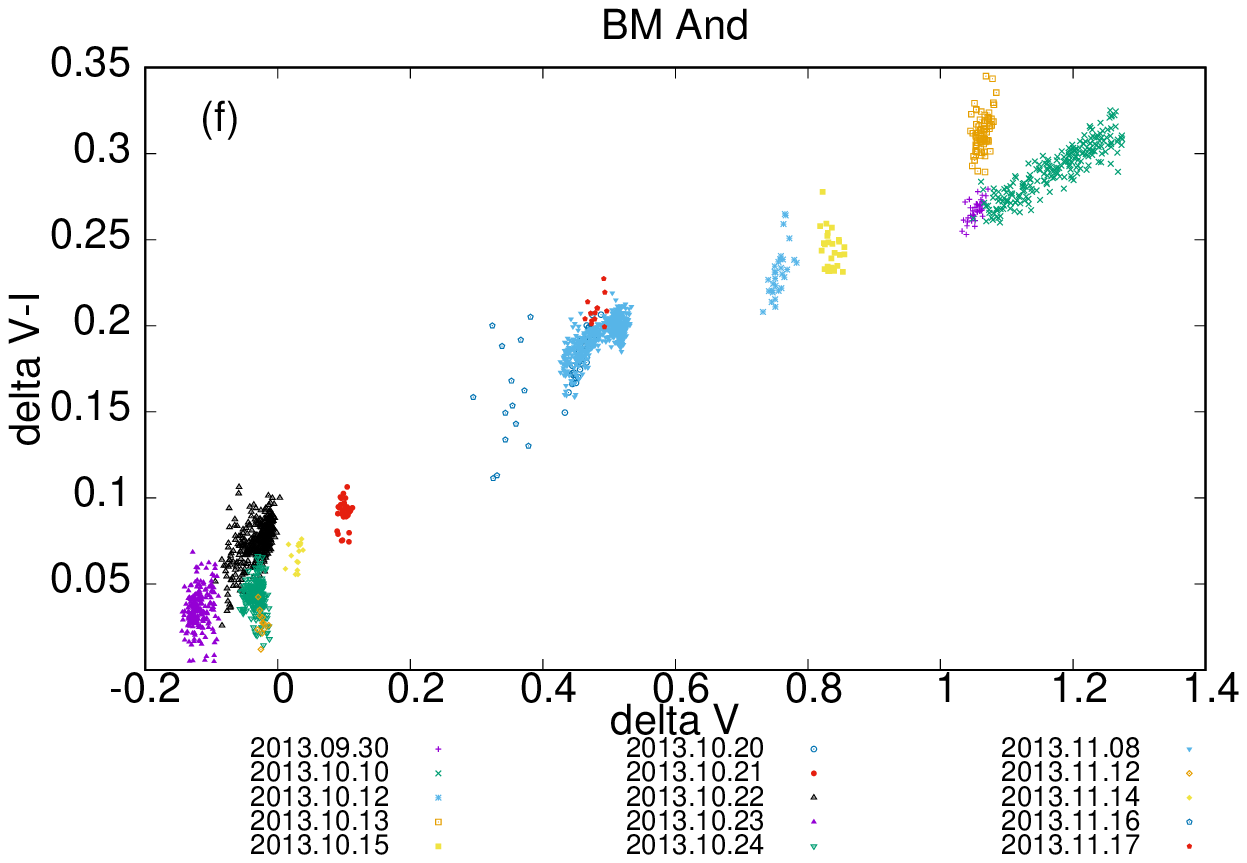}
\end{tabular}}
\FigCap{Colour-magnitude diagrams for VY~Mon, V481~Mon, V687~Mon, LkHa~342 and BM~And.
Data obtained during different nights are marked with different symbols and colours 
(see legends below respective x-axes). 
Only data obtained by means of SBIG camera and filters were used to construct diagrams for 
the first four stars.
Whilst the general trends observed in these diagrams are intrinsic to the stars, the intra-night 
and the nigh-to-night scatter visible in diagrams of first four stars may also be related with uncertainty 
in our colour extinction correction procedure to some extent (see in Section~2).}
\end{figure}

\subsection{Dippers}

\begin{figure}[htb]
\centerline{%
\begin{tabular}{l@{\hspace{0.1pc}}}
\includegraphics[width=.33\linewidth]{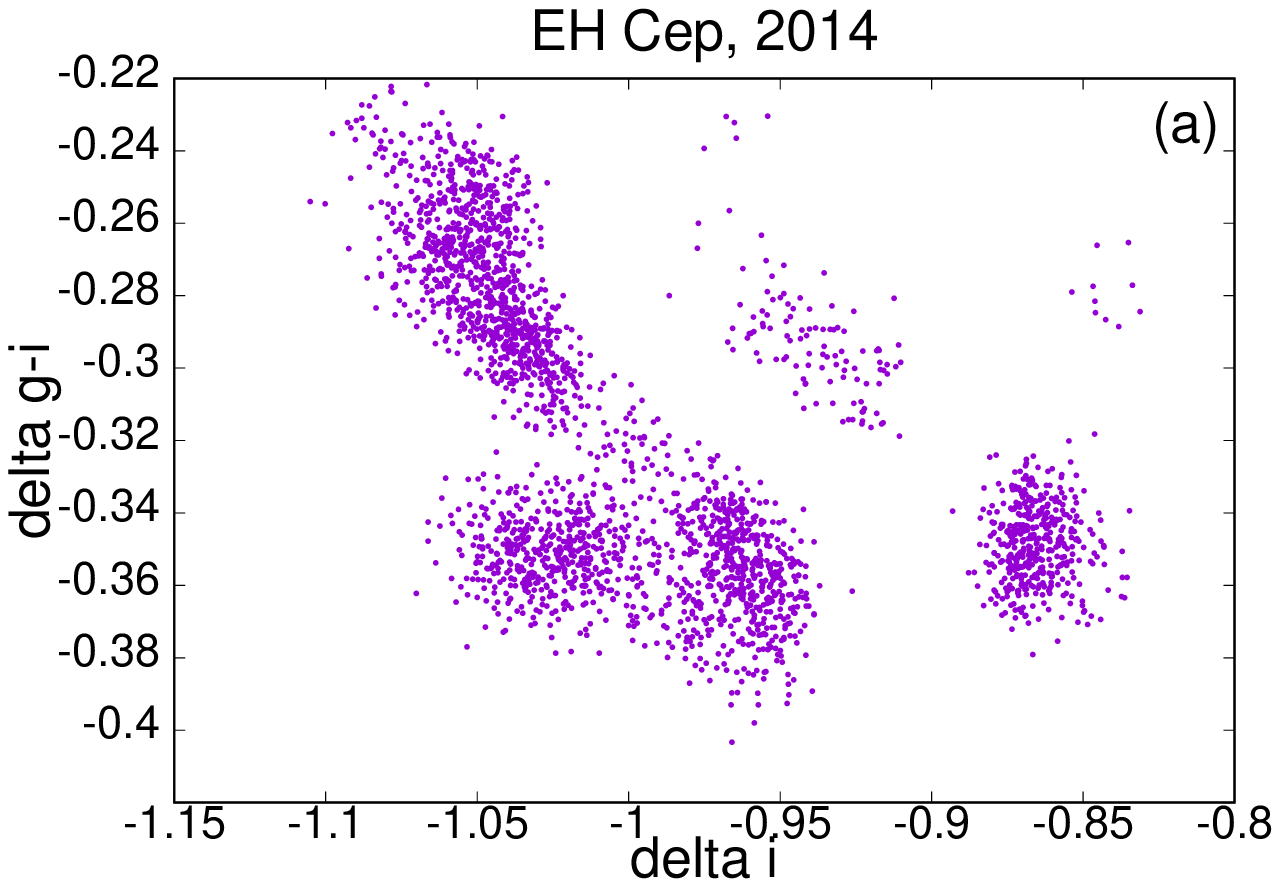}
\includegraphics[width=.33\linewidth]{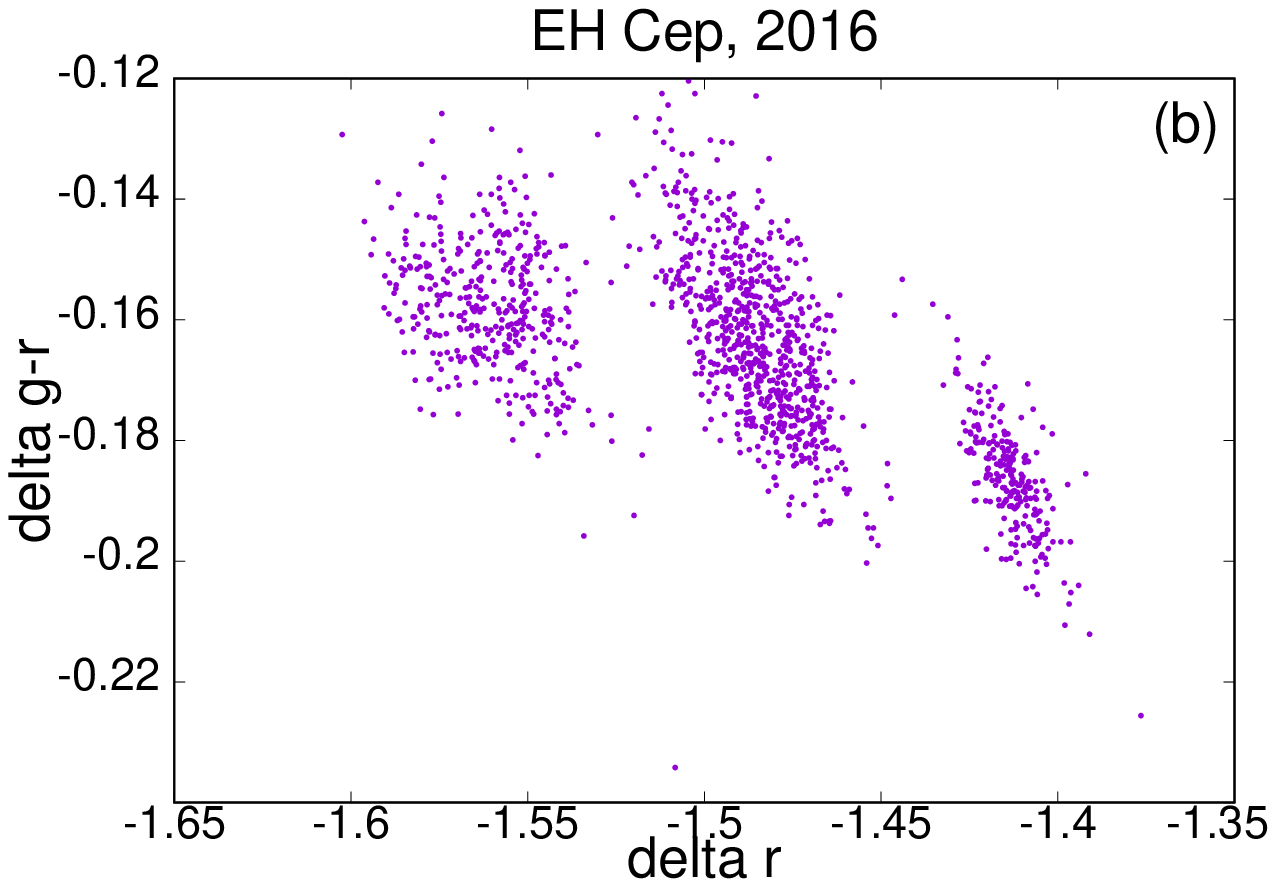}
\includegraphics[width=.33\linewidth]{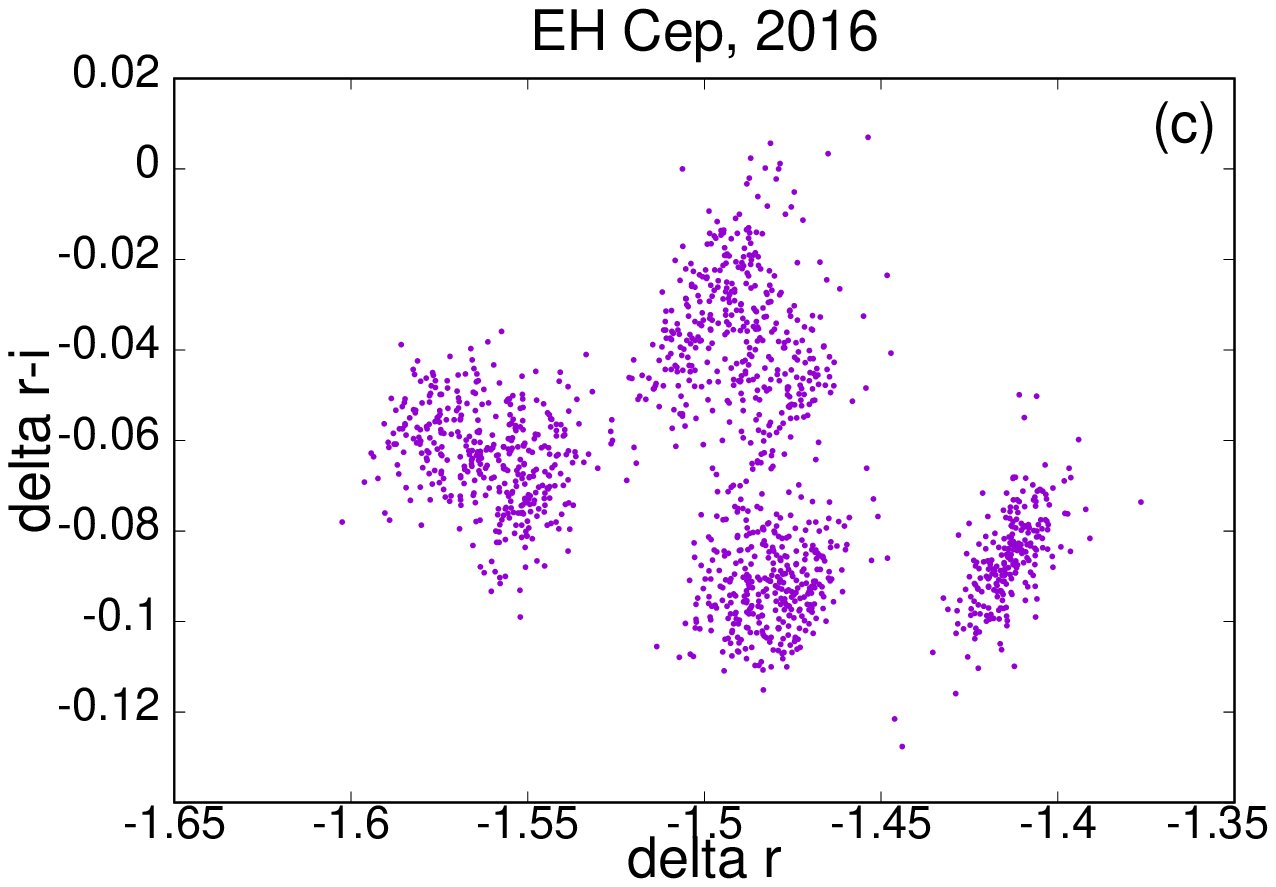}\\  
\includegraphics[width=.33\linewidth]{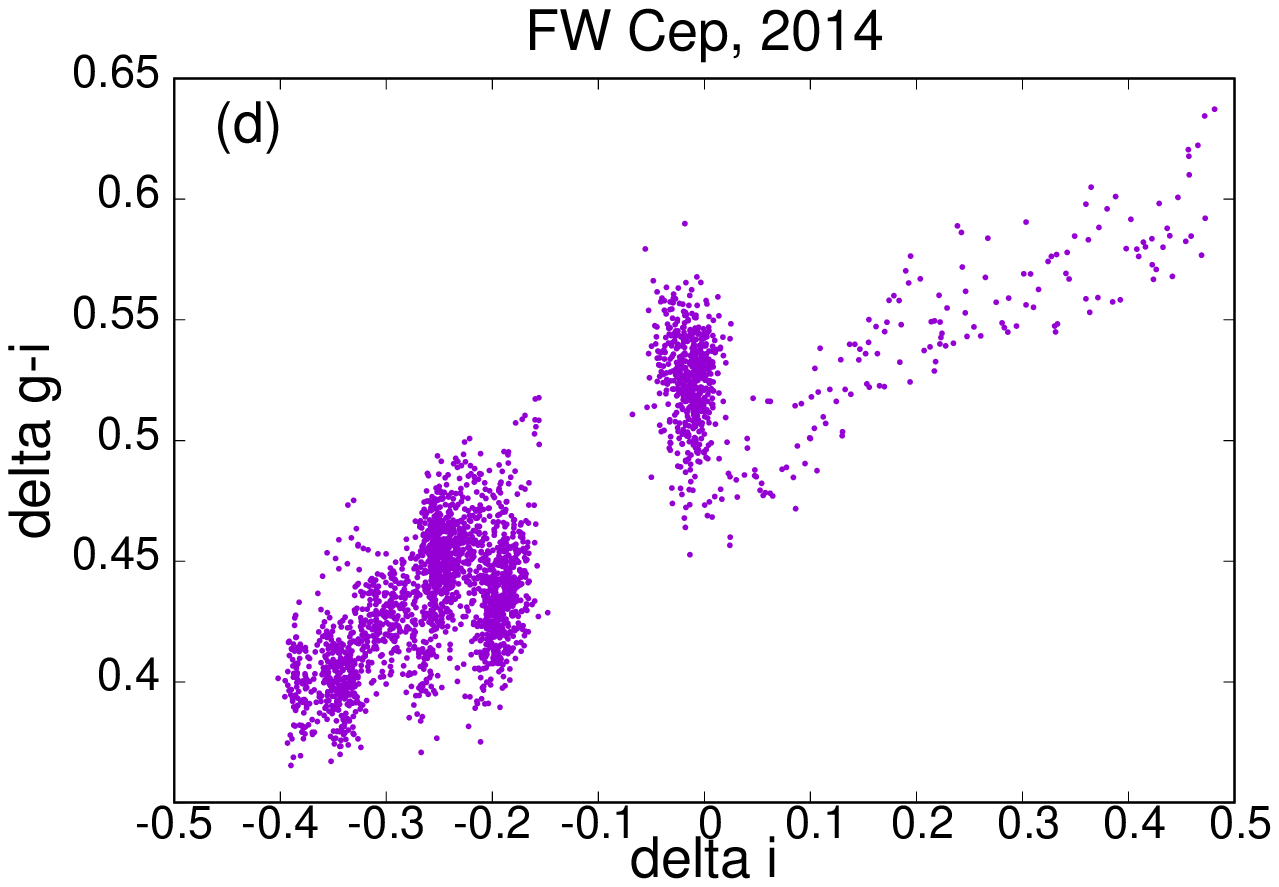}
\includegraphics[width=.33\linewidth]{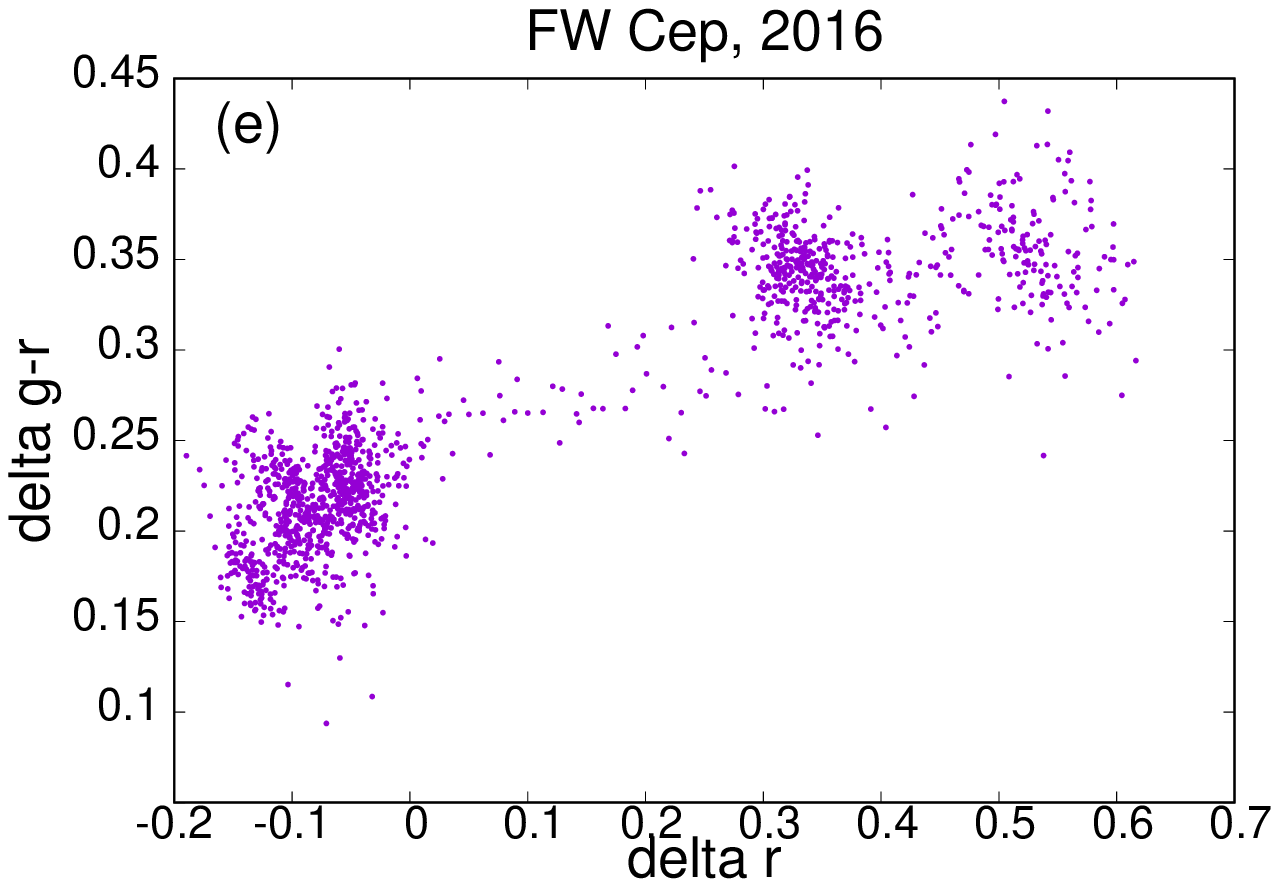}
\includegraphics[width=.33\linewidth]{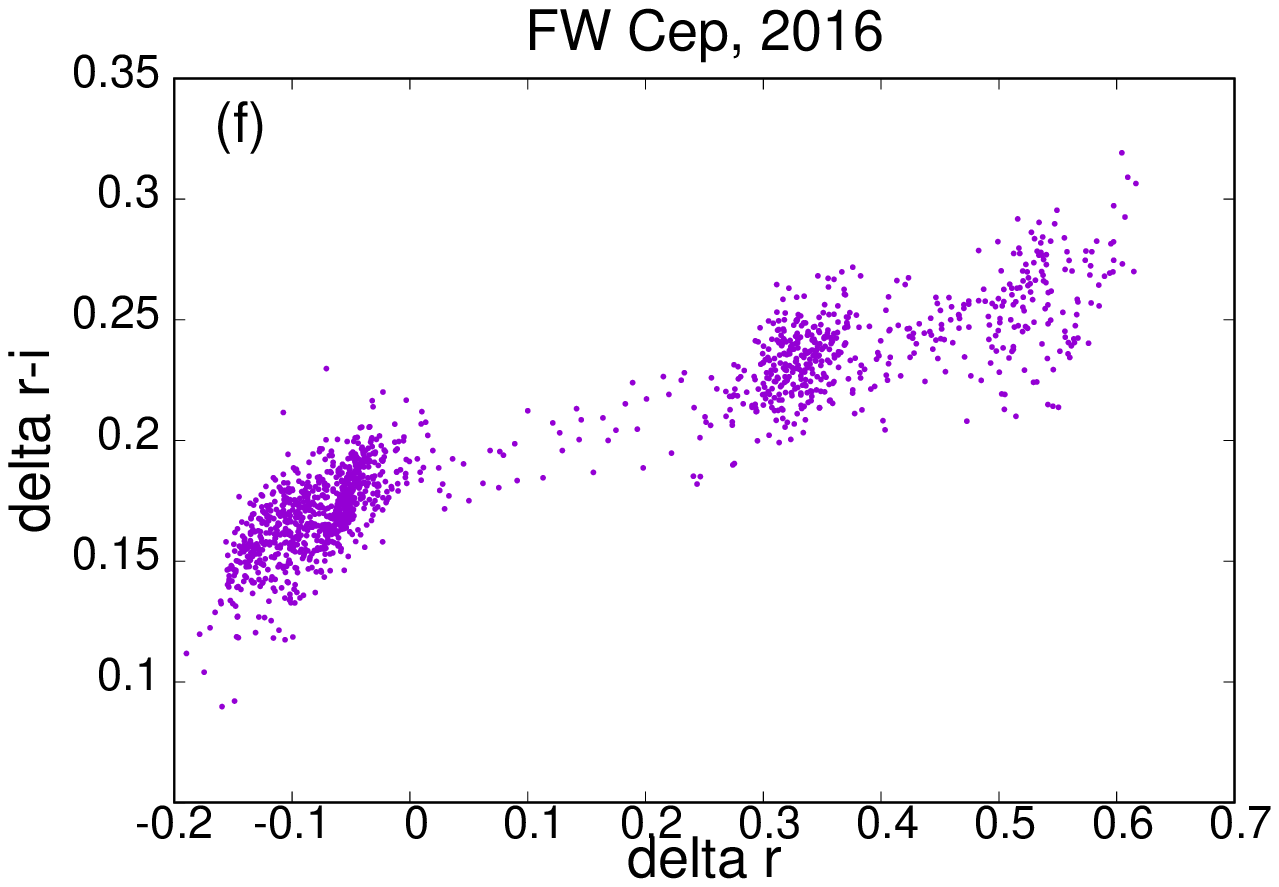}\\
\includegraphics[width=.33\linewidth]{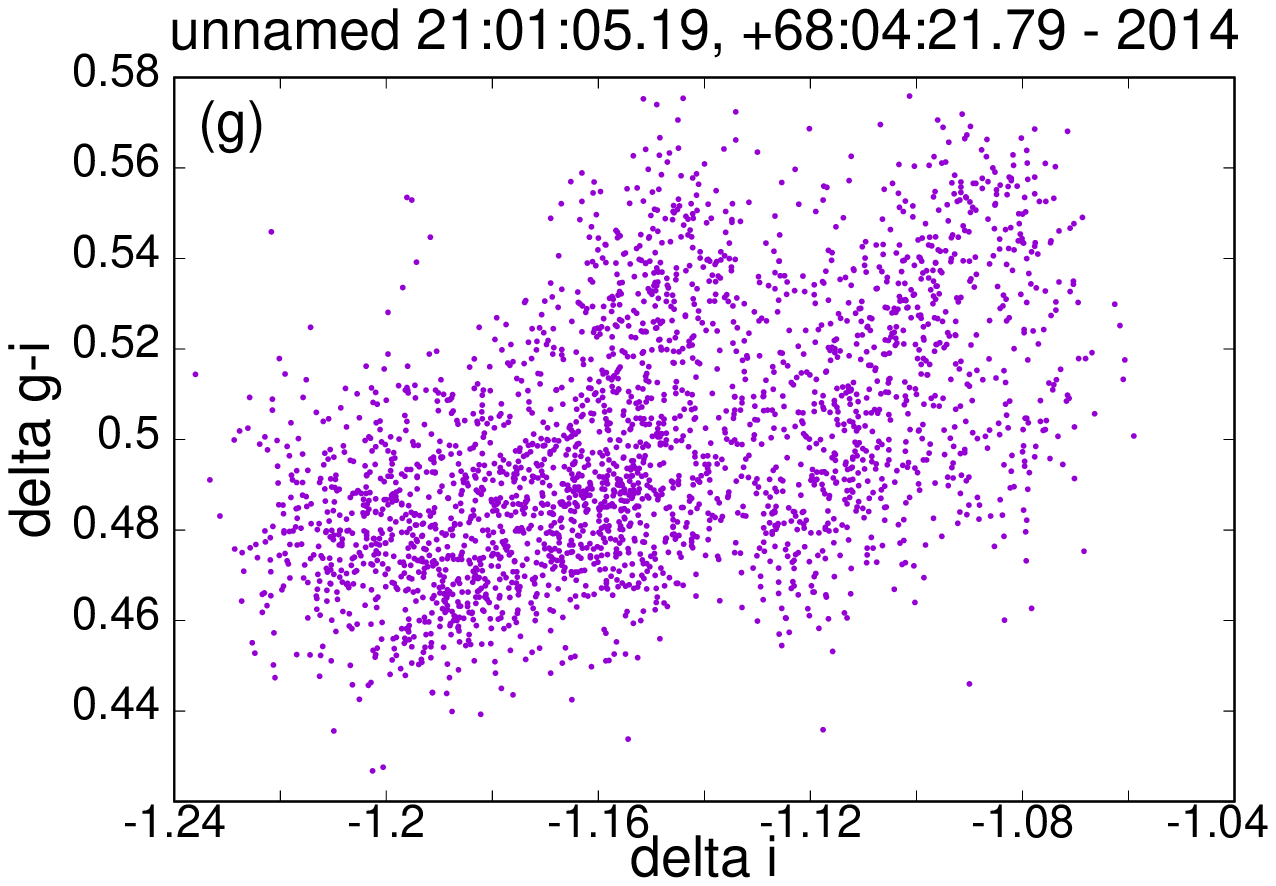}
\includegraphics[width=.33\linewidth]{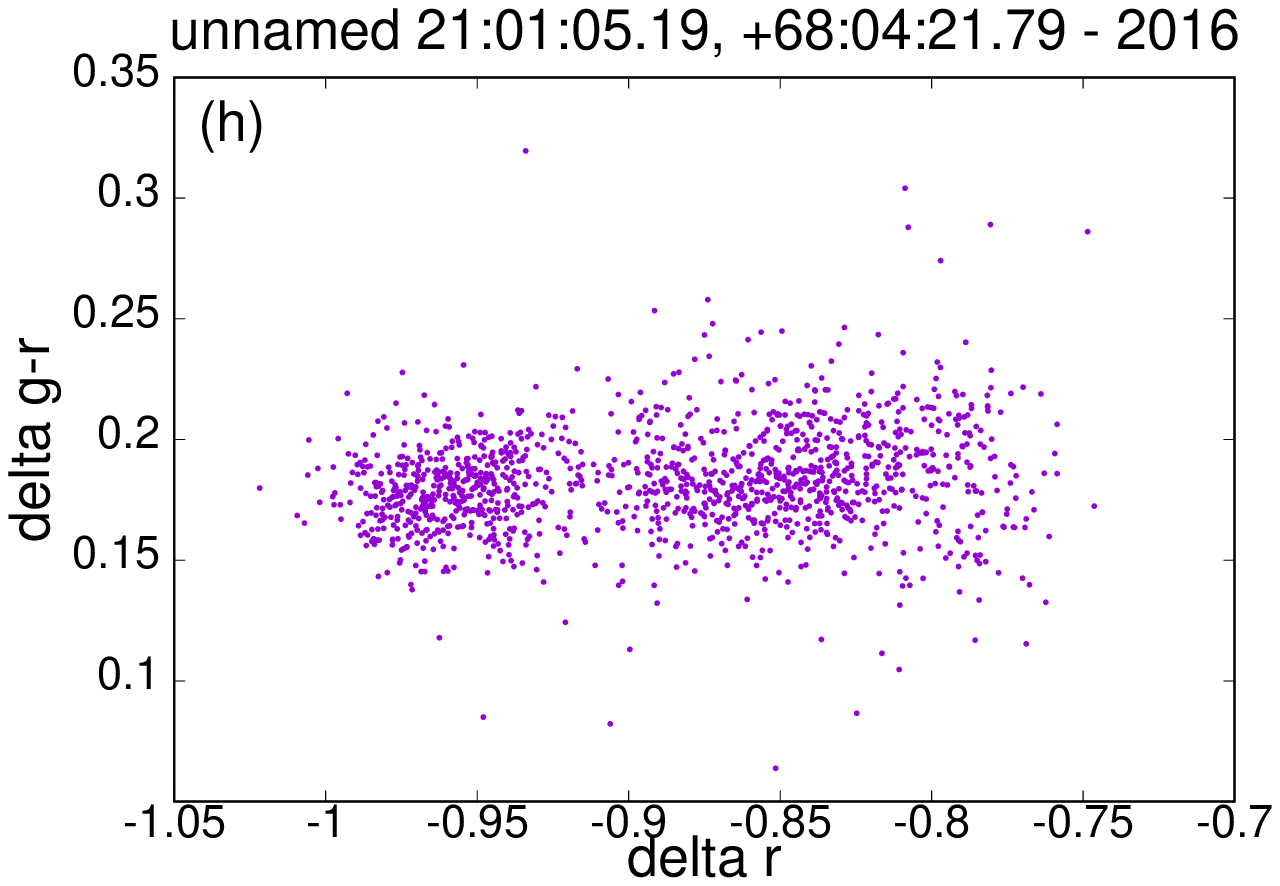}
\includegraphics[width=.33\linewidth]{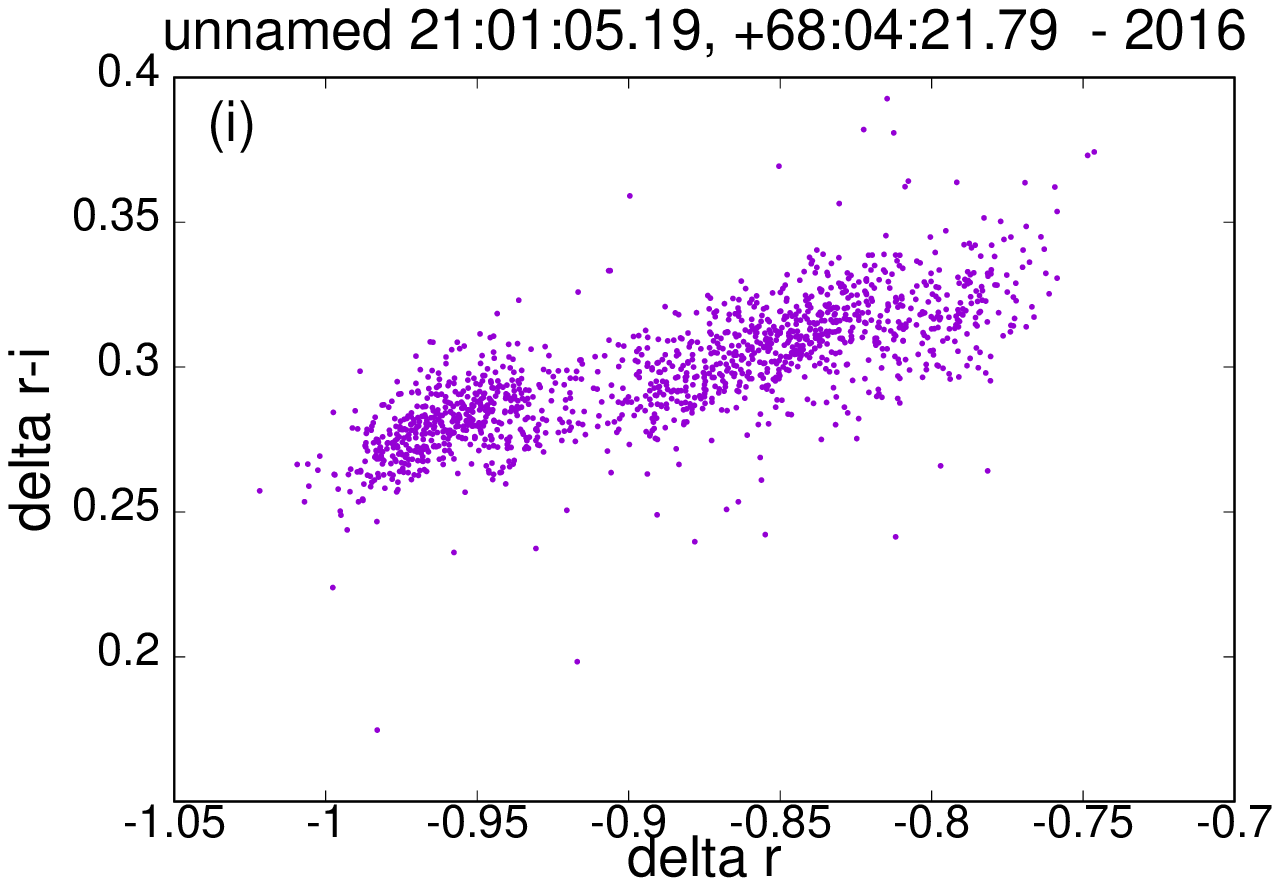}\\
\includegraphics[width=.33\linewidth]{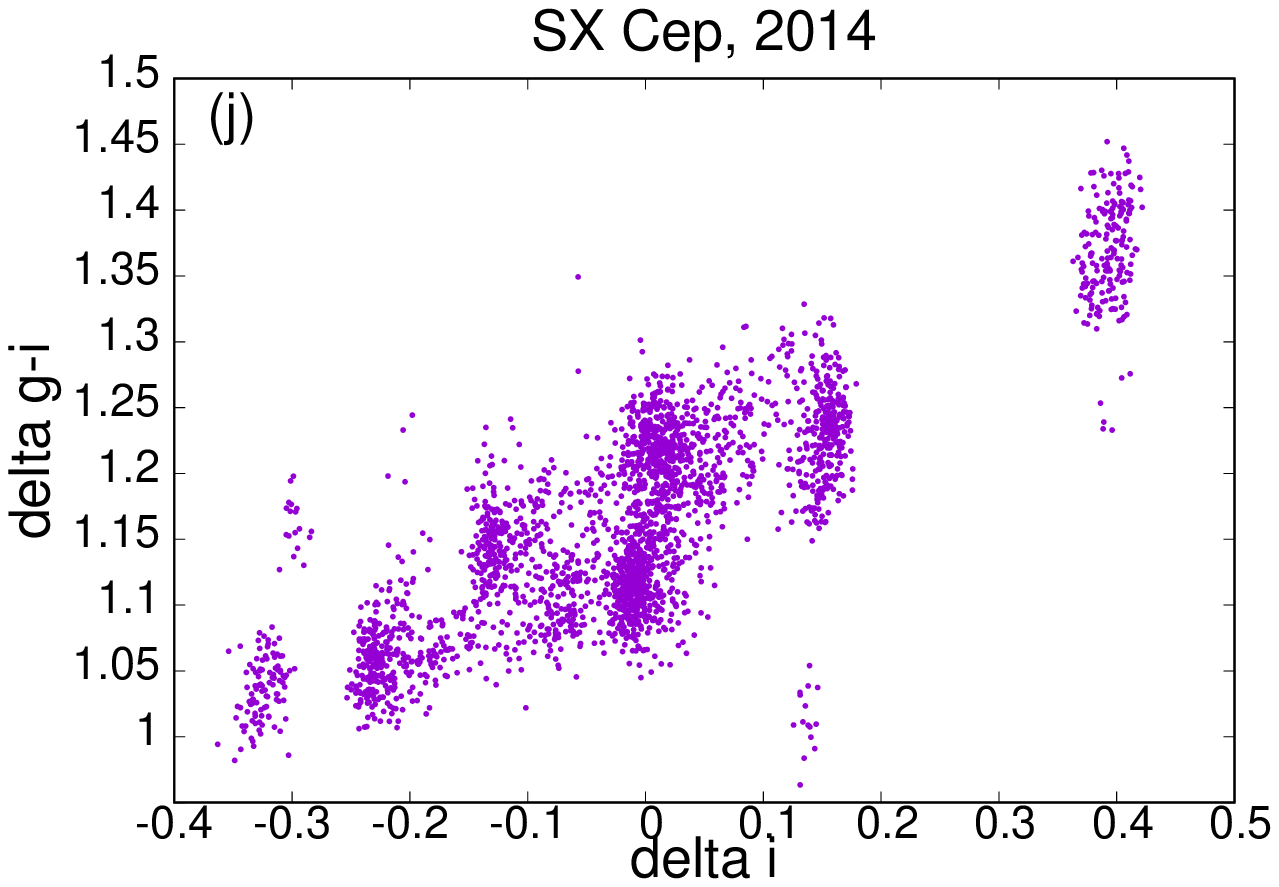} 
\includegraphics[width=.33\linewidth]{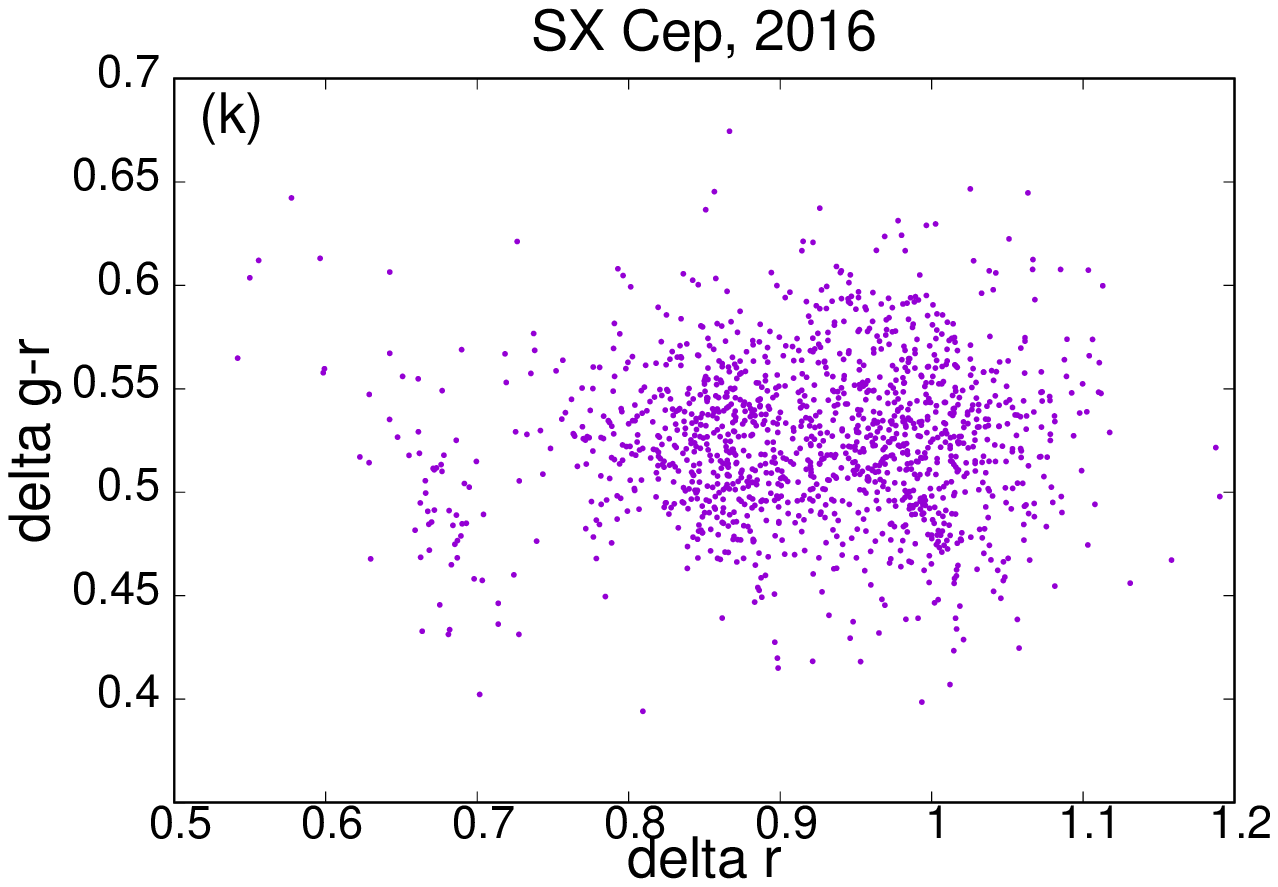}  
\includegraphics[width=.33\linewidth]{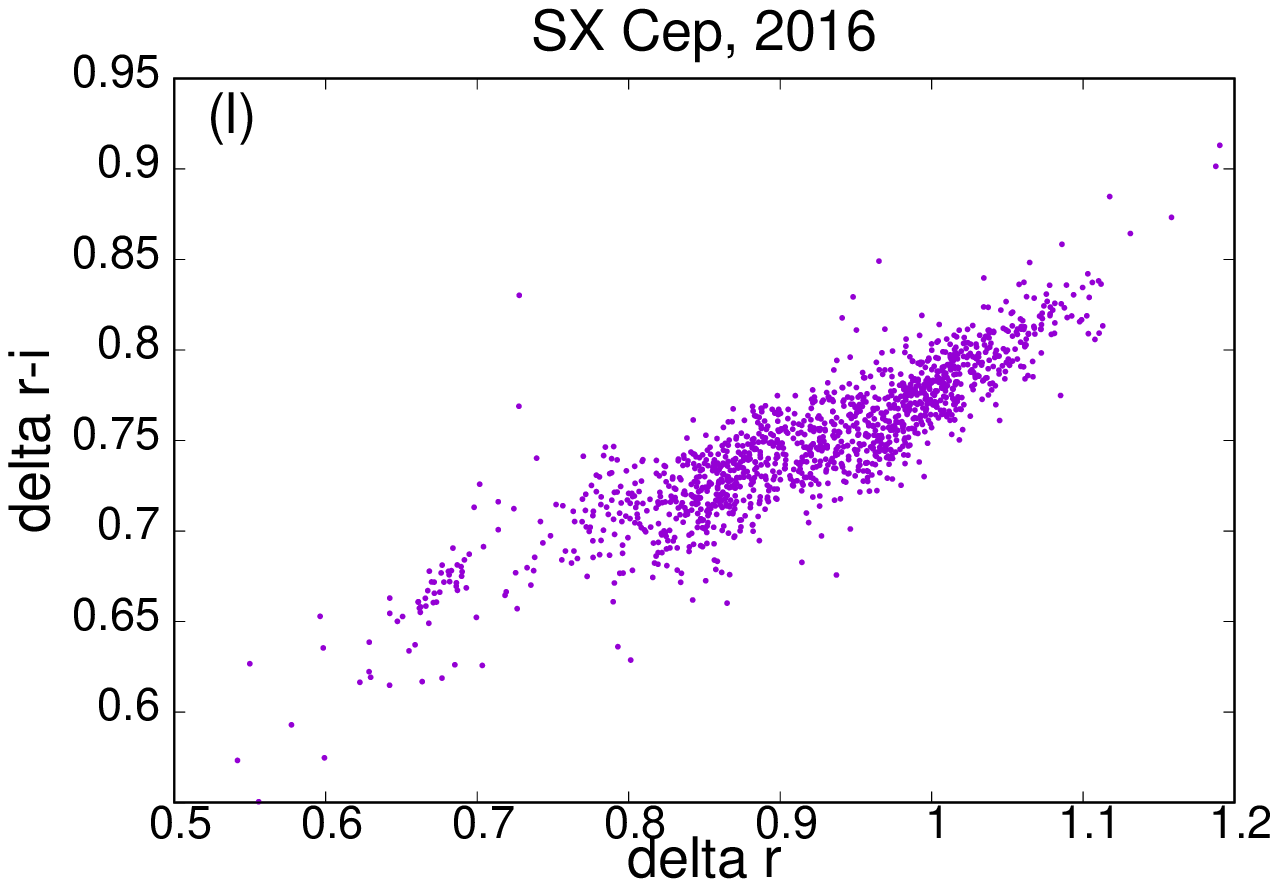}
\end{tabular}}
\FigCap{Colour-magnitude diagrams for EH~Cep, FW~Cep, the unnamed CTTS-candidate and SX~Cep. 
While colours of the last three star are becoming redder when the stars are fainter, it appears 
to be not the case for EH~Cep.}
\end{figure}

The characteristic feature of ''dippers'' are regular or semi-regular large brightness 
drops (e.g. Bouvier et al., 2003). 
The unified models of this phenomenon were most recently presented by McGinnis et al. (2015) 
and Bodman et al. (2017). 
Based on analysis of consecutive occultations observed by {\it CoRoT} spacecraft in a dozen of dippers 
in NGC~2264 SFR and by {\it Kepler-K2} in Upper Sco SFR, the authors obtained 
that the maximum warp height over the disc plane amounts to about 20-30 per cent of the disc radius 
at which it originates, and that it may vary by 10-20 per cent on a time-scale of days. 
It is normally assumed that the major warp causing partial or total occultations of stars arises 
as the result of star-disc interaction through magnetic field (Romanova et al., 2013). 
The optical thickness of the associated funnel flow is subject to change for external observer, 
depending on elongation over the disc plane. 
Similar level of light drop during given minimum observed in all bands indicates that warp is composed 
of grains larger than $1\mu m$ (Bouvier et al., 1999).\\
Particularly interesting objects of this type, which were previously studied by other authors, 
were also observed at the {\it MSO}. 
They are as folows:
\begin{itemize}
\item VY~Mon is the most studied star among those observed in field \#1. 
It is Herbig Ae/Be star and according to Miroschnischenko et al. (1992) the star was also strongly 
variable in the past (at 1.5~mag in Johnson $V$-filter) but the variability scale is strongly increased 
toward shorter wavelengths. 
H-alpha line profiles are variable between P-Cygni type and a single peak, what suggests that high-velocity 
outflows can be important envelope components (Pavlova et al., 2005); this view may be particularly 
true if one takes into account very high mass accretion rate of 10$^{-4}$M$_{\odot}$yr$^{-1}$ 
(Mendigutia et al., 2012). 
This is probably the youngest star investigated here, as its age derived from position 
on H-R diagram and theoretical evolutionary track of pre-main sequence stars does not cross $10^4$ 
years (Manoj et al., 2006).\\ 
Our data gathered for this star are presented in Fig.~6g. 
The colour-magnitude diagram (Fig.~2a) shows unusual trend: the star is becoming bluer when 
the brightness decreases, what is not observed in other dippers in this field (Fig.~2b,c,d), 
nor for BM~And (Fig.~2e,f).
In stars showing significant light variations due to variable extinction 
and without signatures of hot spots any search for rotational period is doomed to failure, 
as in our case. 
\item V521~Cyg appears to be the only ''dipper'' among the stars from field \#2. 
The star was frequently observed in current decade due to close vicinity to one of the most 
recent FU~Ori-type star (FUor) V2493~Cyg (Semkov et al., 2010). 
According to Poljancic Beljan et al. (2014) and Ibryamov et al. (2015, 2016) the major dips 
in V521~Cyg can be folded with the period 503~d.  
Although our observations considerably overlap with the data gathered by these authors, 
we did not observe any of these large-scale dips at the {\it MSO} -- only secondary though 
considerable (1~mag in $V$-filter) light drop at $HJD\approx2456580$ was noticed (Fig.~7a). 
For the rest of the time the star remained in high brightness state and showed only little light 
variations. 
Our attempts to search for periodic signals during the light ''plateau'' have failed.
\item BM~And is the only young star observed in field \#3. 
In the agreement with historical results, our data indicate on large variations (Fig.~7n) 
due to variable extinction, most probably caused by large grains contained in the disc warps (Walker, 1980). 
According to Grinin et al.(1995) the star is observed edge-on. 
Its shows typical behaviour as AA~Tau, i.e. it becomes redder when fainter (Fig.~2e,f). 
\item Field \#5 turned out to be most interesting region monitored during our survey. 
According to Grankin et al. (2007) the blue turnaround of colour-magnitude diagrams at minimum 
brightness of EH~Cep 
suggests light scattering (most efficient in blue band) by circumstellar extinction: 
as the stellar photosphere is partly occulted by circumstellar material (most likely the disc wind), 
the ratio of scattered light to direct light increases, and the system becomes bluer. 
This mechanism could potentially explain, why the range of light variations in EH~Cep observed 
during our survey is smaller in $g'$-band than in $r'$ and $i'$ bands (Fig.~8a,b). 
We stress, that this behavior appears to be always visible in EH~Cep (Fig.~3a,b,c), regardless 
of the brightness of this star. 
Note that this effect was previously observed by us even more clearly in VY~Mon, where massive outflow 
in the form of disc wind is the rule.
\item FW~Cep is classified as CTTS (Kun et al., 2009). 
Although our monitoring in 2014 and 2016 was relatively short, our data show well-defined 
brightness dips, with the decline and rise times of a few hours (Fig.~8e,f). 
The colour-magnitude diagrams show trends typical for AA~Tau-like stars (Fig.~3d,e,f). 
\item Similar well-defined dips were observed almost every single night in 
the bright star from field \#5, which remains unnamed in {\sc Simbad} database, localised 
at $\alpha_{ICRS~2000}=21h01m05.19s$, $\delta_{ICRS~2000}= 68^{\circ}04'21.79''$ (Fig.~9m,n). 
Nevertheless, we decided to present photometry of this star as it shows X-ray and infrared excesses 
typical for other CTTS. 
So far we were only able to establish that colour-magnitude diagrams of this star show behaviour 
typical for ''dippers'' (Fig.~3g,h,i). 
\end{itemize}

\subsection{The search for short-term, small-scale TW~Hya light dips}

In addition to ''dippers'' showing light drops as large as 1-3~mag, 
Siwak et al. (2014) reported {\it MOST} satellite discovery of semi-regular shallow 
($\sim 0.02-0.03$~mag) dips in TW~Hya. 
Intense monitoring of this star both with the {\it MOST} satellite and from 
the ground showed that the high occultation frequency seen in 2011 was rather exception 
than the rule (Siwak et al., 2018). 
We obtained that these dips do occur only in some local brightness 
maxima  and are likely caused by occultations of hot spots by: 
(1)~total or partial absorption of their light by condensed dusty clumps carried toward 
the star within associated accretion tongue, or 
(2)~total or partial absorption of their light by the plasma carried toward the star 
within associated accretion tongue, which becomes optically-thick in $U\&B$-filters. 
The search for similar events in other CTTS to better establish mechanisms leading 
to this phenomenon was one of the main driving forces leading us to undertake this survey. 
As their branch duration times in TW~Hya were of the order of 1-2~min, and the total 
duration times were of 10-20~min we decided to fit single filter change cycle in 30-90~sec, 
depending on stars brightnesses.

Careful inspection all {\it MSO} data obtained for ''accretors'' reveal only two possibilities 
of such events: the first was seen in the light curve of GI Tau, 
at $HJD=2456651.33$ and the second in 2MASS J18290394+0020212
(Figure~1e,f).
The low detection rate is in line with the results for TW~Hya, 
but the event in GI~Tau lasted remarkably long: 
the total duration amounted to almost 3.5~h, but the total flat minimum lasted almost 
1.1~h. 
In the second case, the respective duration times were significantly shorter, 
0.91~h, and 0.36~hr. 
Negative detection of dips lasting for 10-20~min, as in our discovery data in TW~Hya, 
may either suggests that these are extremely rare events, or that they remained undetected 
because of relatively small effective coverage time of every field. 
This search may perhaps be worth of resume in the future, when massive space-based 
continuous monitoring of the large parts of the sky in at least two bands (e.g. Pigulski et al., 2017) 
will become the rule. 

\subsection{Periodic stars}

Data obtained during randomly distributed nights over a few weeks, months, or even during 
different seasons, may allow to determine rotational periods of stars if they posses 
stable spots. 
This can be easily accomplished for WTTS, whose light curves are modulated 
by persistent cold spots (Grankin et al., 2008), and for CTTS accreting in the stable 
and the moderately stable regime, which creates two antipodal hot spots 
(e.g. Herbst et al., 1994; Siwak et al., 2016). 
Nevertheless, this regular variability pattern seen in CTTS can sometimes be disturbed 
by fractional or total occultation of central star by warps in the discs (e.g. SU~Aur 
Cody et al., 2013; Grankin et al., 2018), or by pulsed accretion 
(Tofflemire et al., 2017a, 2017b; Biddle et al., 2018), what significantly limits potential  
of ground-based data for searches for stable periods. 
 
We performed frequency analysis using the method described in Rucinski et al. (2008). 
Surprisingly, in spite of large number of monitored stars regular variability patterns 
turned out to be rather uncommon: only seven periodic stars was found in ''Gulf-Mexico'' 
SFR (field \#2) and another single star was found in field \#15.\\ 
Three out of seven periodic stars from ''Gulf-Mexico'' SFR were previously known to show regular variations: 
we confirmed the value of short period for the CTTS star V1716~Cyg at $4.16\pm0.09$~d 
(ver. $4.1539\pm0.0002$~d in Poljancic Beljan et al., 2014) and for likely WTTS star V1929~Cyg 
at $0.4263\pm0.0015$~d (ver. $0.426257\pm0.000306$~d given by Ibryamov et al., 2015). 
In line with these authors we also obtained the period of $2.42\pm0.05$~d for LkHa~189, which is probably CTTS. 
Contrary to Ibryamov et al. (2015), we fond stable period for V1957~Cyg at $5.2875\pm0.0080$~d, 
yet, the author argue that it may be field M0V dwarf, rather than WTTS.
Except of the above, we also found periodic signals for 2MASS~J20580604 +4349328 (at $1.333\pm0.010$~d), 
2MASS~J20580885+4346598 ($5.85\pm0.01$~d) and the very short $0.3692\pm0.0015$~d well-defined variations 
in the visual component of the likely visual binary star [RGS2011]~J205745.44 +434845.1. 
All these three stars are classified as young stellar objects in {\sc Simbad} database, are well-visible 
on X-ray and infrared images, yet, the short 0.37~d period of the last star strongly suggests that it may be WTTS. 
Spectroscopic observations of this star are strongly encouraged. 
If the WTTS membership is confirmed, it would be one of the fastest rotating WTTS ever observed.\\ 
We also note that neither our nor Ibryamov et al. (2015, 2016) data do not confirm 
the period of 2.08~d for LkHa~191, reported by Artemenko et al. (2012).
Interestingly, in spite of stable behavior of V1716~Cyg in our data set (see above), 
Findestein et al. (2013) classified this star as a burster: 
the first event of inhomogeneous accretion lasted 5-20 days while the second was found 35 days 
later and lasted for about 3 days. 

The last star among our targets showing regular variations is V395~Cep, also known as AS~507 
and BD +73~1031. 
It is classified as CTTS with the moderate value of the mass accretion rate 
$3\times10^{-9}$ M$_{\odot}$ yr$^{-1}$ (Kun et al., 2009; Frasca et al., 2018).
It was previously found as periodic star (at 3.432~d) by Chugainov et al. (1995) and our data 
phased with this value fully support their result. 
Stability of this value since almost three decades indicates that it is strictly related with stellar 
rotation, most probably with a hot spot produced by stable funnel flow. 
Similar long-term stable modulation was also noticed for IM~Lupi (Siwak et al., 2016) 
and for SU~Aur (Cody et al., 2013).

We also noticed slight excesses of power in frequency spectra of a handful of other young stars 
from fields \#7 and \#9, but data phased with respective periods do not show any regular patterns. 
Among them are LkHa~168 and LkHa~172 observed by Ibraymov et al. (2018), who were also unable 
to find any periodic behavior. 
These and other similar stars showing little but eventually false-positive signatures of periodic behaviour 
are usually classified as HAEBE or partly embedded young stellar objects (YSO). 
We stress that even the precise {\it MOST} satellite observations of HD~37806 and AB~Aur did not 
firmly reveal any periodic variability (Rucinski et al., 2010; Cody et al., 2013). 
In addition, although the light curves appear to be similar to these of CTTS, amplitudes of light variations 
are considerably scaled down, to the level impossible for detailed investigation from the ground.

\begin{figure}[htb]
\centerline{%
\begin{tabular}{l@{\hspace{0.5pc}}}
\includegraphics[width=.5\linewidth]{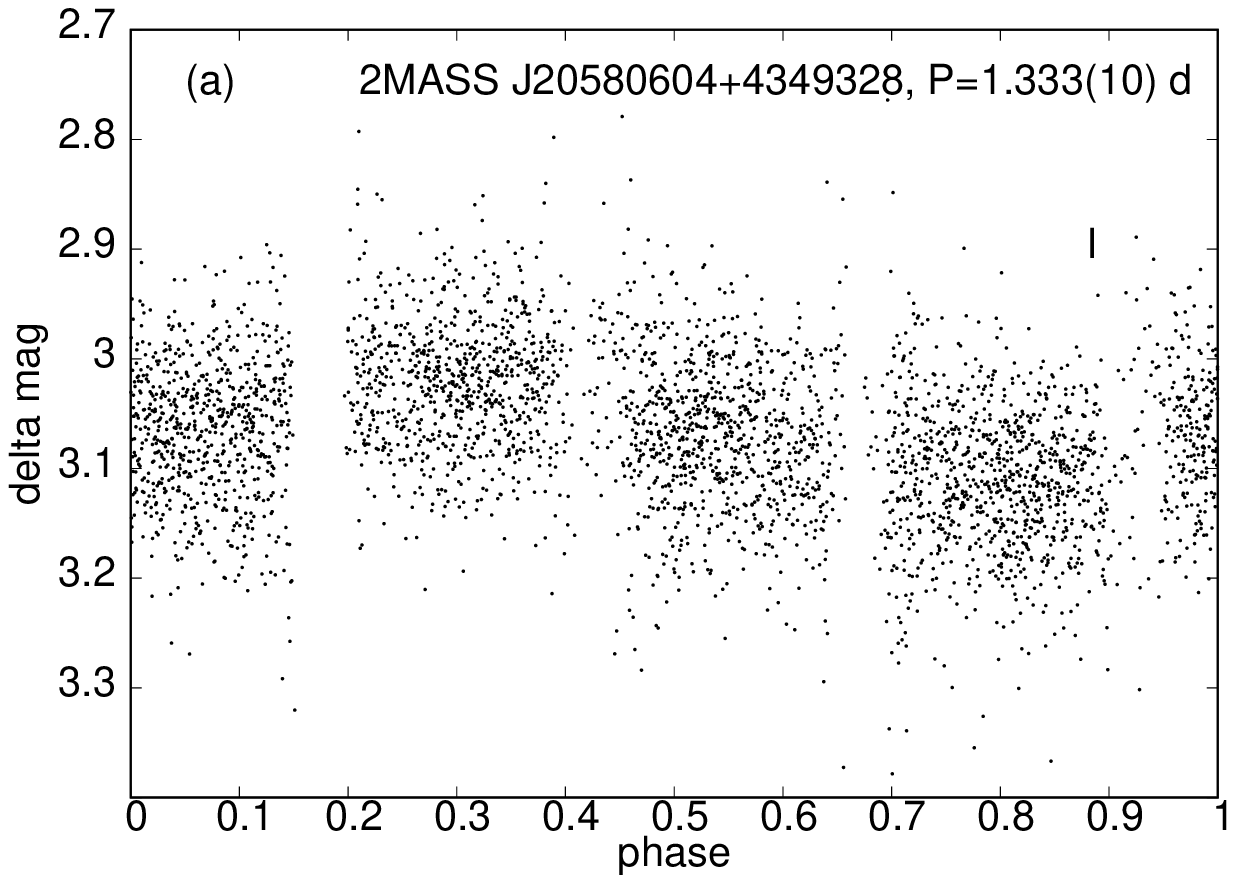} 
\includegraphics[width=.5\linewidth]{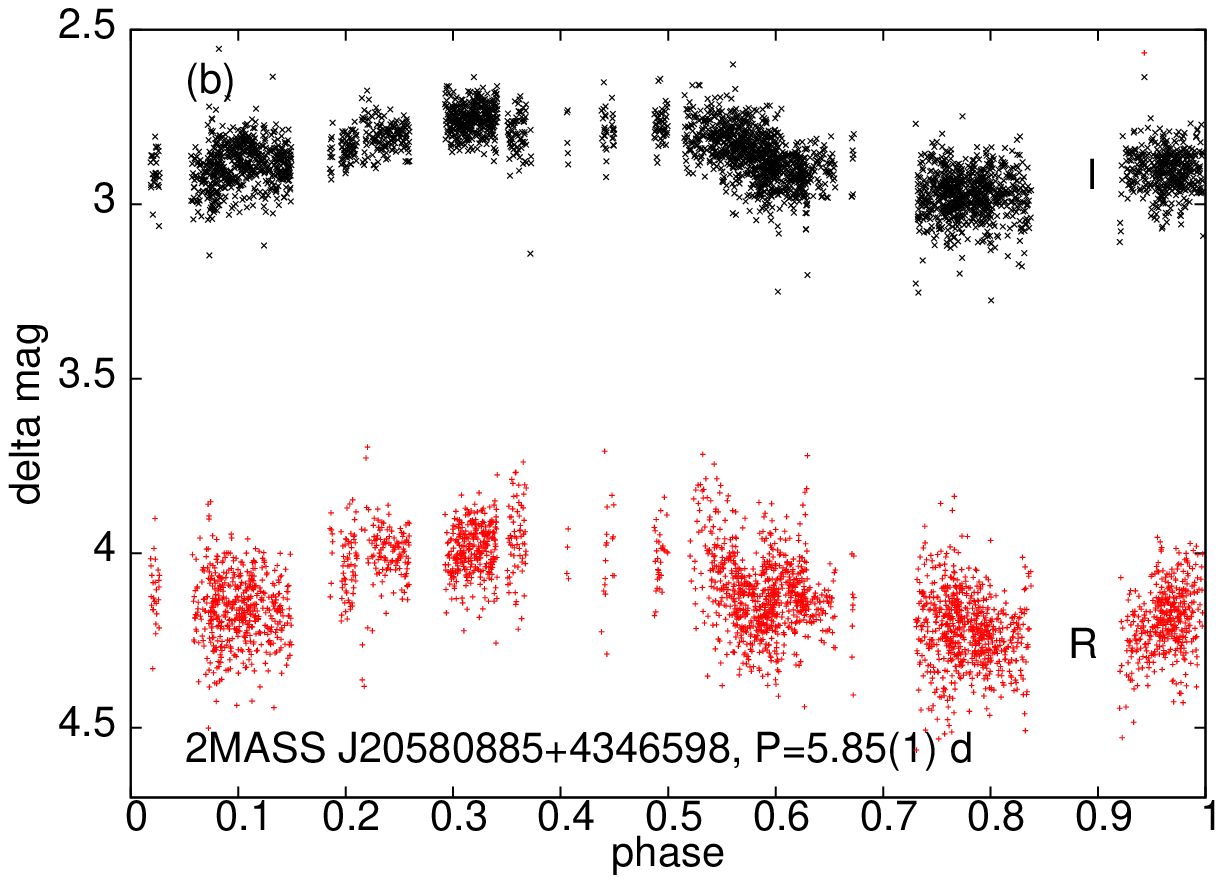}\\
\includegraphics[width=.5\linewidth]{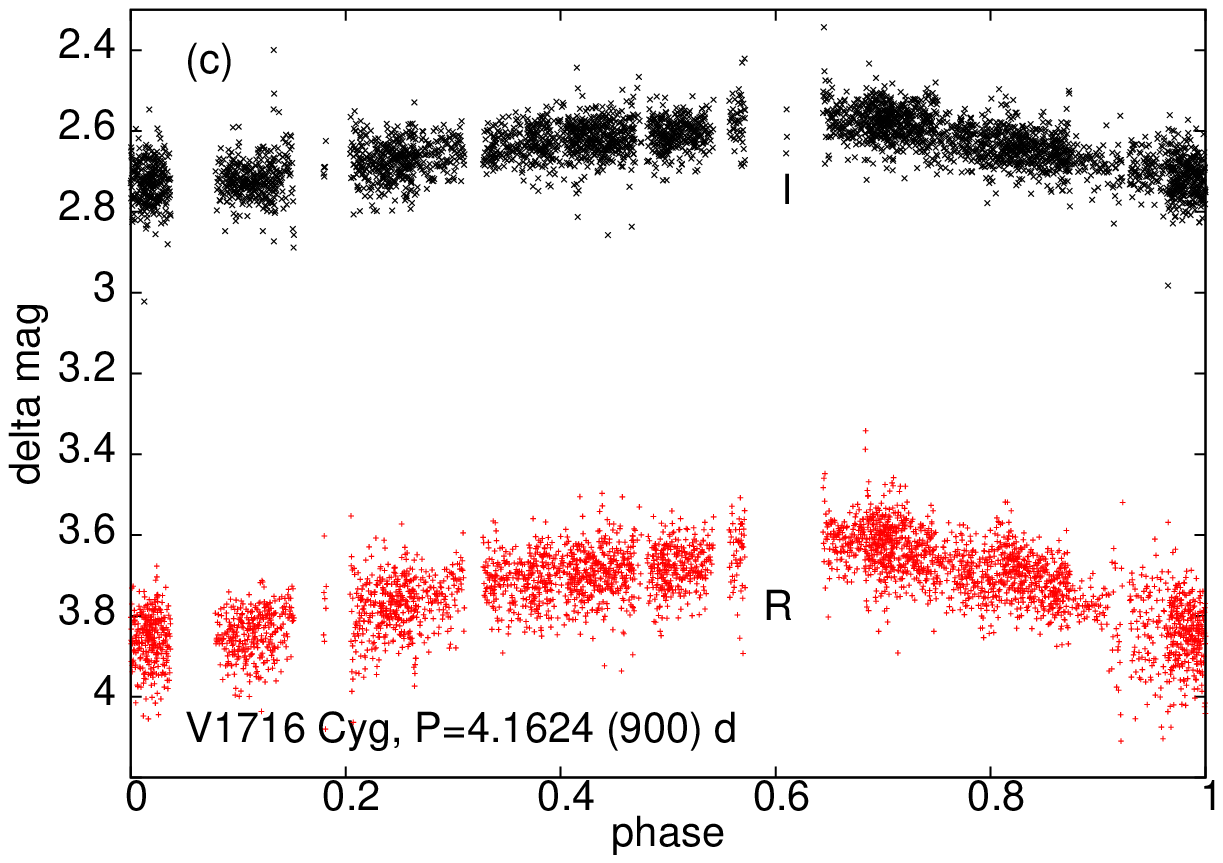}  
\includegraphics[width=.5\linewidth]{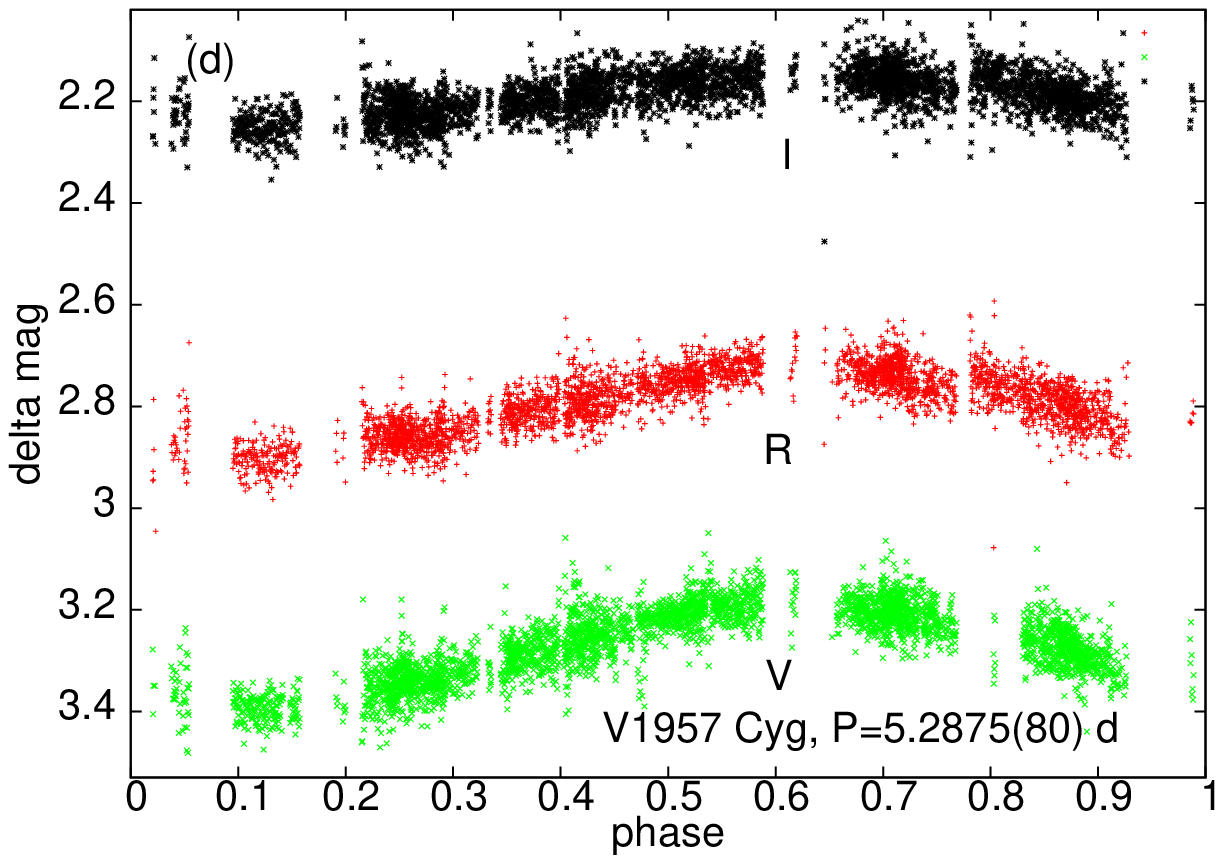}\\
\includegraphics[width=.5\linewidth]{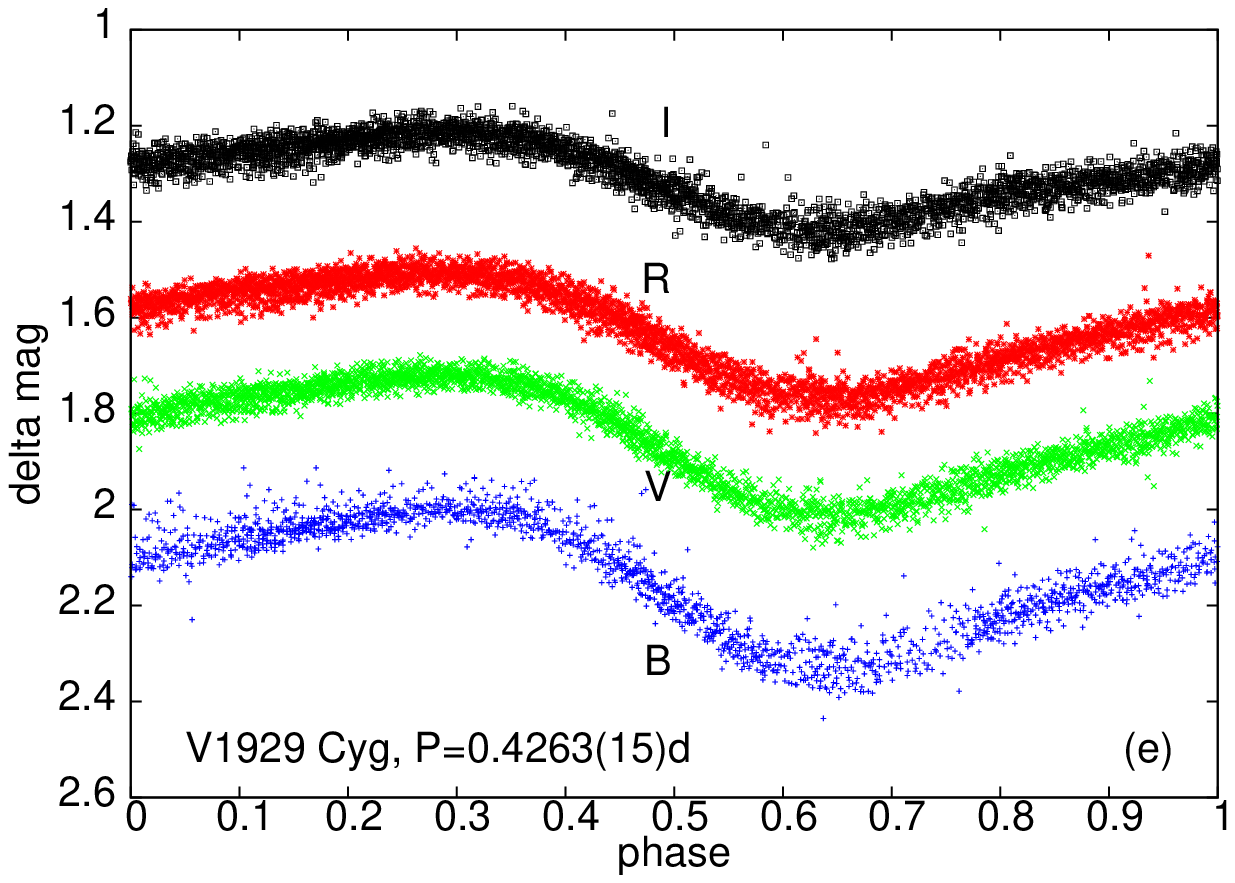}
\includegraphics[width=.5\linewidth]{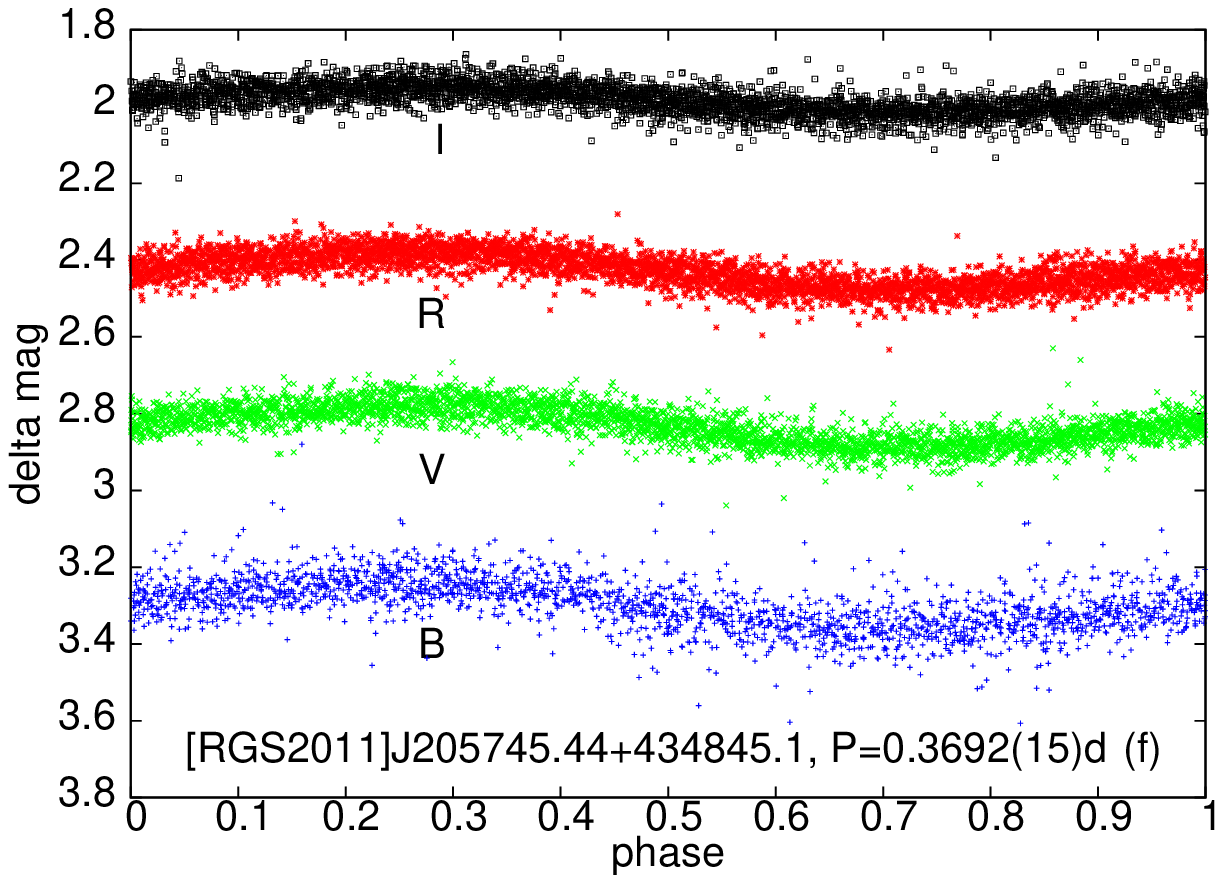}
\end{tabular}}
\FigCap{Periodic T~Tauri-type stars in ''Gulf-Mexico'' SFR.}
\end{figure}

\subsection{Stars with uncertain causes of variability}

\begin{figure}[htb]
\centerline{%
\begin{tabular}{l@{\hspace{0.1pc}}}
\includegraphics[width=.5\linewidth]{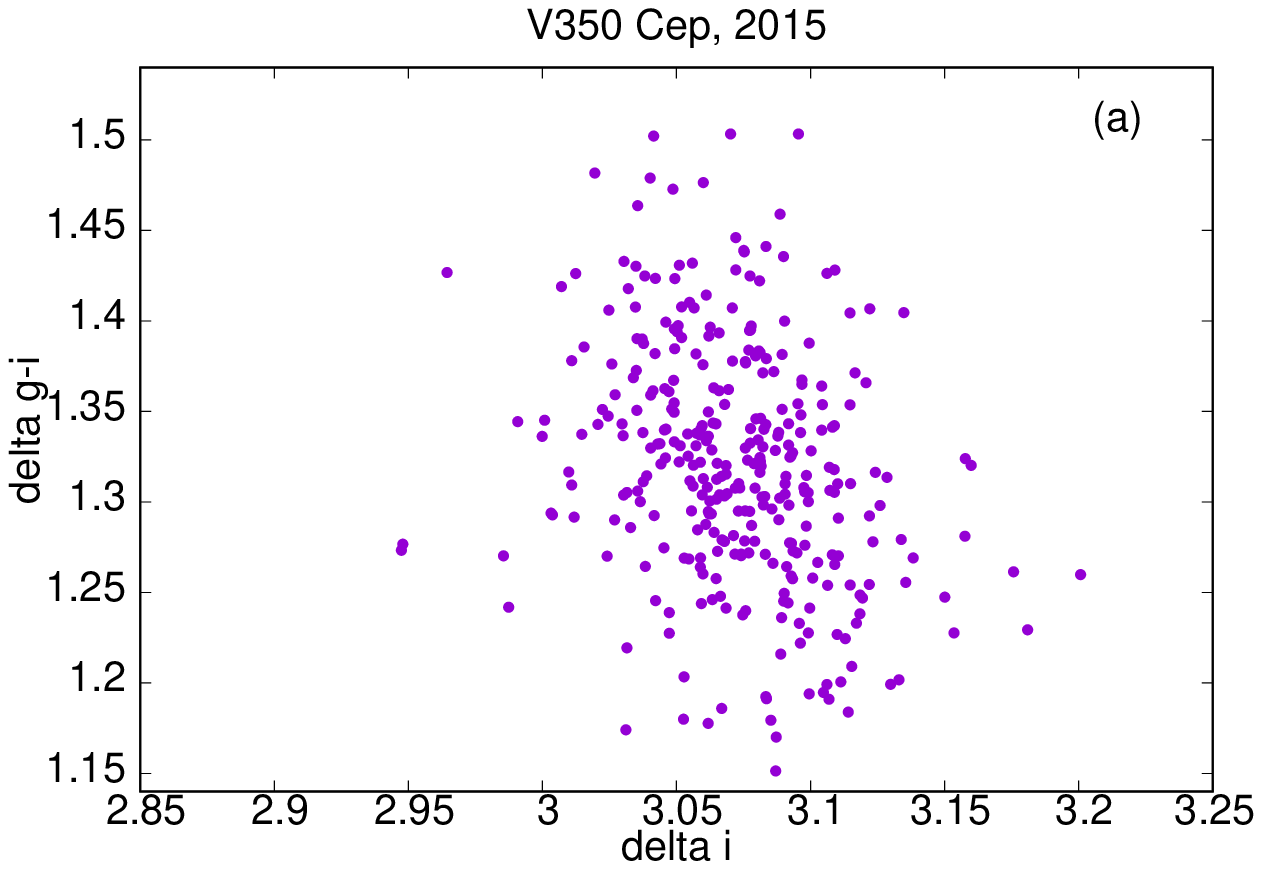}  
\includegraphics[width=.5\linewidth]{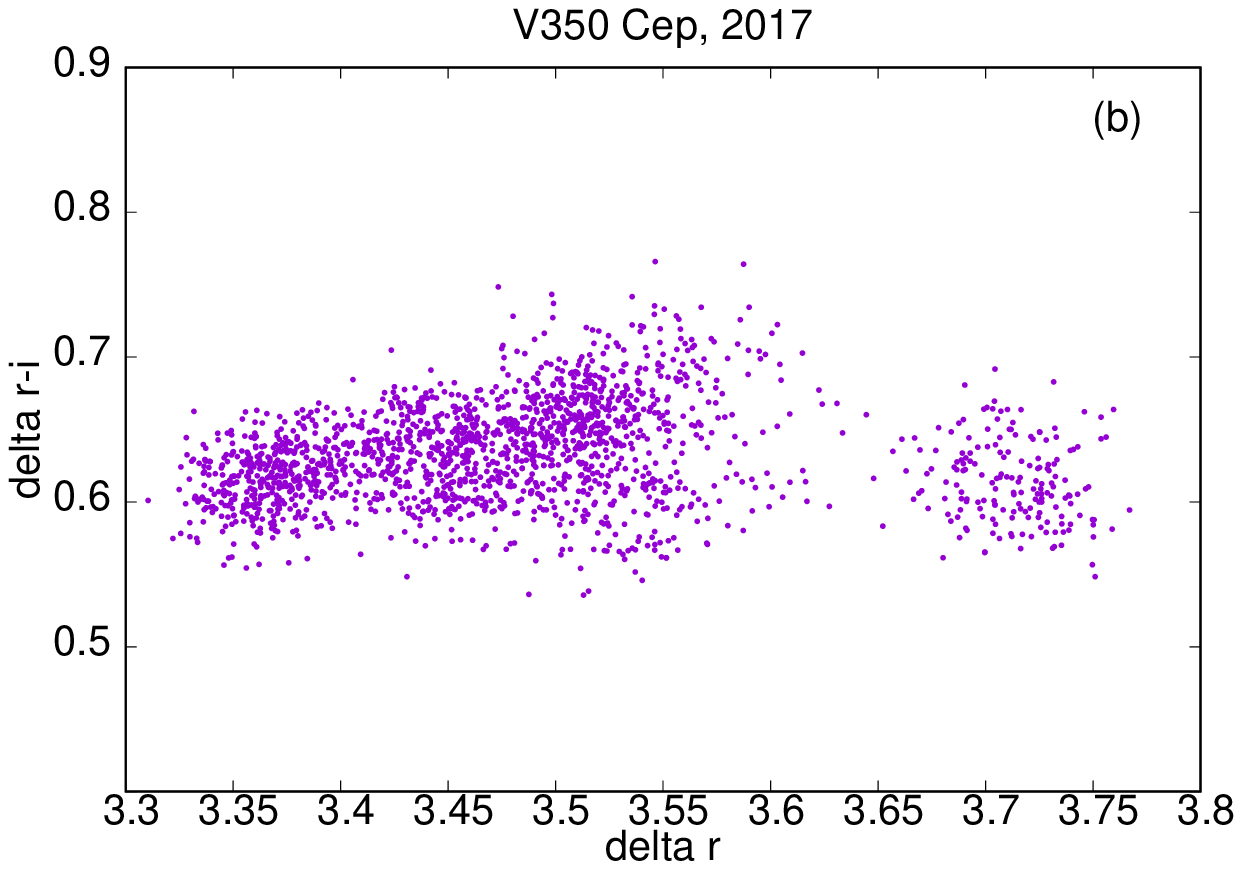}
\end{tabular}}
\FigCap{Colour-magnitude diagrams for V350~Cep. 
}
\end{figure}

As mentioned above, many stars classified as YSO and HAEBE exhibit amplitudes 
of brightness changes to small to be detaily investigated from the ground. 
This is in accord with Herbst \& Shevchenko (1999), who stated that HAEBE 
often show little and (mostly) irregular variations beyond the major flux dips (if any), 
which are occurring in long time scales. 
They conclude, that these stars evidently do not possess either the large, stable cool 
spots or persistent hot spots associated with strong surface magnetic fields 
and magnetospheric accretion mechanism, which is the rule for CTTS.\\
Except of these stars, we observed several stars with considerable light variations and yet the 
exact mechanism leading to these light changes remains unknown:
\begin{itemize}
\item SX~Cep and 2MASS~J21014358+6809361 (field \#5) are classified 
as WTTS (Kun et al., 2009) and show significant variations in time scales of hours and days. 
The second star showed also one stellar flare during the 2016 run (Fig.~9j).  
However, we failed to find any stable periods both in separate and combined 2014 and 2016 data sets. 
The colour-magnitude diagrams of both stars (see in Fig.~3j,k,l for SX~Cep) show similar amplitudes 
of light variations in $g$ and $r$ filters. 
This may suggests presence of hot spot(s) with a temperature only slightly higher than 
the temperature of the photosphere, similarly as for MN~Lup (Strassmeier et al., 2005). 
If it is the case, perhaps CTTS classification would be more apropriate for these stars.

\item Variability of 2MASS J21012706+6810381 and 2MASS J21011252+6810- 195 from the same 
field may appear to be due to variable extinction, but our data are insufficient to confirm 
this view -- stellar rotation with hot spots may also be responsible for observed pattern.
For the same reason, we are unable to conclude what is the mechanism leading to variability 
in LkHa~428, where we observe combined light of two components of this young wide binary star. 

\item V350~Cep was preliminarily classified as EXor or FUor (Ibryamov et al., 2014), but this was 
later put in doubts by Dahm et al. (2015) and Jurdana-Sepic et al. (2018).  
The last authors classified it as ''dipper'', but also this classification was later questioned 
by Semkov et al. (2017).\\ 
Data gathered during our survey are too poor to resolve this ambiguity. 
The colour-magnitude diagrams constructed from all available data do not show any significant 
trends (Fig.~5a,b). 
\end{itemize} 

\section{Summary}

We performed the first high-cadence survey of pre-main sequence stars dedicated to study variety 
of rapid light variations simultaneously in at least two visual bands\footnote{Questions regarding 
this manuscript and requests for presented light curves and/or calibrated frames for independent 
measurements can be send to the first author of this paper at any time.}.
Although we did not find any signs of planetary transits, 
the data proved to be useful for determination of rotational periods of some stars and for 
studies of mechanisms causing observed light variations. 
The most important findings are as follows:
\begin{itemize}
\item We determined rotational periods for four new stars 
in ''Gulf-Mexico'' SFR and confirmed the period stability in V395~Cep, i.e. that value 
determined more than two decades ago (Chugainov et al., 1995) is strictly related to the rotational 
period of the star. 
\item We found FU~Cep to be one of the most intriguing accretors in our sample, as it showed 
rare event of oscillatory-like behavior in the local maximum, similar as TW~Hya (Siwak et al., 2018). 
These oscillations were only seen in the blue band and arose right with 
the appearance of the hot spot causing associated light maximum. 
We do not indicate any particular mechanism that could firmly explain this variability, 
but rather urge theorists to undertake theoretical studies of this issue (see e.g. de~S{\'a} et al., 2019). 
\item Most of our targets firmly classified as ''dippers'' show nearly ''achromatic'' 
variability, i.e. most likely caused by large grains, as in AA~Tau (Bouvier et al., 1999). 
These stars are usually getting bluer when their brightness increases.
Yet, EH~Cep and VY~Mon show inverse relationships -- they are becoming bluer when their brightness decreases. 
This indicates significant radiation scattering (most efficient on short wavelengths) 
when the stars are obscured likely by disc wind, consisting of individual atoms 
and molecules rather than dusty grains. 
\end{itemize}
Analysis of the {\it MSO} data gathered during this survey is not complete yet. 
We do not present here the results obtained for about 80 CTTS in young cluster IC~348, which will 
be the subject of separate publication. 
Similarly, our data obtained for two FUors will be analysed together with {\it TESS} light curves 
gathered during Cycle~2. 
Certainly, the massive inflow of uninterrupted light curves gathered by this satellite for huge amount 
of young stars may also reveal planetary transits, perhaps similar to CVSO 30b, the phenomenon 
strongly desired but undetected during our modest survey.
We stress that despite the fact that {\it TESS} observations may sharpen the view on mechanisms 
causing variability in the aforementioned stars, it is limited to members of rather close 
and uncrowded star forming regions. 
Therefore ground-based monitoring is still mandatory for investigation of crowded fields, like these 
in Cygnus and Cepheus constellations.

\MakeTable{rccccc}{12cm}{Log of observations. Field names, dates of observations 
and filters used during given observation are listed in consecutive columns.}
{\hline
Field No. & Comp. stars    & Date [yyyy.mm.dd] &  Filters  & Date [yyyy.mm.dd] & Filters \\ \hline
{\#  1}   & 3UC~202-059221 & 2013.02.11        & BVRI      & 2014.02.06  &  VI     \\
0.9~d     & 3UC~201-058983 & 03.28             & BVRI      & 02.07       &  VI     \\   
          &     & 03.02             & BVRI      & 02.08       &  VI     \\
          &     & 03.04             & BVRI      & 02.10       &  VI     \\
          &     & 03.05             & BVRI      & 2015.03.07  & I       \\   
          &     & 03.06             & BVRI      & 03.08       & I       \\   
          &     & 03.16             & BVRI      & 03.09       & I       \\    
          &     & 03.17             & BVRI      & 03.10       & I       \\ 
          &     & 12.27             & BVI       & 03.16       & I       \\
          &     & 2014.01.07        & BVI       & 03.17       & I       \\   
          &     & 02.03             & VI        & 03.20       & I       \\
          &     & 02.05             & VI        &             &         \\ \hline
{\#2}     & 3UC~268-201607 & 2013.07.18 & BVRI      & 2013.08.25  & BVRI    \\
0.96~d    & 3UC~268-201639 & 07.19      & BVRI      & 08.27       & BVRI    \\ 
          & 3UC~268-201660 & 07.20      & BVRI      & 09.03       & BVRI    \\ 
          &     & 07.21             & BVRI      & 09.04       & BVRI    \\  
          &     & 07.22             & BVRI      & 09.05       & BVRI    \\
          &     & 07.23             & BVRI      & 09.06       & BVRI    \\ 
          &     & 07.24             & BVRI      & 09.07       & BVRI    \\ 
          &     & 07.25             & BVRI      & 09.08       & BVRI    \\
          &     & 07.26             & BVRI      & 09.10       & BVRI    \\
          &     & 07.27             & BVRI      & 09.13       & BVRI    \\
          &     & 07.28             & BVRI      & 09.14       &  VRI    \\
          &     & 08.01             & BVRI      & 09.30       &  VRI    \\
          &     & 08.02             & BVRI      & 10.10       &  VRI    \\
          &     & 08.03             & BVRI      & 10.11       &  VRI    \\
          &     & 08.04             & BVRI      & 10.12       &  VRI    \\
          &     & 08.05             & BVRI      & 10.13       &  VRI    \\
          &     & 08.06             & BVRI      & 10.14       &  VRI    \\
          &     & 08.07             & BVRI      & 10.15       &  VRI    \\
          &     & 08.08             & BVRI      & 10.17       &  VRI    \\
          &     & 08.15             & BVRI      & 10.19       &  BVRI   \\
          &     & 08.16             & BVRI      & 10.20       &  BVRI   \\
          &     & 08.17             & BVRI      & 10.21       &  BVRI   \\
          &     & 08.18             & BVRI      & 10.25       &  BVRI   \\
          &     & 08.22             & BVRI      & 10.26       &  BVRI   \\
          &     & 08.23             & BVRI      & 10.27       &  BVRI   \\
          &     & 08.24             & BVRI      & 10.28       &  BVRI   \\ \hline
{\#3}     & TYC~3642-1917-1 & 2013.09.10        &  VRI        & 2013.10.22  &  BVI    \\
0.48~d    & 3UC~277-285405  & 09.30             &  BVRI       & 10.23       &  BVI    \\ 
          &     & 10.10             &  BVI      & 10.24       &  BVI    \\
          &     & 10.12             &  BVI      & 11.08       &  BVI    \\
          &     & 10.13             &  BVI      & 11.12       &  BVI    \\
          &     & 10.15             &  BVI      & 11.14       &  BVI    \\
          &     & 10.20             &  BVI      & 11.16       &  BVI    \\
          &     & 10.21             &  BVRI     & 11.17       &  BVI    \\ \hline
}

\MakeTable{rccccc}{12cm}{Table 5b - continuation.}
{\hline
Field  & Comp. stars & Date [yyyy.mm.dd]& Filters & Date [yyyy.mm.dd] & Filters \\ \hline
{\#4}  & 3UC~229-025810 & 2013.11.07       &  BVI    & 2013.12.24        &  BVI    \\    
0.83~d & TYC~1829-186-1 & 11.08            &  BVI    & 12.27             &  BVI    \\ 
       &       & 12.03            &  BVI    & 12.28             &  BVI    \\ 
       &       & 12.12            &  BVI    & 12.31             &  BVI    \\  
       &       & 12.13            &  BVI    & 2014.01.02        &  BVI    \\
       &       & 12.18            &  BVI    & 01.03             &  BVI    \\  
       &       & 12.19            &  BVI    & 01.06             &  BVI    \\
       &       & 12.23            &  BVI    & 01.07             &  BVI    \\ \hline   
{\#5}  & 3UC~317-077414 & 2014.08.02       &  gri    & 2014.11.02        &  gi     \\
0.93~d & 3UC~317-077395 & 09.19            &  gri    & 11.23             &  gi     \\ 
       & 3UC~317-077386 & 09.20            &  gri    & 2016.09.26        &  gri    \\
       &       & 10.27            &  gi     & 09.27             &  gri    \\
       &       & 10.28            &  gi     & 09.29             &  gri    \\
       &       & 10.29            &  gi     & 09.30             &  gri    \\
       &       & 10.30            &  gi     & 10.01             &  gri    \\
       &       & 10.31            &  gi     &                   &         \\  \hline 
{\#6}  & GSC~3997-1833  & 2014.11.03       &  gri    &  10.30            & gri     \\   
0.75~d & GSC~3997-2030  & 11.09            &  gri    &  11.02            & gri     \\
       & 3UC~298-177839 & 12.04            &  gri    &  11.03            & gri     \\
       & 3UC~298-177787 & 12.05            &  gri    &  11.08            & gri     \\
       &       & 2015.10.27       &  gri    &  11.24            & gri     \\
       &       &  10.29           &  gri    &  11.25            & gri     \\ \hline           
{\#7}  & TYC~3179-198-1 & 2015.08.13       &  gi     & 2015.10.10        &  gi     \\ 
0.49~d & USNO-A2 1275-14250279 & 09.15            &  gi     & 10.24             &  gi     \\ 
       &       & 09.16            &  gi     & 10.29             &  gi     \\ 
       &       & 09.17            &  gi     & 10.30             &  gi     \\ 
       &       & 09.21            &  gi     & 11.02             &  gi     \\
       &       & 09.22            &  gi     & 11.03             &  gi     \\
       &       & 10.04            &  gi     & 11.04             &  gi     \\
       &       & 10.05            &  gi     & 11.06             &  gi     \\ \hline
{\#8}  & 2MASS~J0418423+281140 & 2015.12.19       &  gri    & 2016.12.20        &  gri    \\    
0.90~d & JH~161& 12.20            &  gri    & 12.21             &  gri    \\ 
       &       & 12.29            &  gri    & 12.30             &  gri    \\ 
       &       & 12.30            &  gri    & 12.31             &  gri    \\  
       &       & 2016.12.05       &  gri    & 2017.01.01        &  gri    \\   
       &       & 12.07            &  gri    &                   &         \\ \hline
{\#9}  & TYC~3160-1814-1 & 2016.07.10       &  gri    & 2016.08.17        &  gri    \\              
1.24~d & 3UC~265-198648   &      07.11       &  gri    &      08.18        &  gri    \\
       &       &      07.12       &  gri    &      08.19        &  gri    \\
       &       &      07.22       &  gri    &      08.28        &  gri    \\
       &       &      07.24       &  gri    &      08.31        &  gri    \\
       &       &      07.27       &  gri    &      09.07        &  gri    \\
       &       &      07.29       &  gri    &      09.12        &  gri    \\
       &       &      07.30       &  gri    &      09.13        &  gri    \\
       &       &      08.02       &  gri    &      09.15        &  gri    \\
       &       &      08.04       &  gri    &      09.21        &  gri    \\
       &       &      08.08       &  gri    &      09.22        &  gri    \\ \hline               
}

\MakeTable{rccccc}{12cm}{Table 5c - continuation.}
{\hline
Field   & Comp. stars & Date [yyyy.mm.dd]& Filters &Date [yyyy.mm.dd]& Filters \\ \hline
{\#10}  & USNO-A2 0975-01667849 & 2016.12.29       &  gri  & 2017.02.13 &  gri      \\ 
0.80~d  & USNO-A2 0975-01682657 & 2017.01.19       &  gri  &      02.15 &  gri      \\
        &      &      01.22     &  gri  &      02.25        &  gri      \\
        &      &      01.28     &  gri  & 2018.01.14        &  gri      \\
        &      &      02.11     &  gri  &      01.15        &  gri      \\
        &      &      02.12     &  gri  &      01.16        &  gri      \\ \hline
{\#11}  & TYC~4274-1993-1       & 2015.09.16   & gi  & 2015.11.06   & gi \\  
0.70~d  & USNO-A2 1500-08311095 &      09.17   & gi  &      11.08   & gi \\
        &                       &      09.21   & gi  & 2017.08.30   & gri \\
        &                       &      10.10   & gi  &      08.31   & gri \\
        &                       &      10.24   & gi  &      09.30   & gri \\ 
        &                       &      10.29   & gi  &      10.01   & gri \\   
        &                       &      10.30   & gi  &      10.18   & gri \\
        &                       &      11.02   & gi  &      11.04   & gri \\      
        &                       &      11.03   & gi  &      11.05   & gri \\  
        &                       &      11.04   & gi  &      12.13   & gri \\ \hline 
{\#12}  & USNO-A2 0900-12955282  & 2017.03.31 &  gri  & 2017.06.09        &  gri      \\
0.45~d  & USNO-A2 0900-12947374  & 04.01      &  gri  &  07.06            &  gri      \\ 
        & USNO-A2 0900-12917818  & 04.02      &  gri  &  07.19            &  gri      \\
        &      & 04.09          &  gri  &  07.25            &  gri      \\
        &      & 04.30          &  gri  &  08.01            &  gri      \\
        &      & 05.01          &  gri  &  08.02            &  gri      \\
        &      & 05.27          &  gri  &  08.05            &  gri      \\
        &      & 06.08          &  gri  &  08.09            &  gri      \\ \hline
{\#13}  & TYC~1318-514-1 & 2017.01.19     &  gri  & 2017.03.24            &  gri      \\
0.42~d  &      &      01.20     &  gri  &  03.27            &  gri      \\
        &      &      01.22     &  gri  &  03.28            &  gri      \\
        &      &      01.28     &  gri  &  03.31            &  gri      \\
        &      &      02.12     &  gri  &  04.01            &  gri      \\
        &      &      02.13     &  gri  &2018.02.19         &  gri      \\   
        &      &      02.25     &  gri  &                   &           \\ \hline
{\#14}  & XEST~15-OM-160 & 2018.02.16   &  gri      & 2018.11.29        &  BVI      \\
0.48~d  & XEST~15-OM-056 &      03.01   &  gri      &      11.30        &  BVI      \\
        &                &      03.03   &  gri      &                   &           \\ \hline 
{\#15}  & TYC~4490-1027-1 & 2018.11.05  &  vby      & 2018.11.10&  vby      \\
0.72~d  & TYC~4490-1174-1 & 11.07       &  vby      &  11.11    &  vby      \\ 
        & TYC~4490-1241-1 & 11.08       &  vby      &  11.12    &  vby      \\ \hline\hline
{\#16}  & TYC~4467-312-1  & 2017.07.18  &  BV       & 2017.07.21&  BV      \\
        & TYC~4466-384-1  & 07.19       &  BV       &  07.23    &  BV      \\ 
        &                 & 07.20       &  BV       &           &          \\ \hline
{\#17}  & HD~174587       & 2017.07.29  &  BV       & 2018.08.05&  BV      \\
        & TYC~1026-2089-1 & 07.30       &  BV       &  08.07    &  BV      \\ 
        &    & 07.31      & BV          &  08.08    &  BV      \\        
        &    & 08.01      & BV          &  08.09    &  BV      \\ \hline
{\#18}  & HD~189851       &2017.09.04   &  BV       & 2017.09.29&  BV      \\
        & HD~189823       &09.07        &  BV       &  09.30    &  BV      \\ 
        &    & 09.27      & BV          &  10.01    &  BV      \\        
        &    & 09.28      & BV          &           &          \\ \hline
{\#19}  & BD+68~1118      & 2017.08.14  &  BV       & 2017.08.26&  BV      \\
        & TYC~4461-1154-1 & 08.20       &  BV       &  08.28    &  BV      \\ 
        &    & 08.23      & BV          &  09.04    &  BV      \\        
        &    & 08.24      & BV          &           &          \\ \hline
{\#20}  & BD+61~155       & 2017.10.15  &  BV       & 2017.10.17&  BV      \\
        & TYC~4020-1206-1 & 10.16       &  BV       &           &          \\ \hline
              
}

\begin{figure}[]
\centerline{%
\begin{tabular}{l@{\hspace{0.1pc}}}
\includegraphics[width=.5\linewidth]{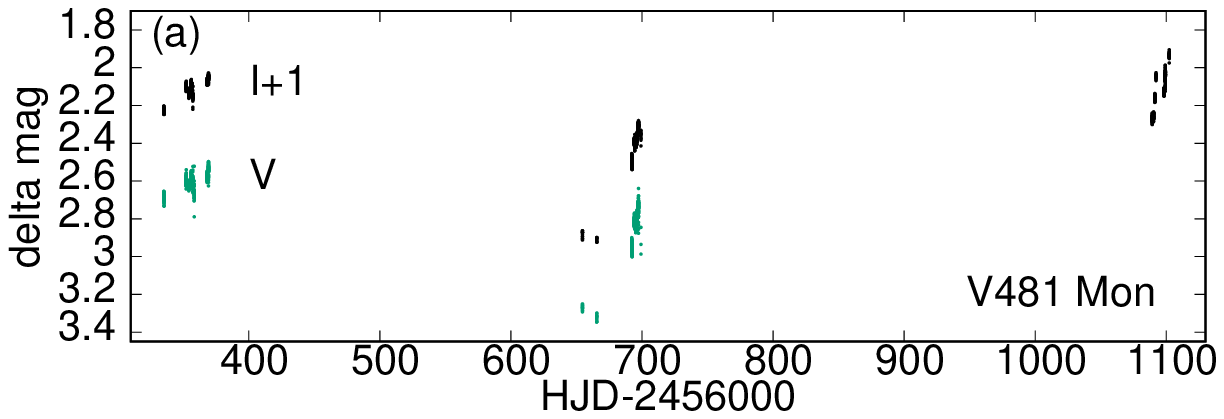} 
\includegraphics[width=.5\linewidth]{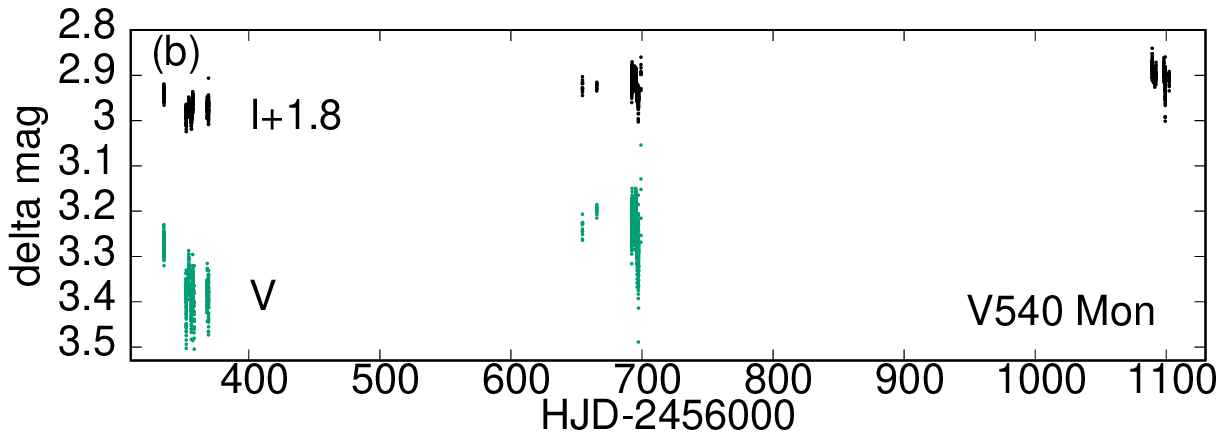}\\
\includegraphics[width=.5\linewidth]{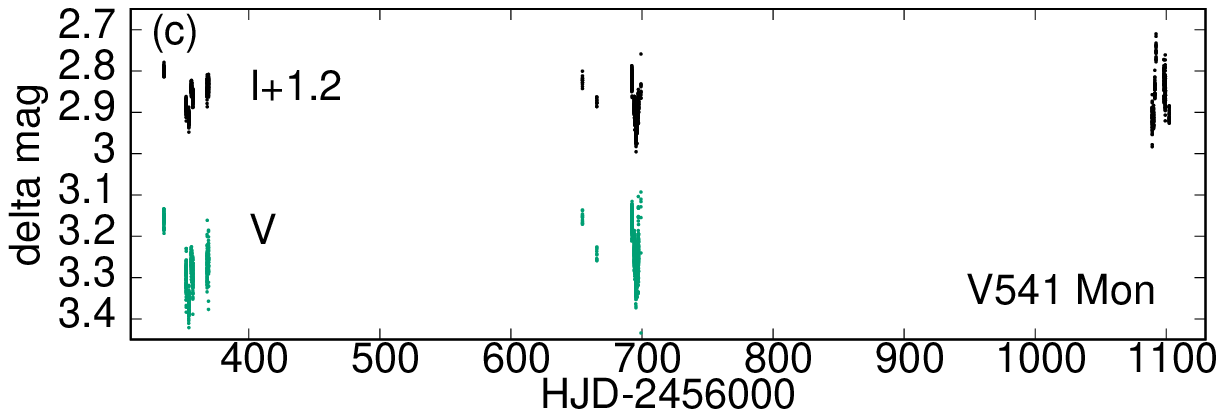}  
\includegraphics[width=.5\linewidth]{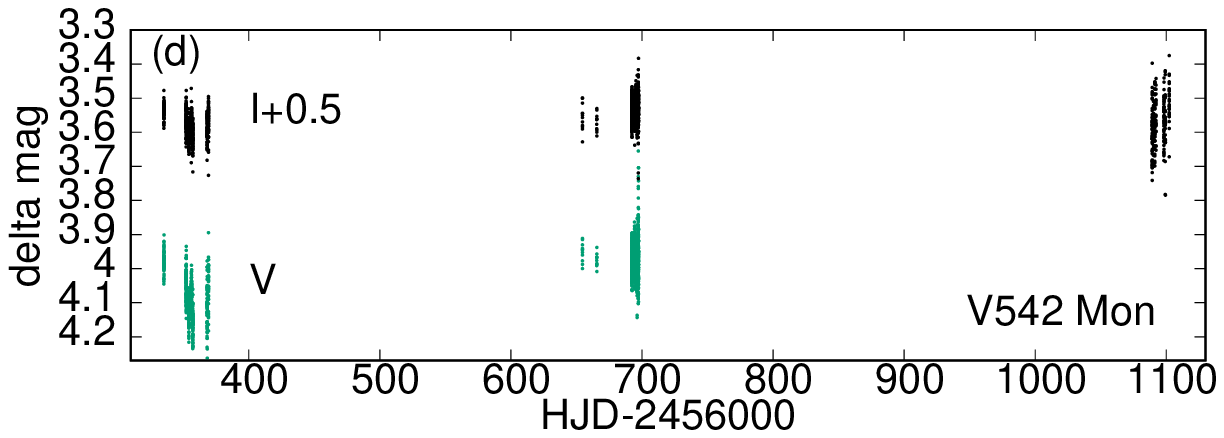}\\ 
\includegraphics[width=.5\linewidth]{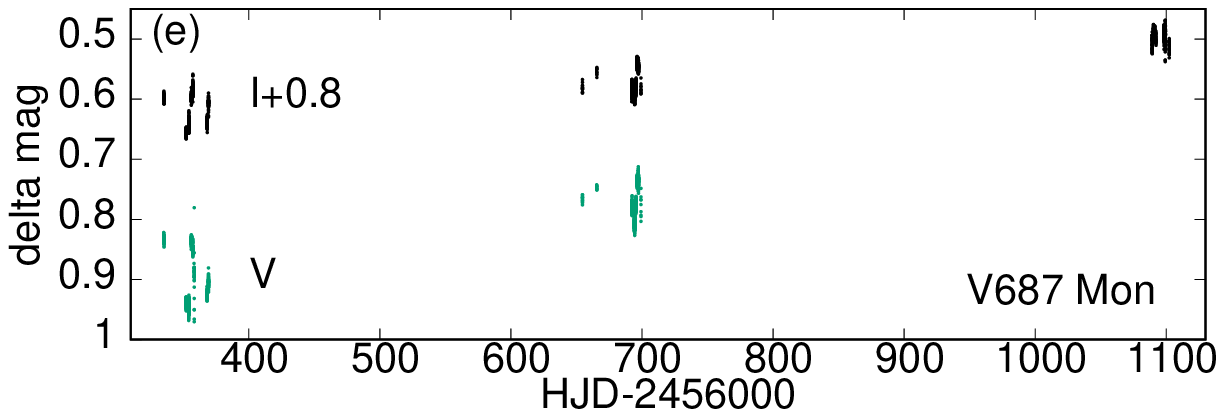}
\includegraphics[width=.5\linewidth]{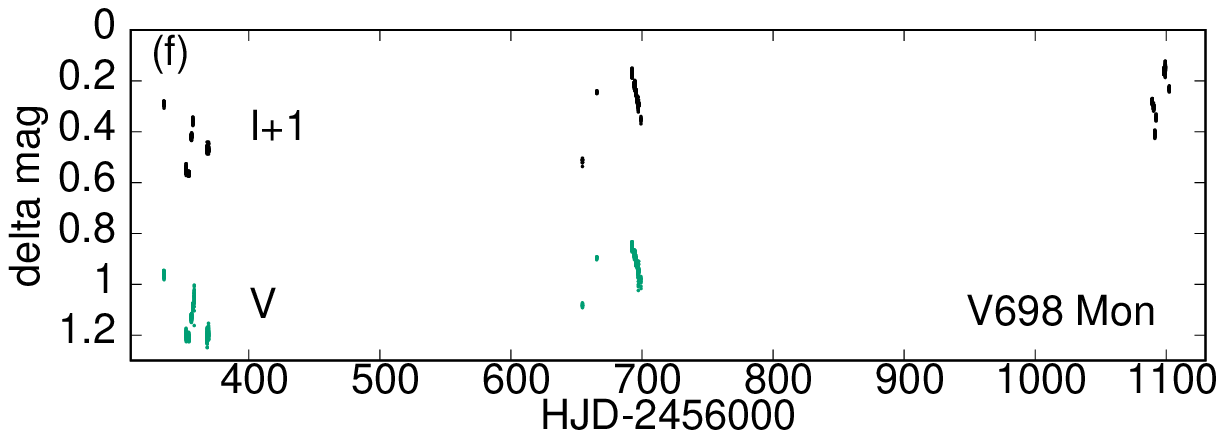}\\
\includegraphics[width=.5\linewidth]{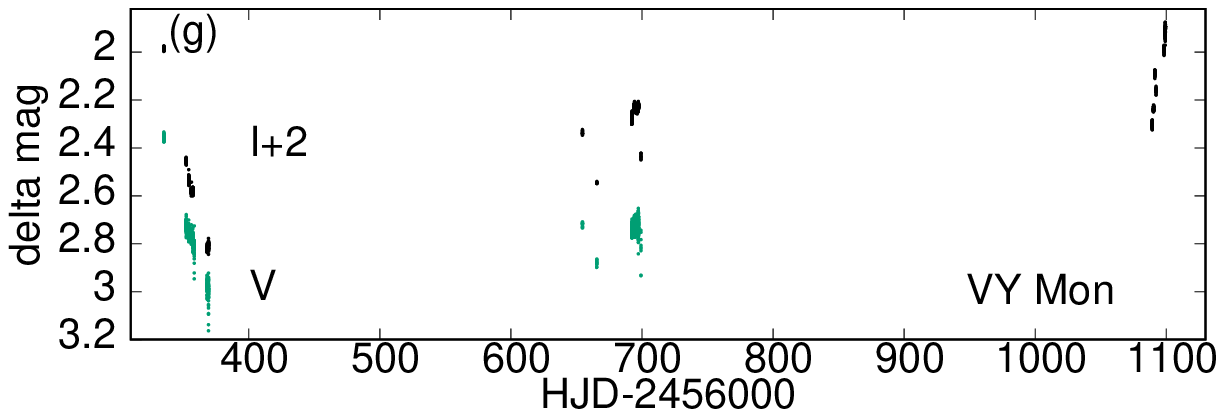}
\includegraphics[width=.5\linewidth]{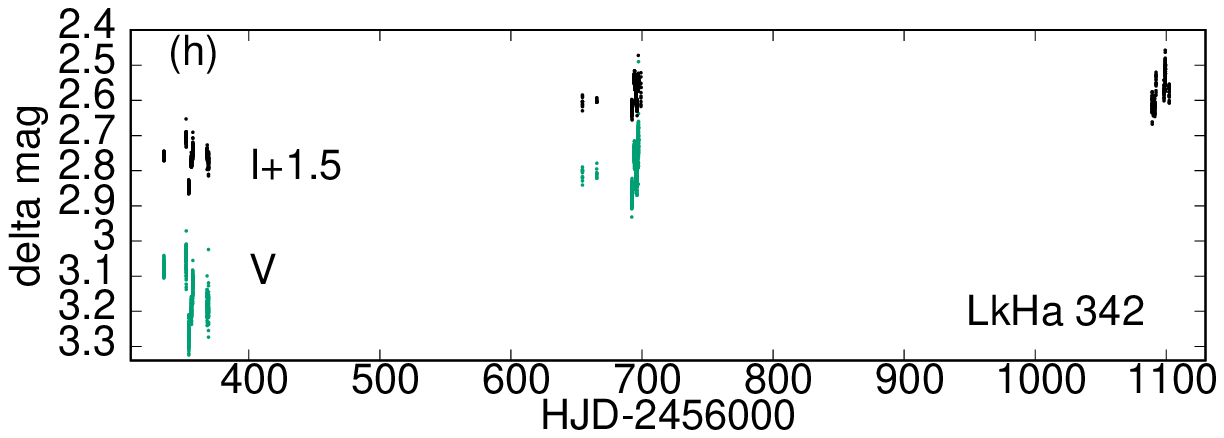}\\
\includegraphics[width=.5\linewidth]{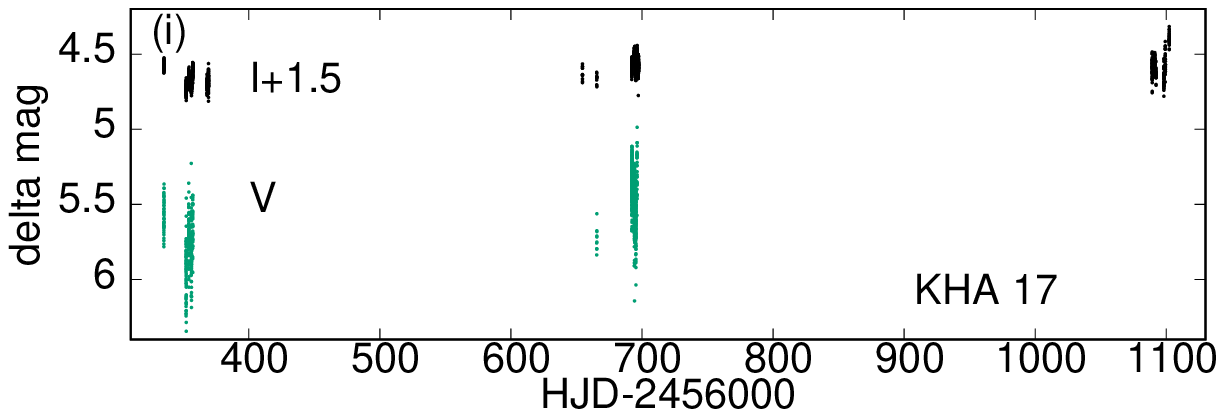}
\includegraphics[width=.5\linewidth]{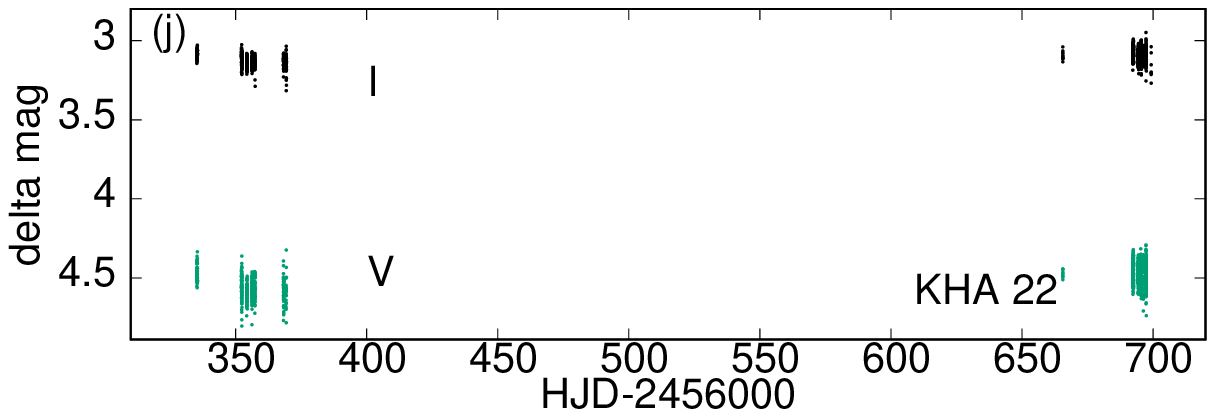}\\
\includegraphics[width=.5\linewidth]{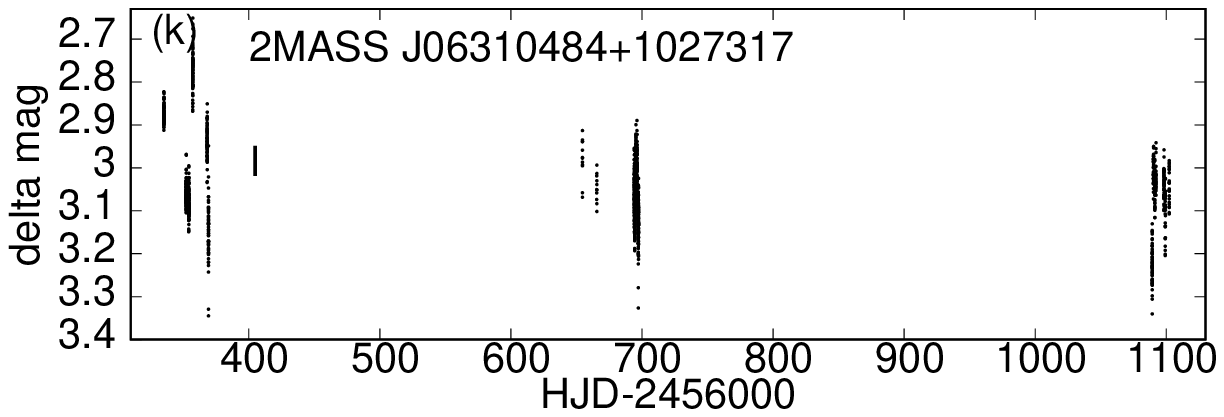}
\includegraphics[width=.5\linewidth]{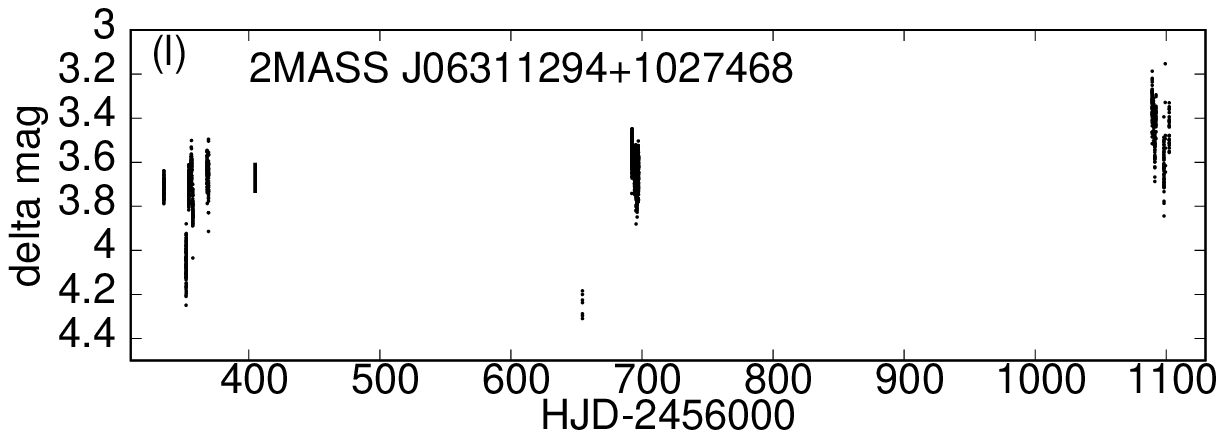}\\
\includegraphics[width=.5\linewidth]{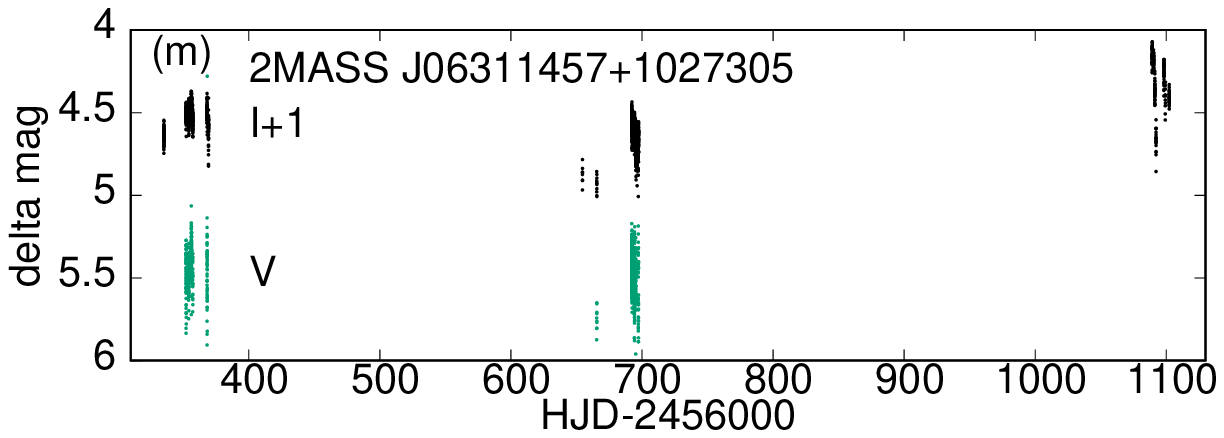}
\end{tabular}}
\FigCap{Results of 2013, 2014 and 2015 runs for variable stars from field \#1. Observations in $V$-filter 
were obtained during the first two seasons only. In addition, extraction of photometry in this band was 
impossible during nights with poor seeing conditions and/or bright Moon. 
Only data for stars, which evidently show any kind of variability, are shown.}
\end{figure}

\begin{figure}[]
\centerline{%
\begin{tabular}{l@{\hspace{0.1pc}}}
\includegraphics[width=.5\linewidth]{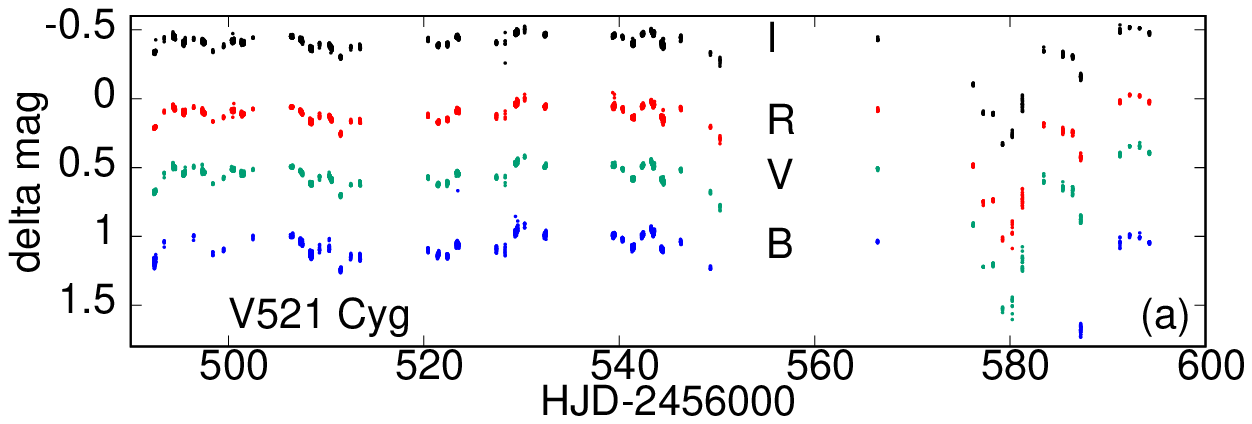} 
\includegraphics[width=.5\linewidth]{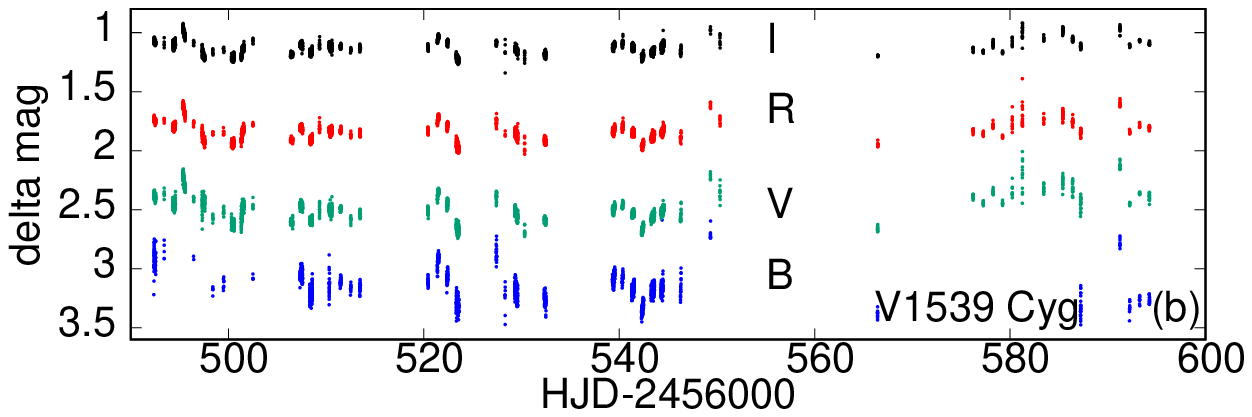}\\   
\includegraphics[width=.5\linewidth]{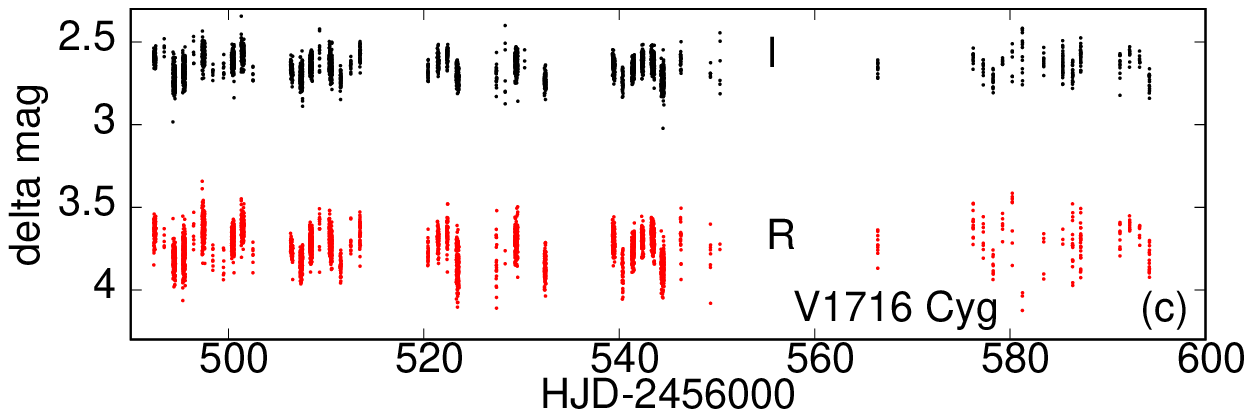}
\includegraphics[width=.5\linewidth]{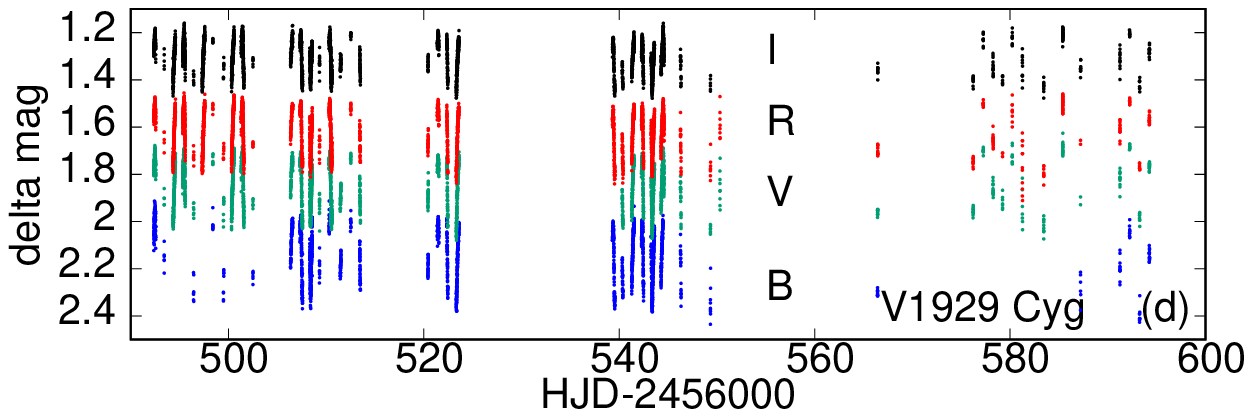}\\  
\includegraphics[width=.5\linewidth]{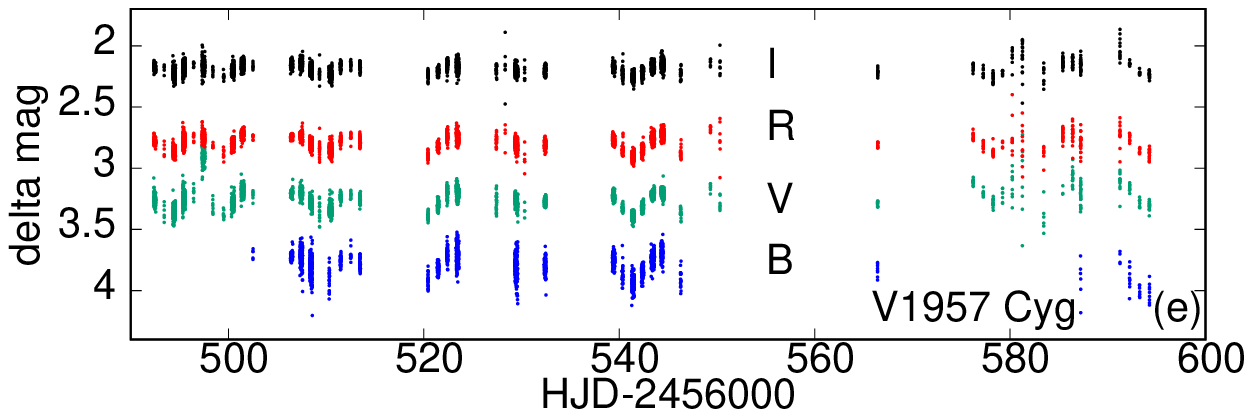}
\includegraphics[width=.5\linewidth]{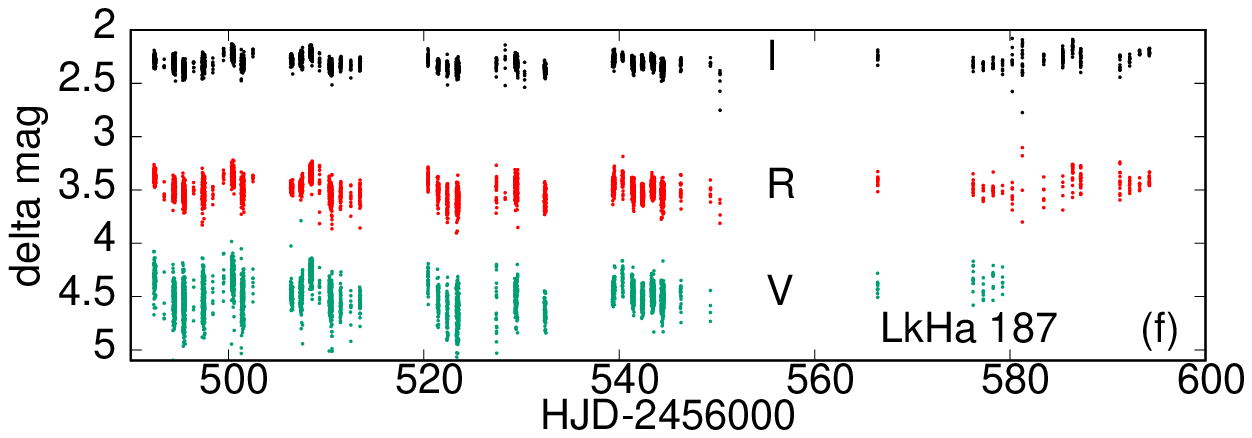}\\  
\includegraphics[width=.5\linewidth]{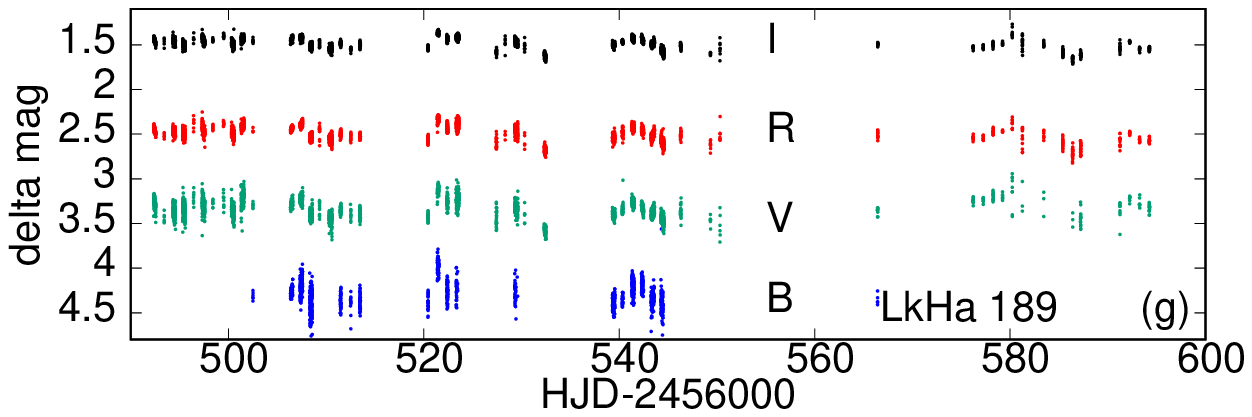}
\includegraphics[width=.5\linewidth]{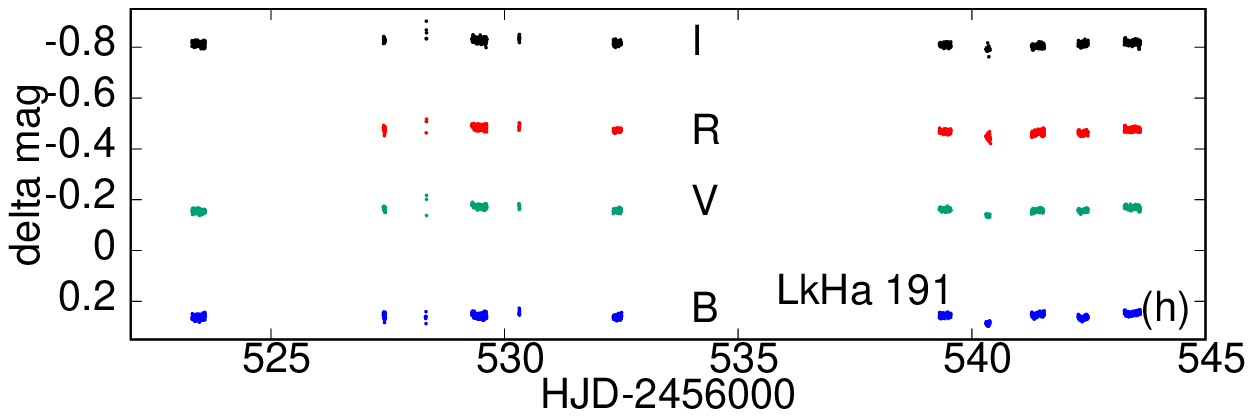}\\
\includegraphics[width=.5\linewidth]{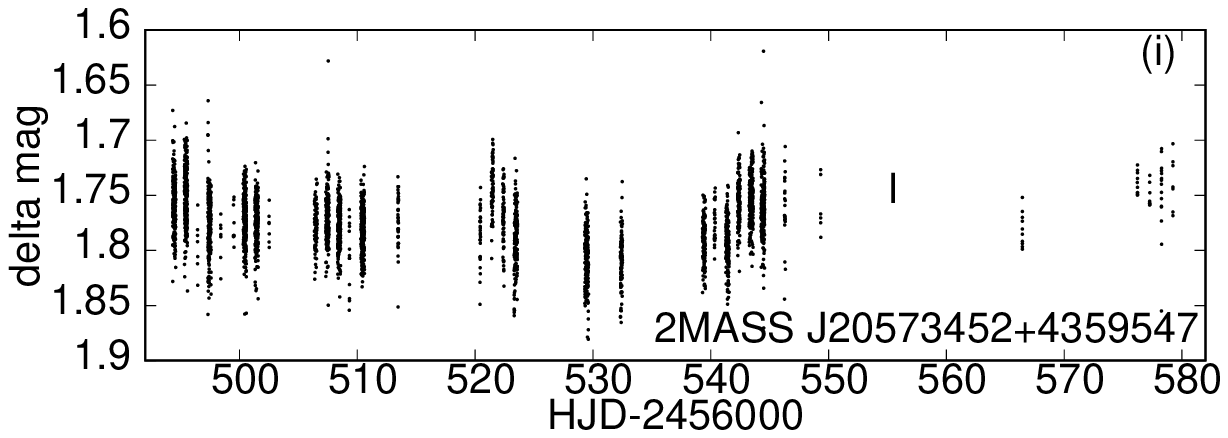}
\includegraphics[width=.5\linewidth]{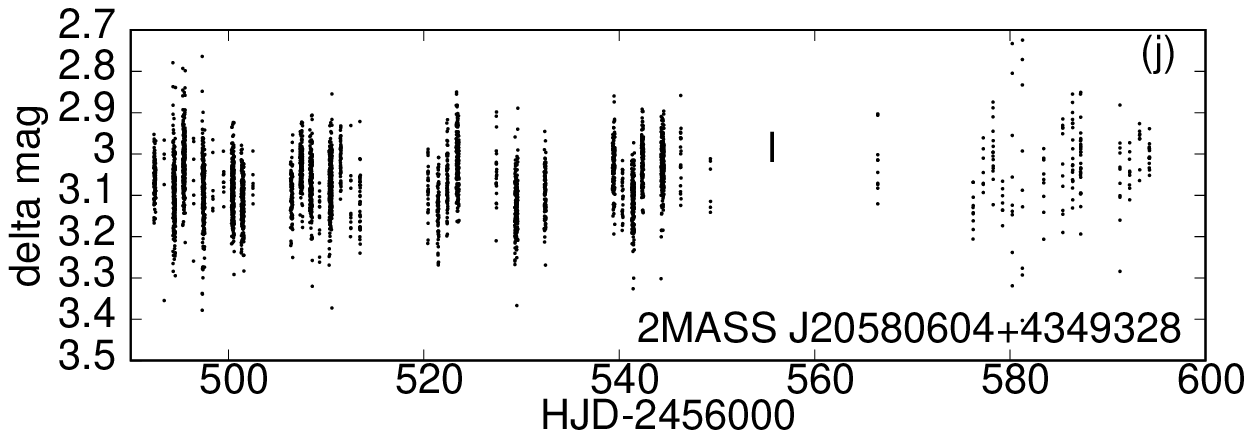}\\
\includegraphics[width=.5\linewidth]{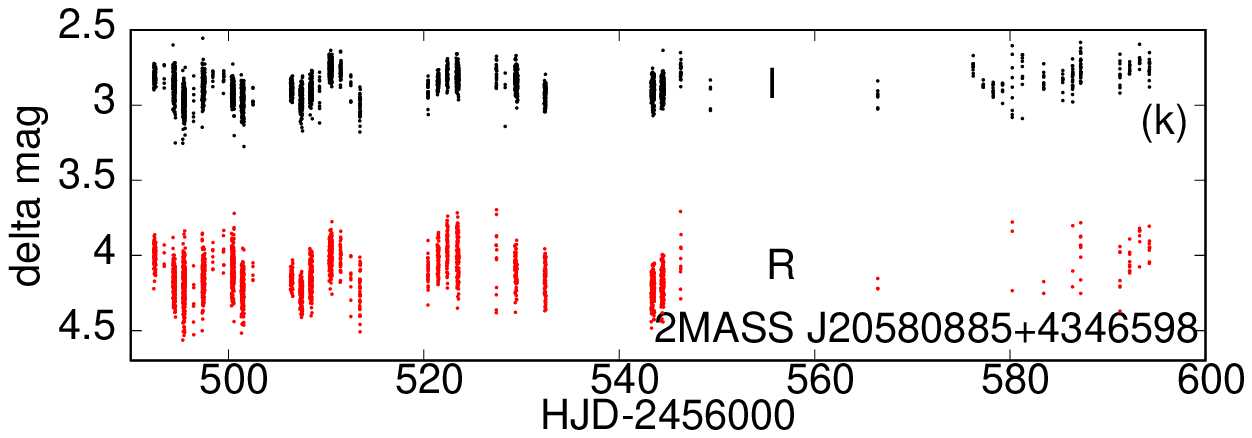}
\includegraphics[width=.5\linewidth]{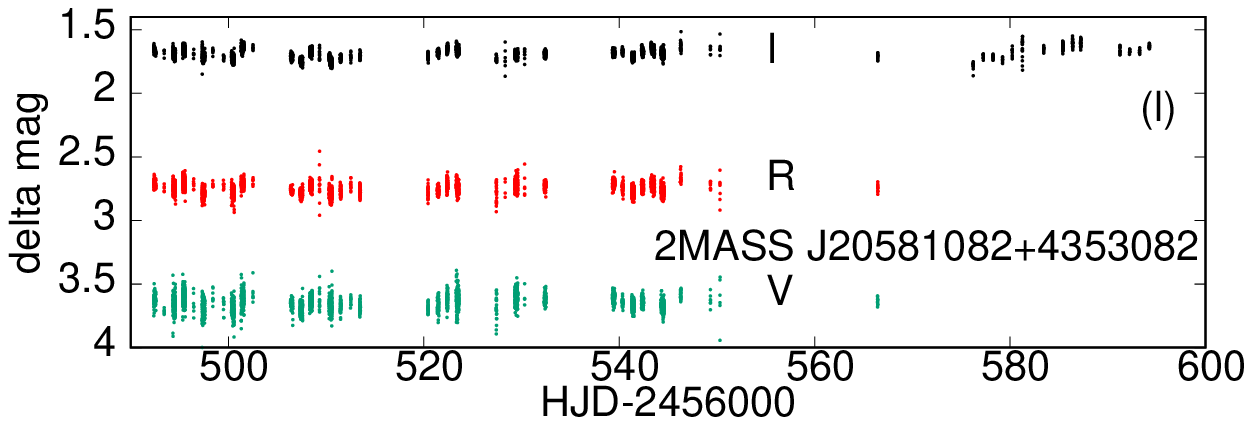}\\
\includegraphics[width=.5\linewidth]{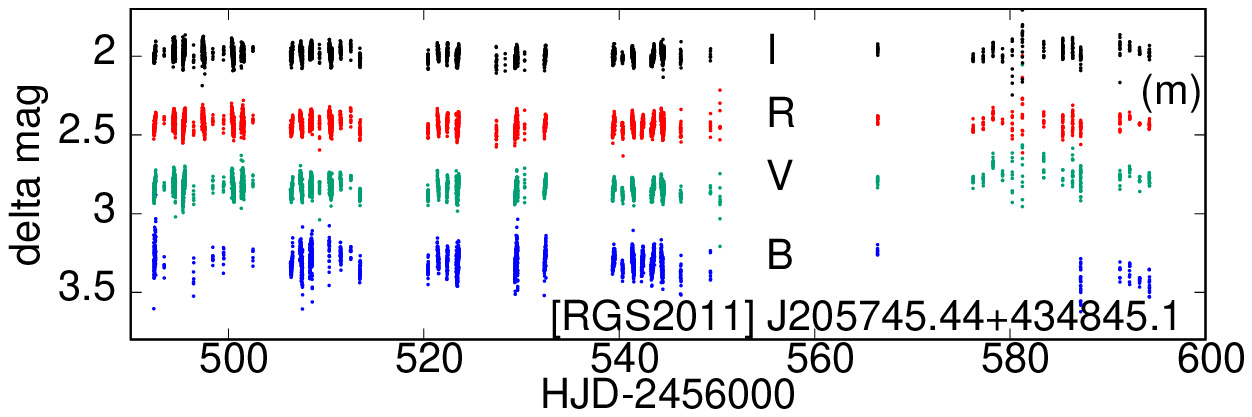}
\includegraphics[width=.5\linewidth]{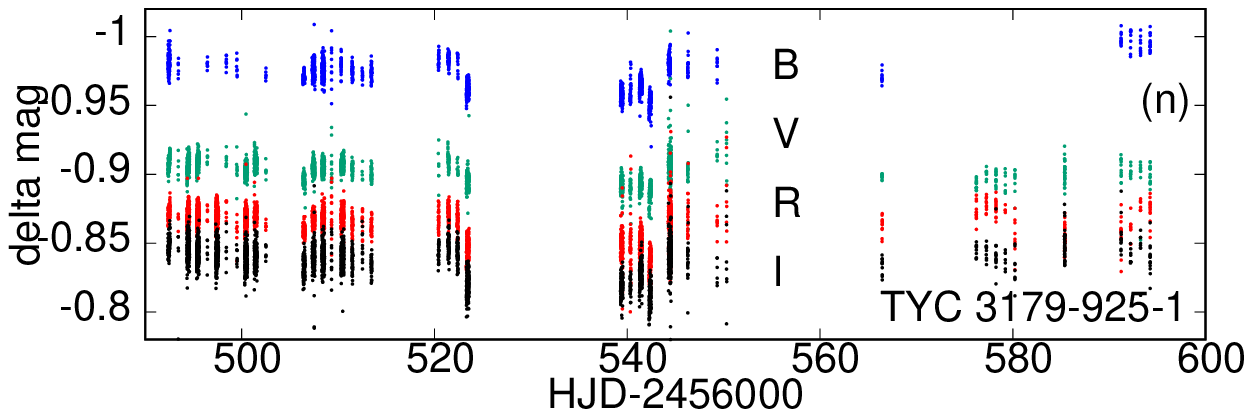}\\
\includegraphics[width=.5\linewidth]{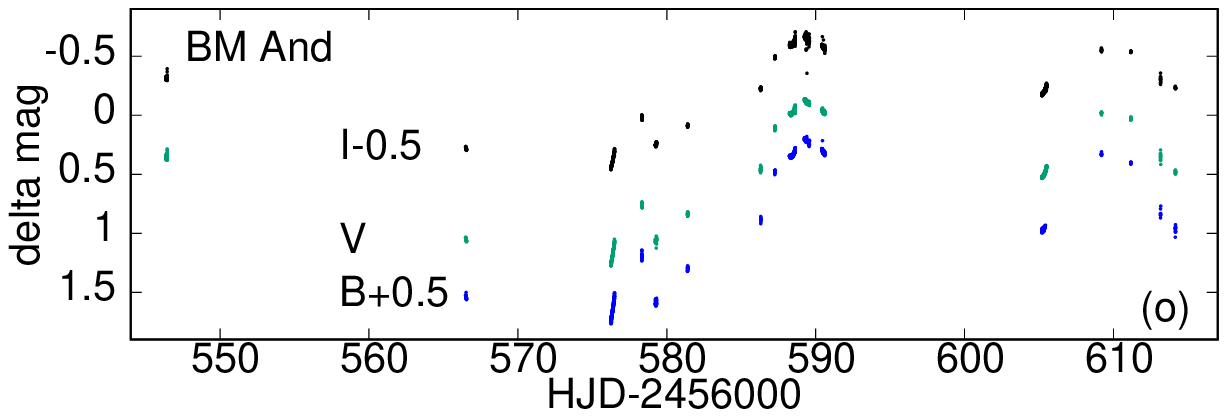}
\end{tabular}}
\FigCap{Results for young variable stars from field \#2 (''Gulf Mexico'') and BM~And (field \#3).}
\end{figure}

\begin{figure}[]
\centerline{%
\begin{tabular}{l@{\hspace{0.1pc}}}
\includegraphics[width=.5\linewidth]{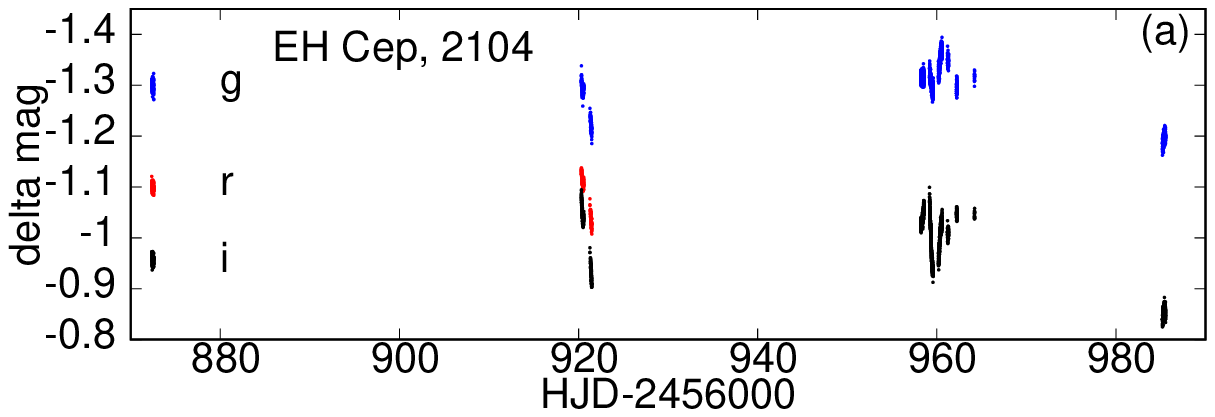} 
\includegraphics[width=.5\linewidth]{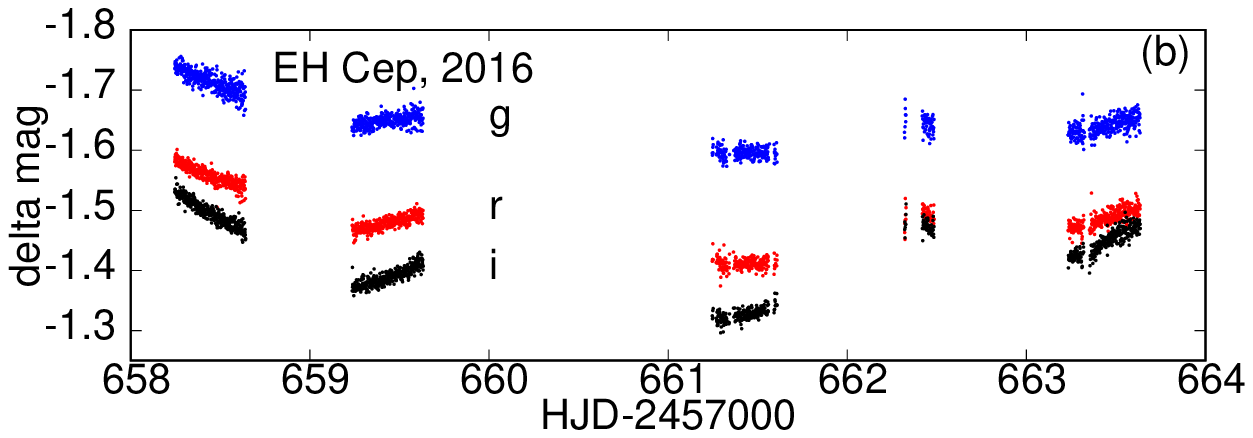}\\
\includegraphics[width=.5\linewidth]{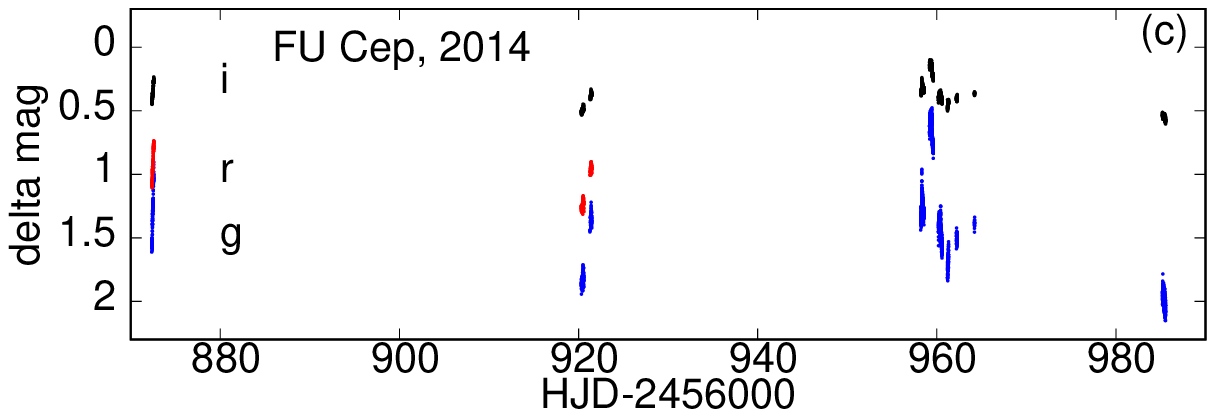}  
\includegraphics[width=.5\linewidth]{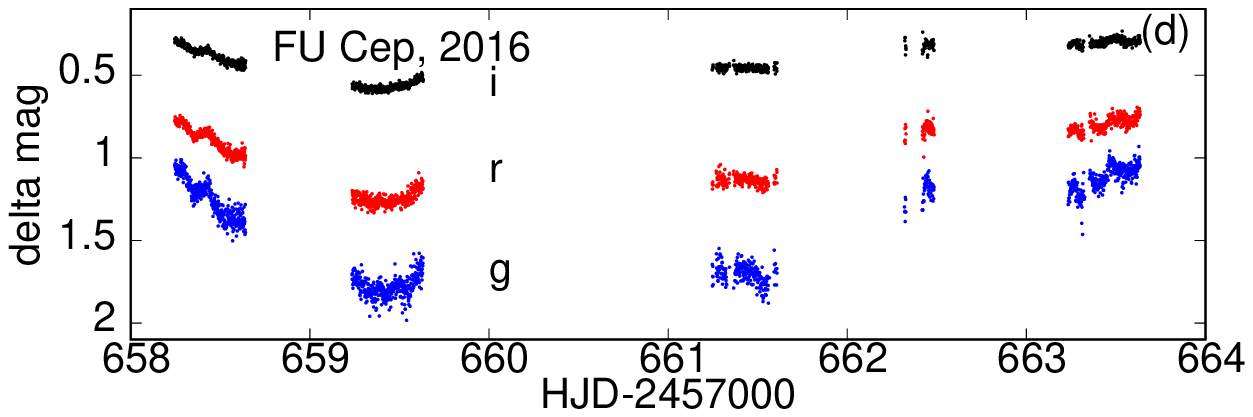}\\ 
\includegraphics[width=.5\linewidth]{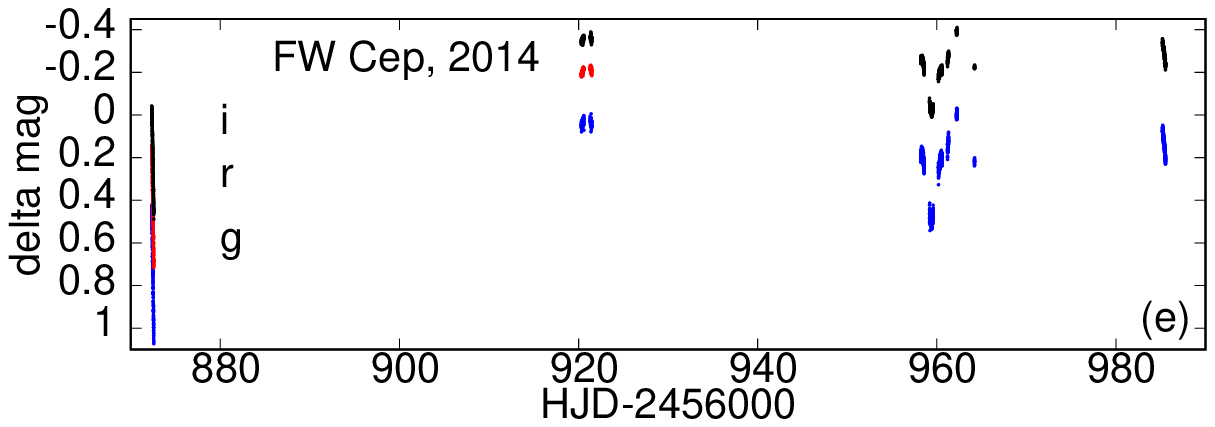}
\includegraphics[width=.5\linewidth]{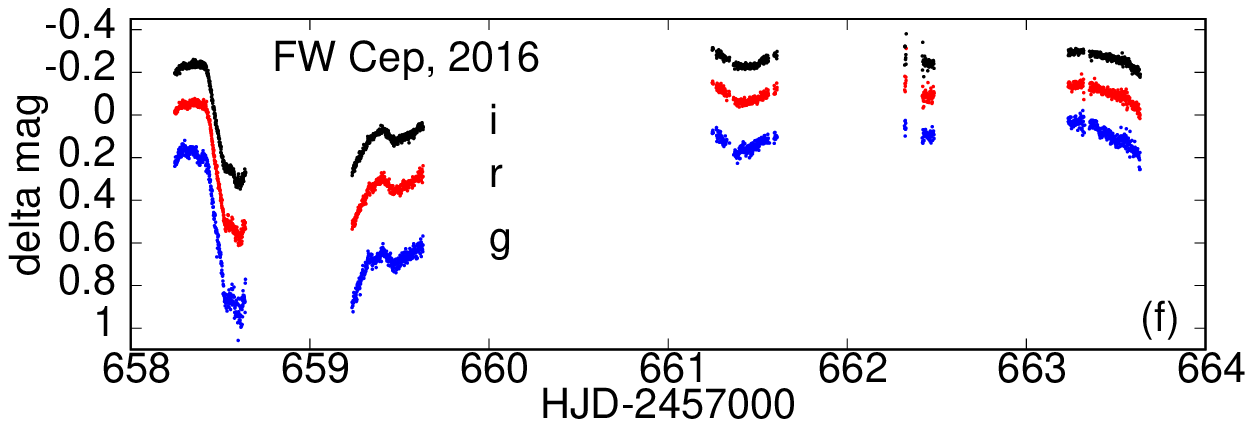}\\
\includegraphics[width=.5\linewidth]{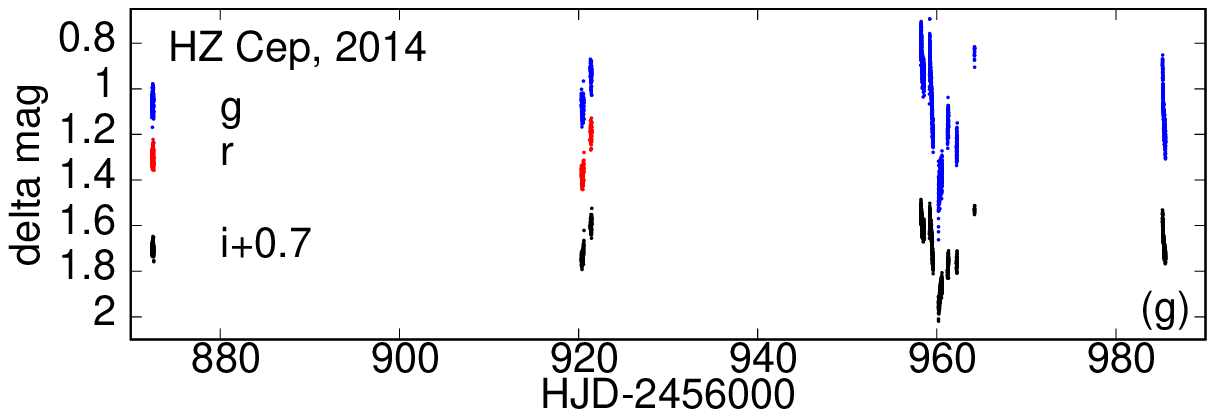}
\includegraphics[width=.5\linewidth]{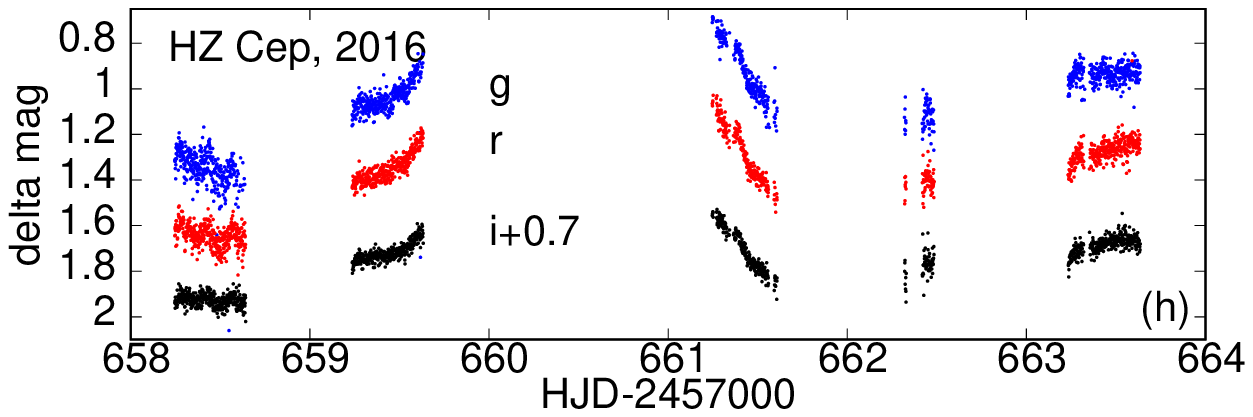}\\
\includegraphics[width=.5\linewidth]{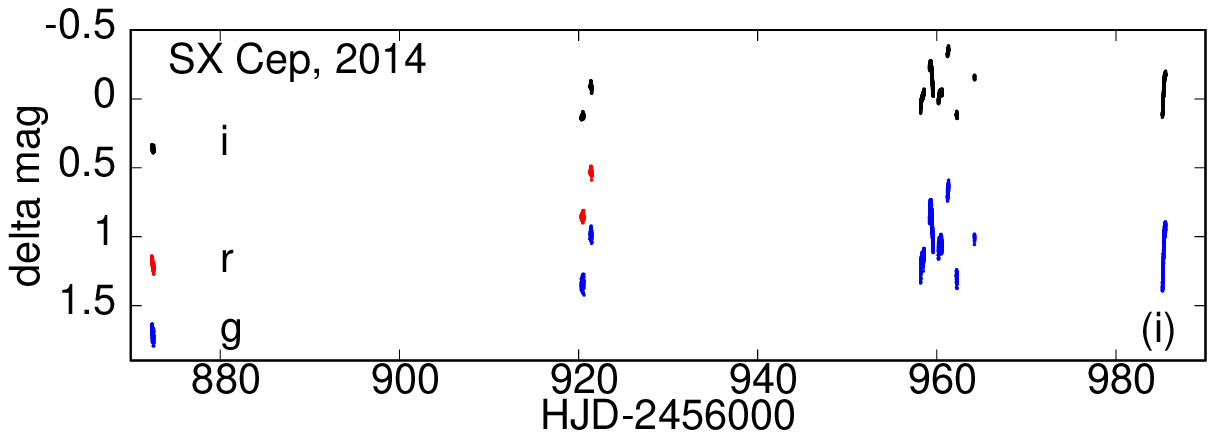} 
\includegraphics[width=.5\linewidth]{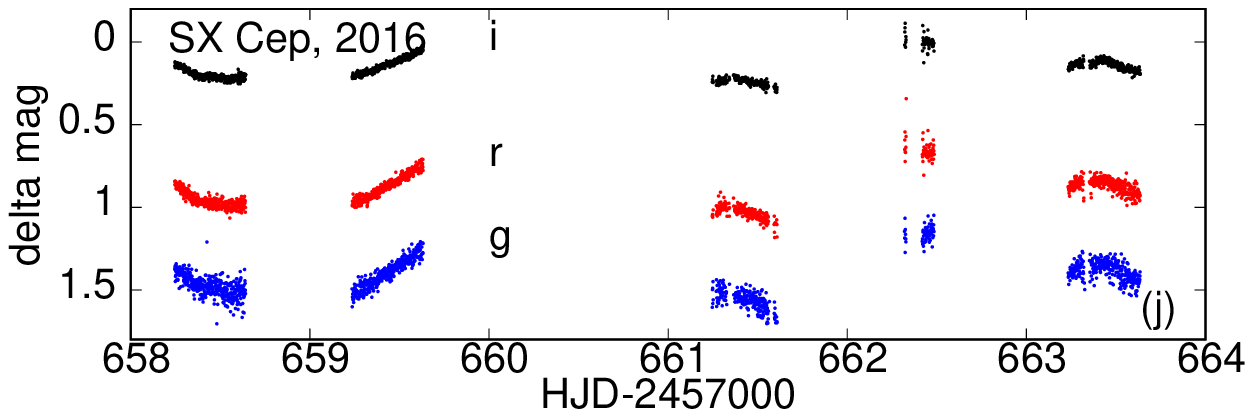}\\
\includegraphics[width=.5\linewidth]{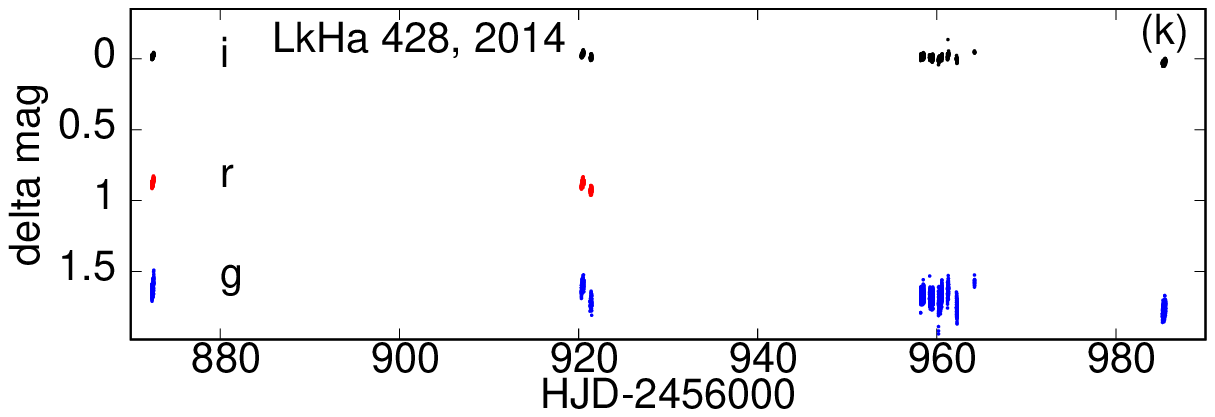}
\includegraphics[width=.5\linewidth]{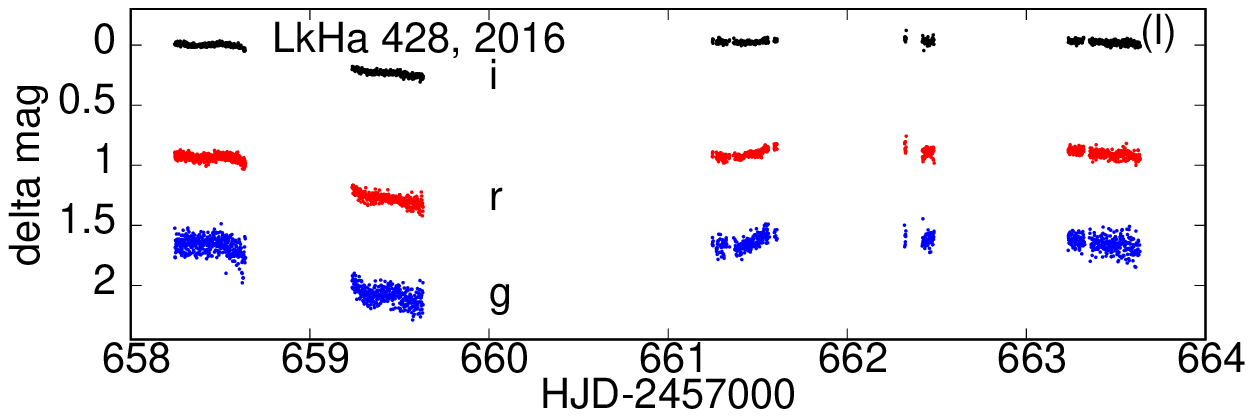}\\
\includegraphics[width=0.5\linewidth]{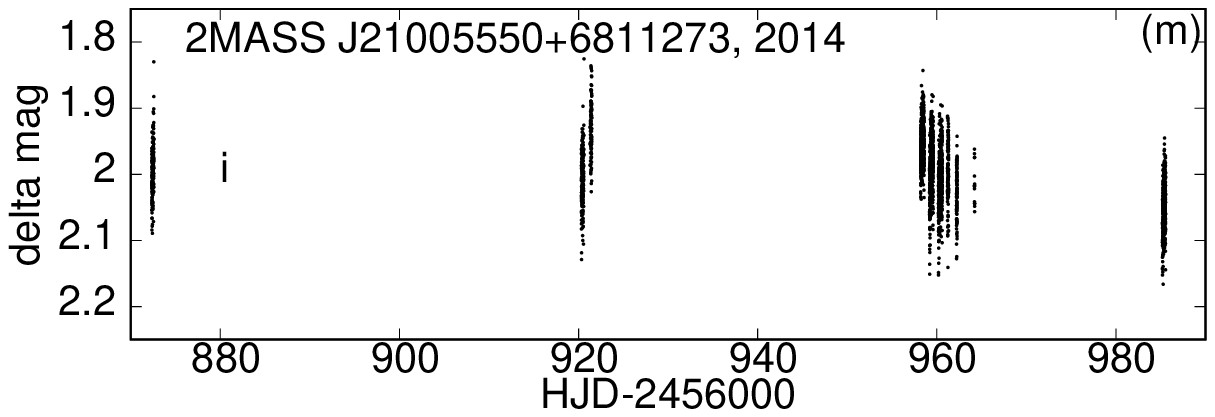}
\includegraphics[width=0.5\linewidth]{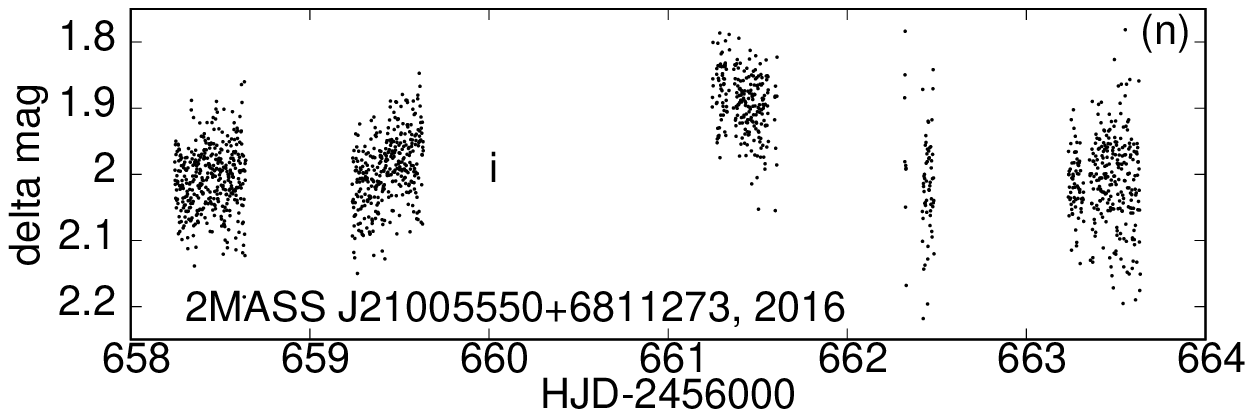}\\
\includegraphics[width=0.5\linewidth]{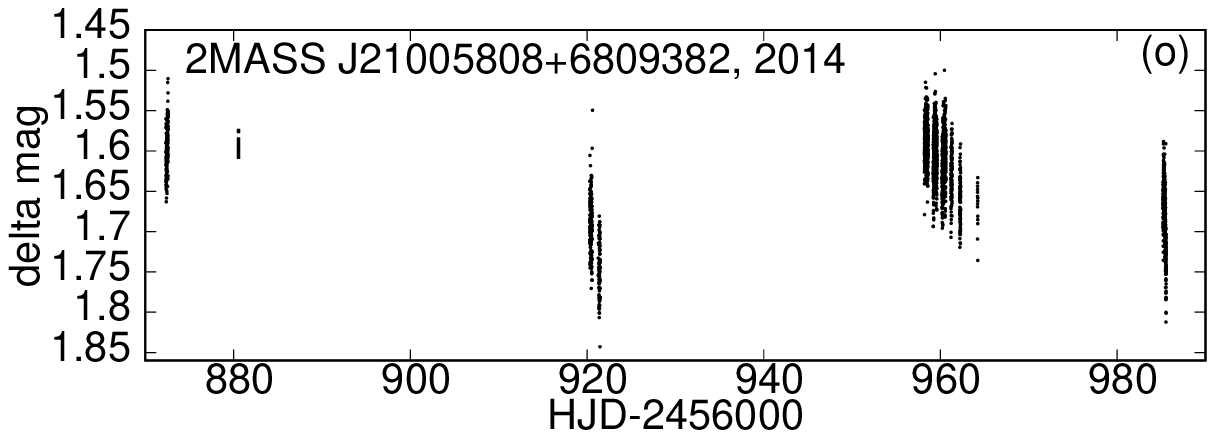}
\includegraphics[width=0.5\linewidth]{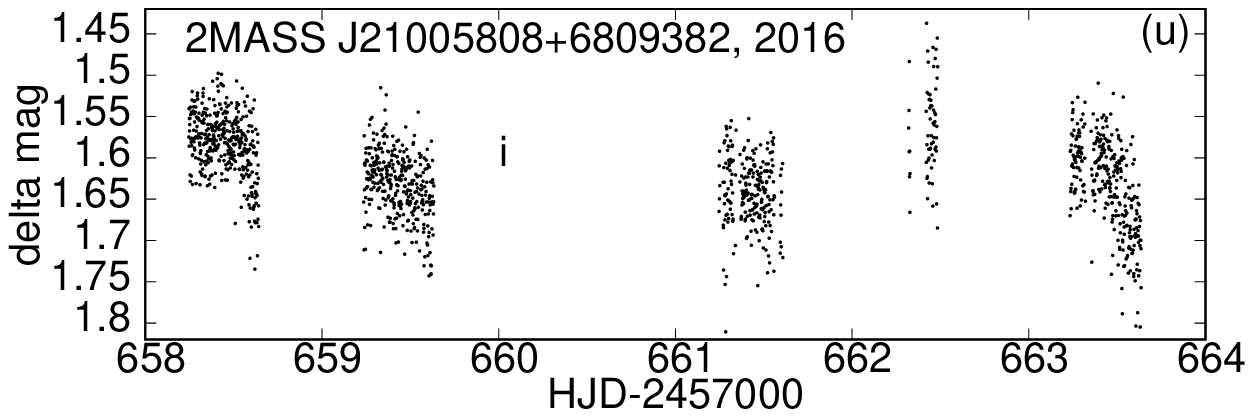} 
\end{tabular}}
\FigCap{Results of 2014 and 2016 observations of young variable stars from field \#5 (NGC~7023).}
\end{figure}

\begin{figure}[]
\centerline{%
\begin{tabular}{l@{\hspace{0.1pc}}}
\includegraphics[width=0.5\linewidth]{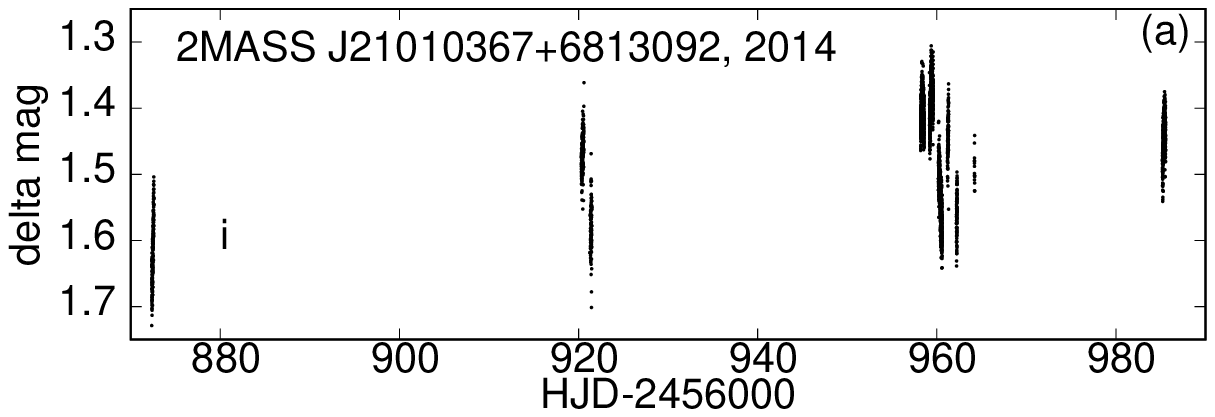}  
\includegraphics[width=0.5\linewidth]{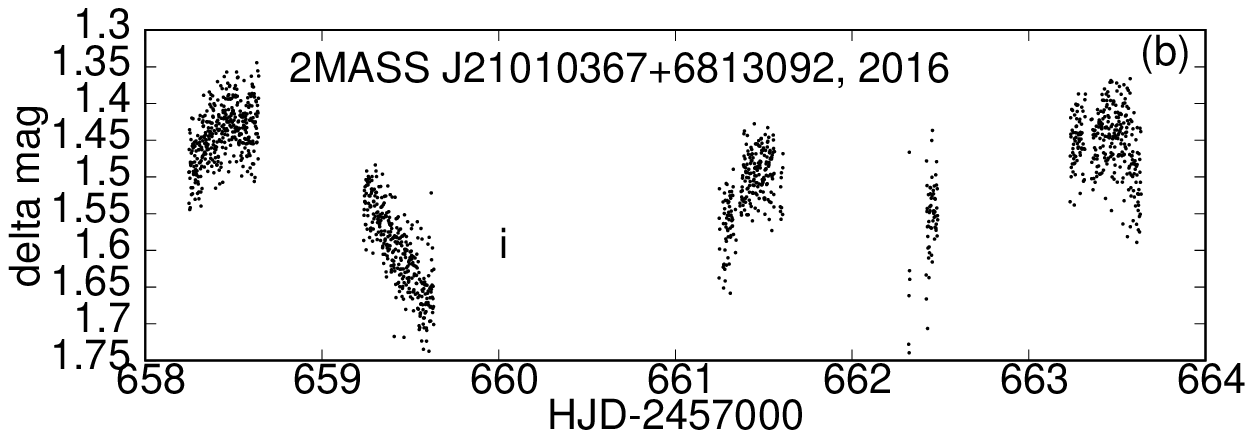}\\
\includegraphics[width=0.5\linewidth]{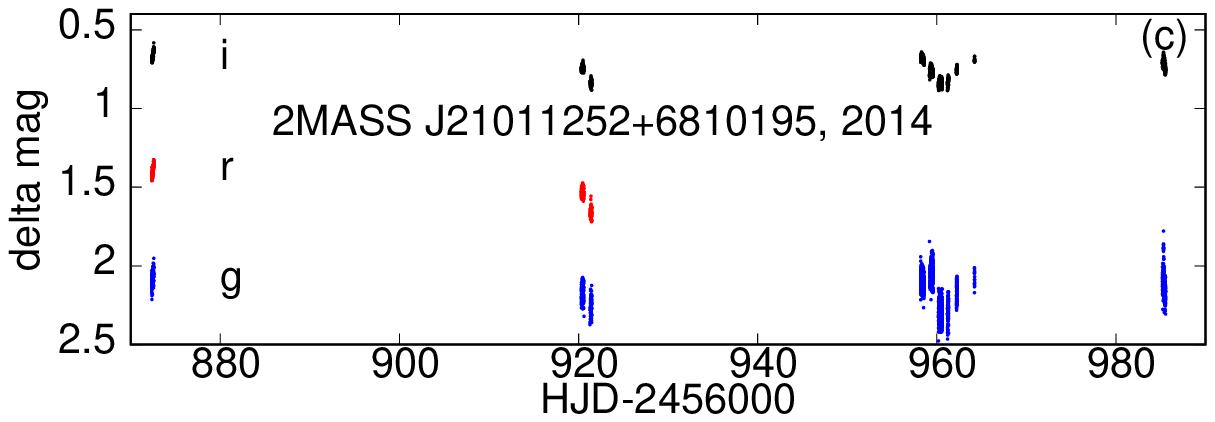} 
\includegraphics[width=0.5\linewidth]{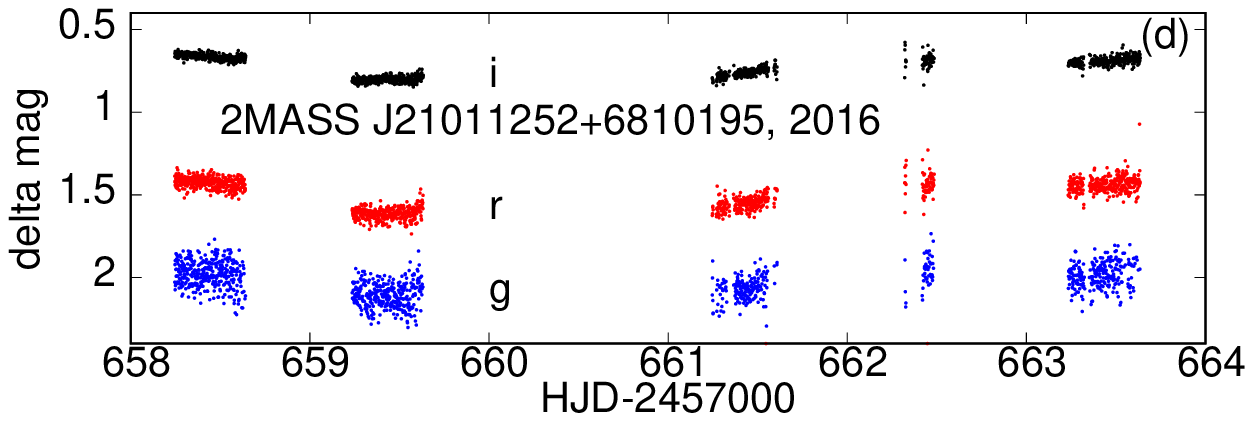}\\
\includegraphics[width=0.5\linewidth]{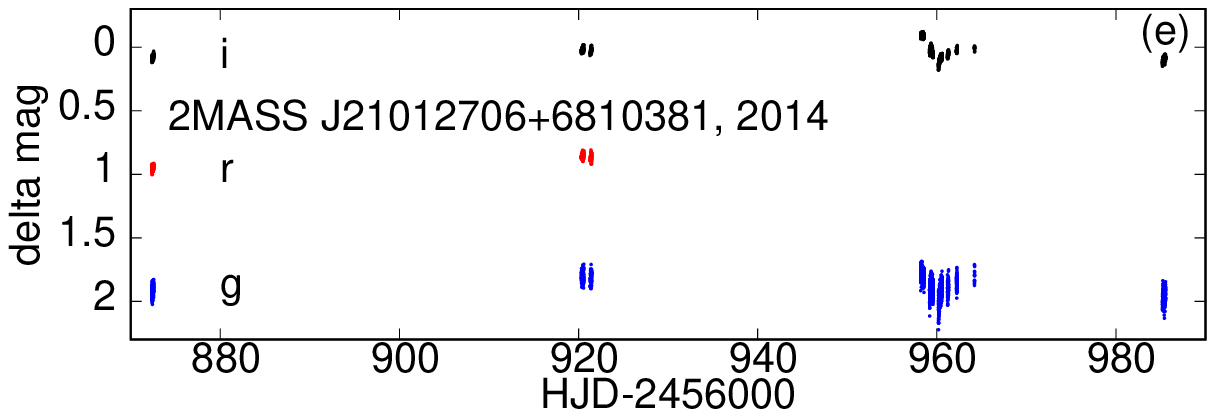}
\includegraphics[width=0.5\linewidth]{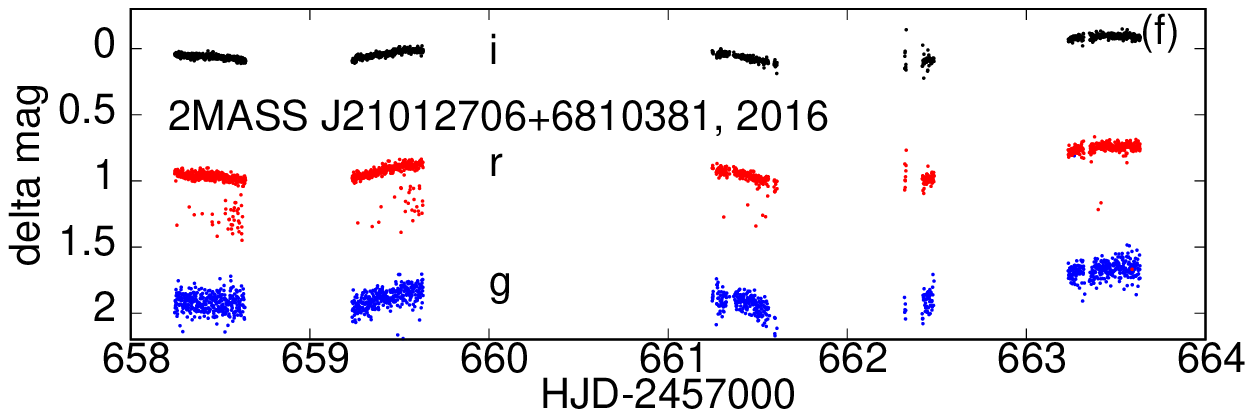}\\
\includegraphics[width=0.5\linewidth]{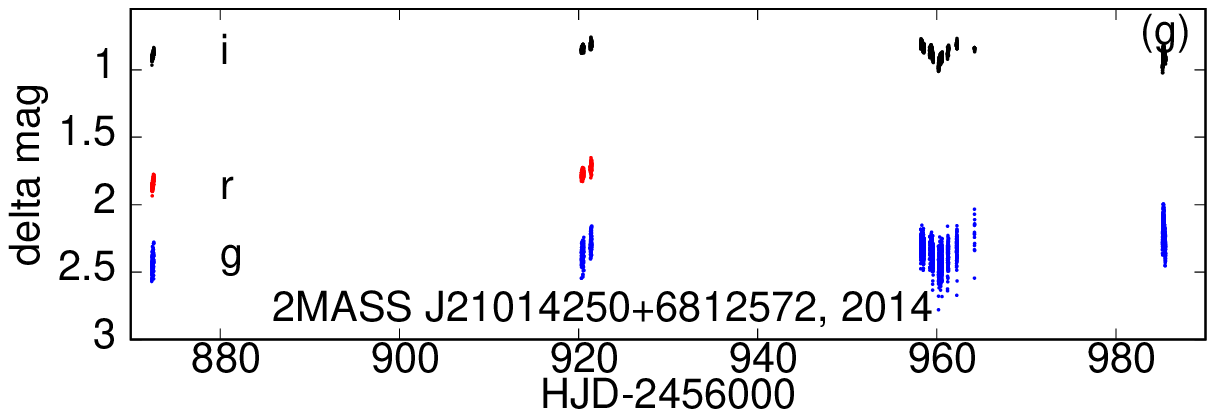}
\includegraphics[width=0.5\linewidth]{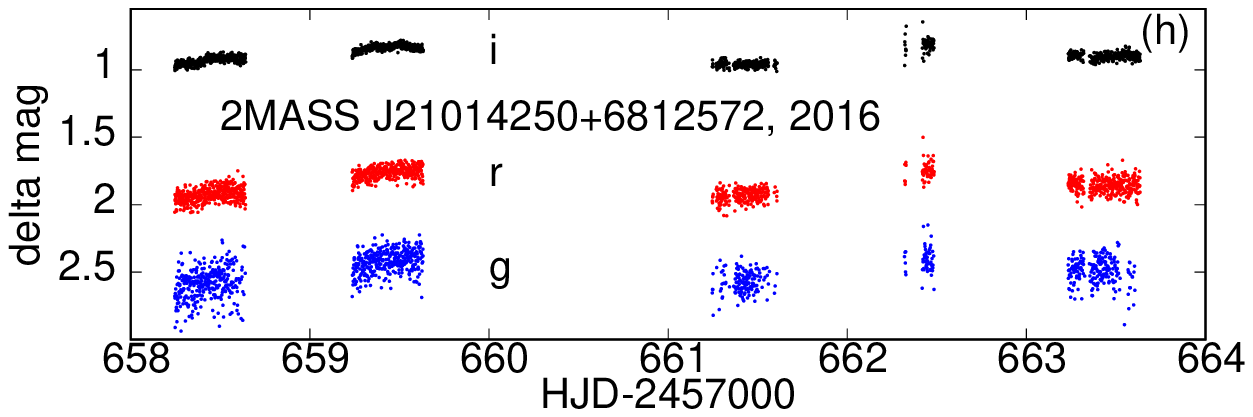}\\
\includegraphics[width=0.5\linewidth]{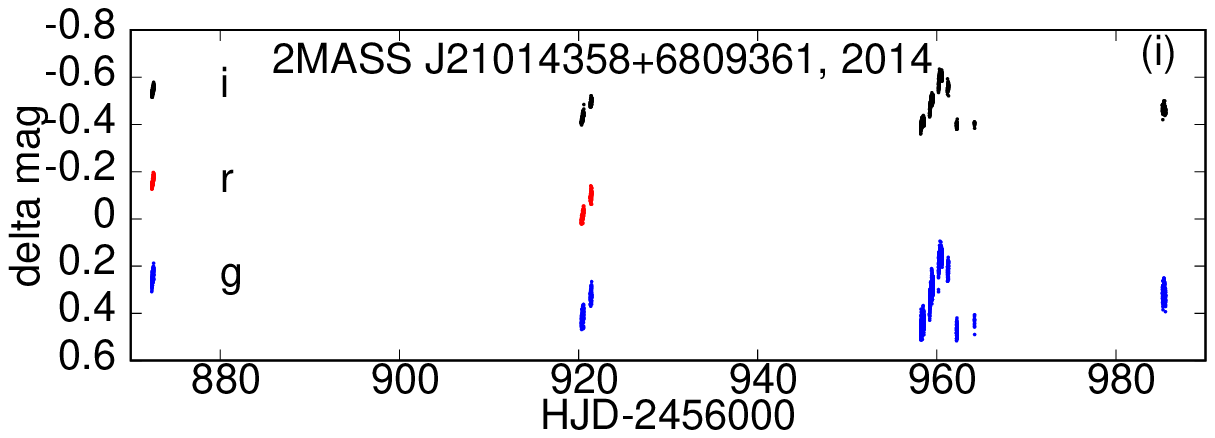}
\includegraphics[width=0.5\linewidth]{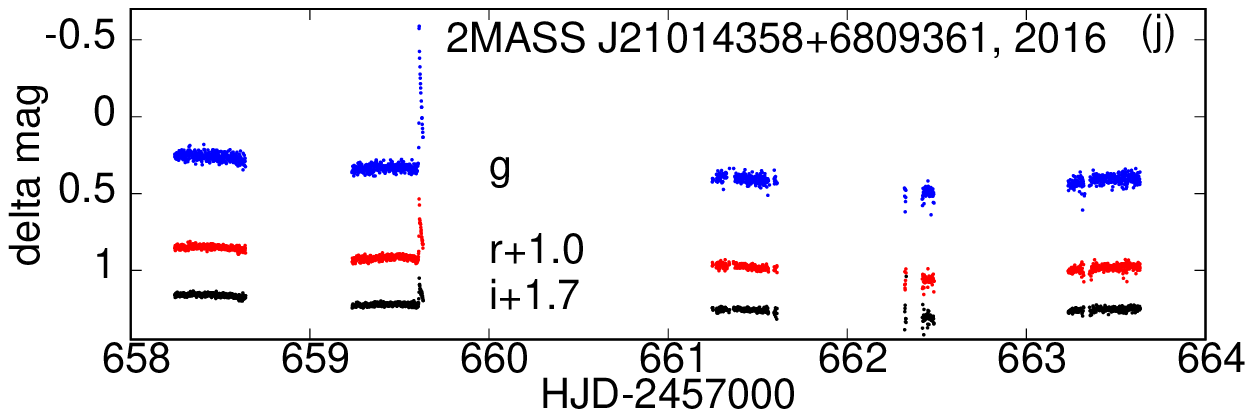}\\
\includegraphics[width=0.5\linewidth]{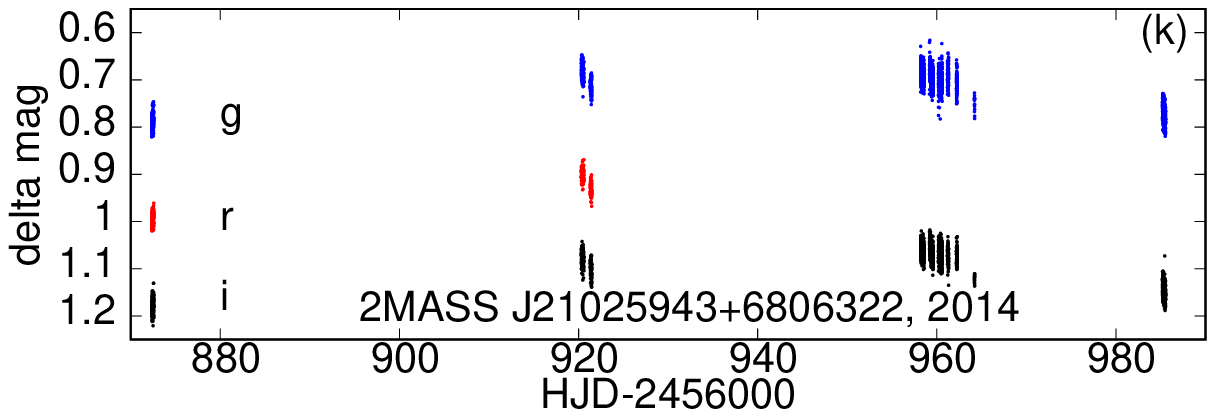}
\includegraphics[width=0.5\linewidth]{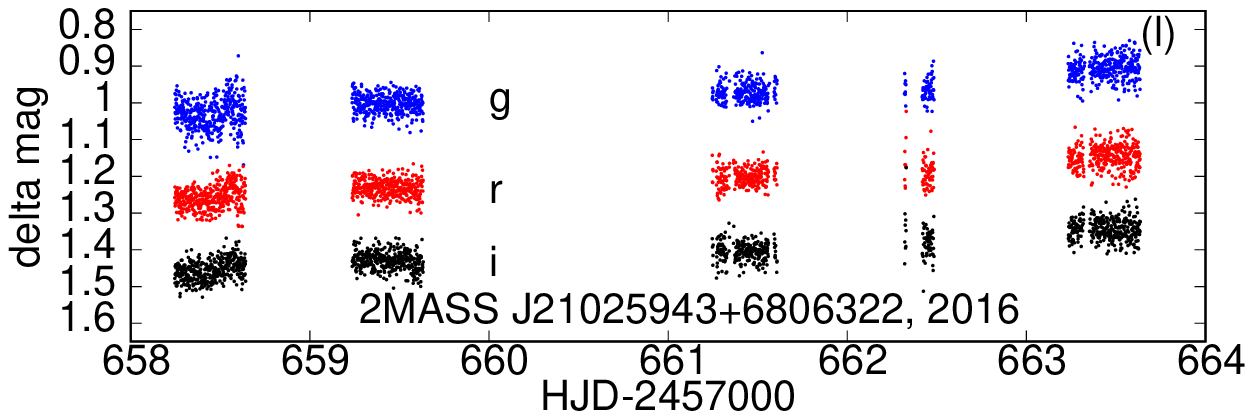}\\
\includegraphics[width=0.5\linewidth]{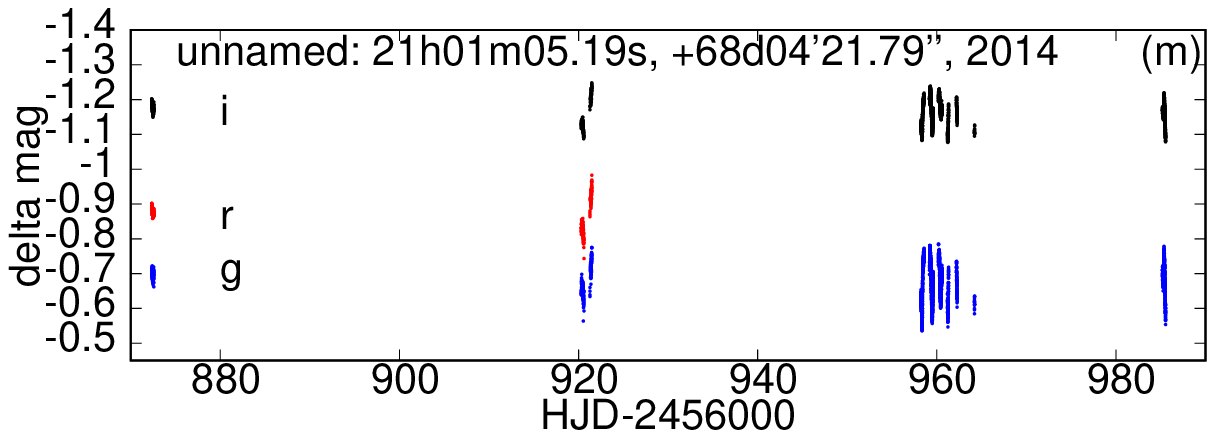}
\includegraphics[width=0.5\linewidth]{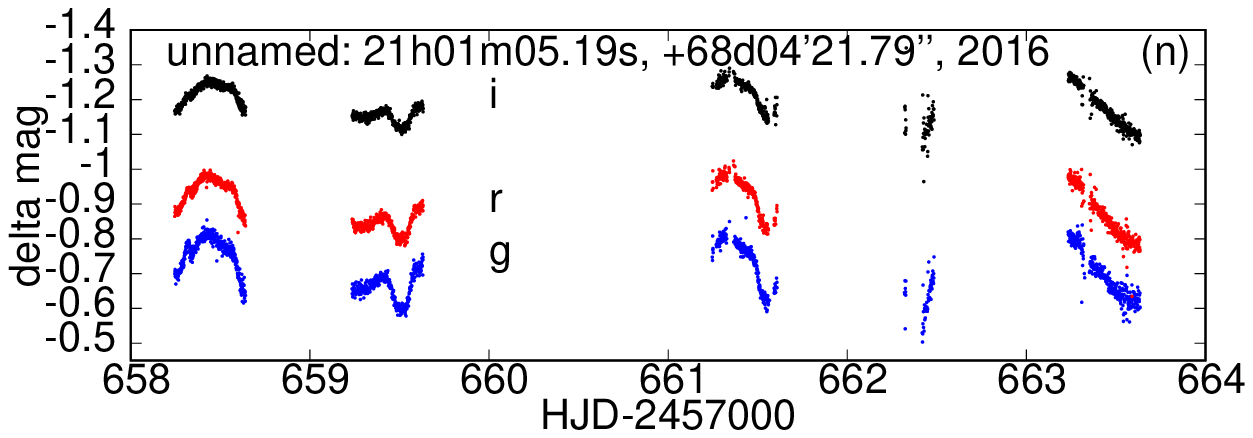}\\
\includegraphics[width=0.5\linewidth]{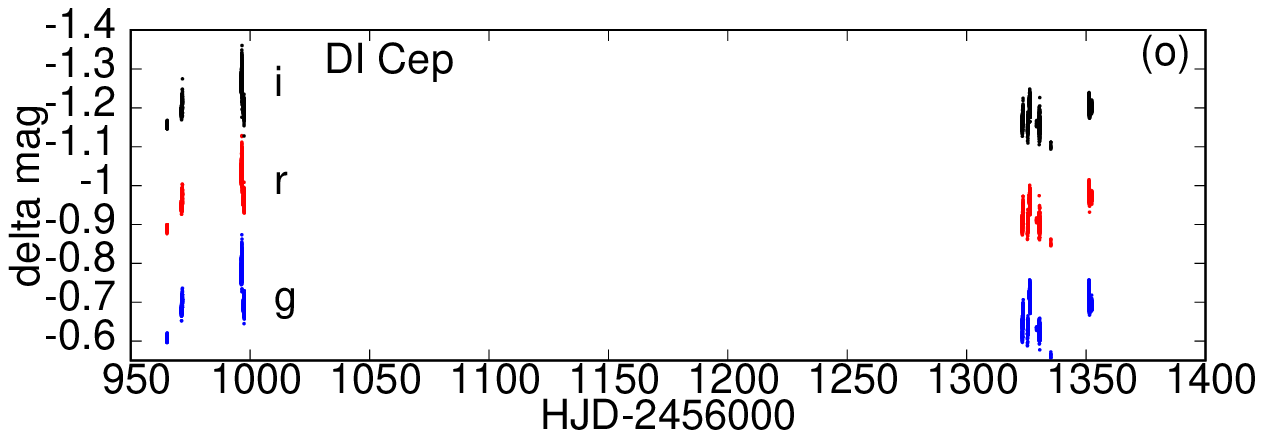}
\includegraphics[width=0.5\linewidth]{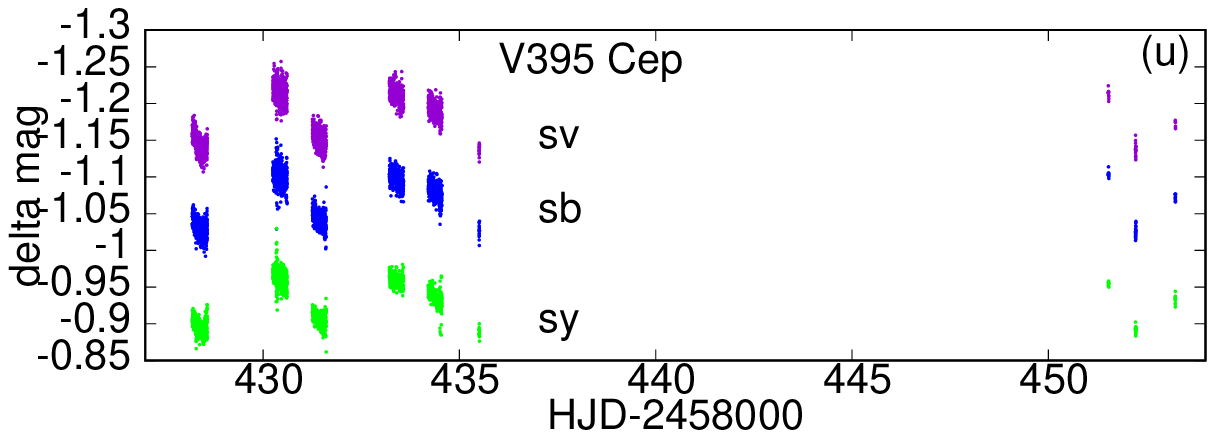}
\end{tabular}}
\FigCap{Results of 2014 and 2016 observations of young variable stars from fields \#5 (NGC~7023), 
\#6 (DI Cep), and 2018 observations of V395 Cep (field \#15).}
\end{figure}

\begin{figure}[]
\centerline{%
\begin{tabular}{l@{\hspace{0.1pc}}}
\includegraphics[width=0.5\linewidth]{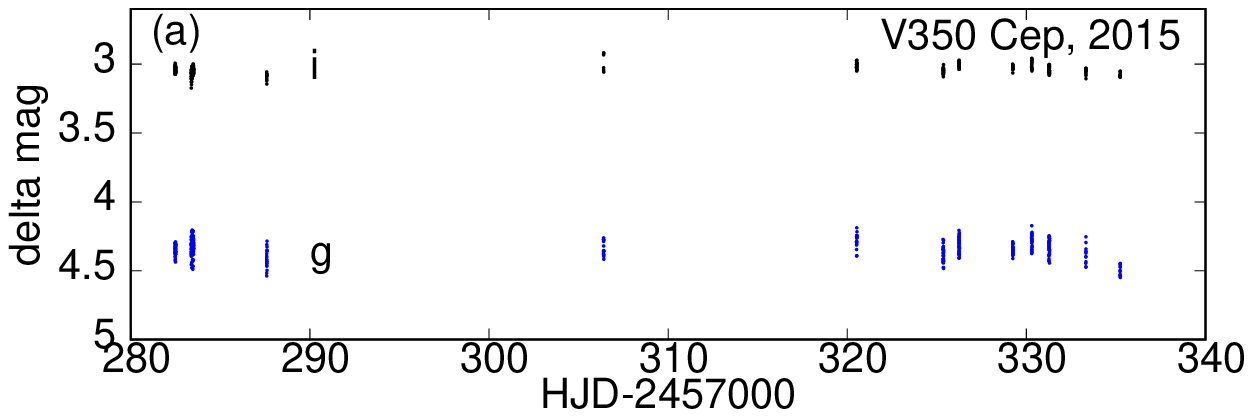}
\includegraphics[width=0.5\linewidth]{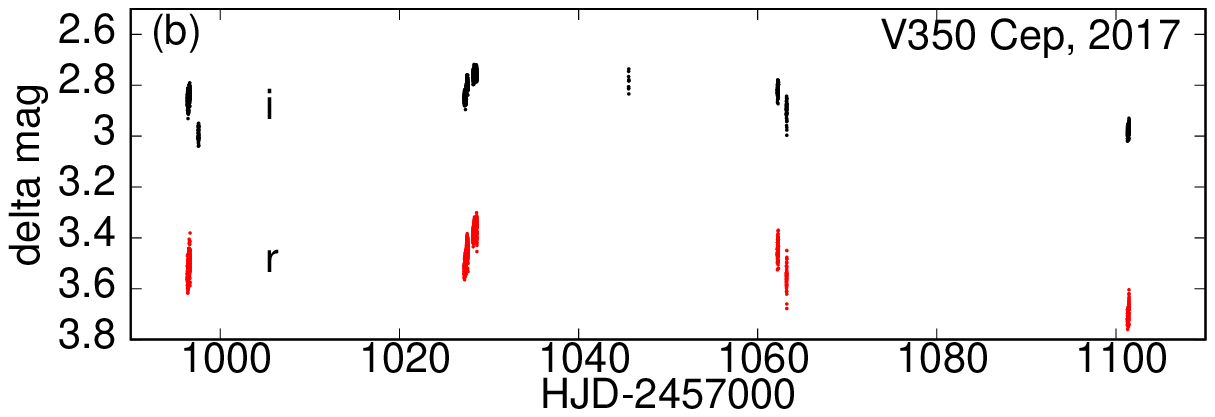}\\
\includegraphics[width=0.5\linewidth]{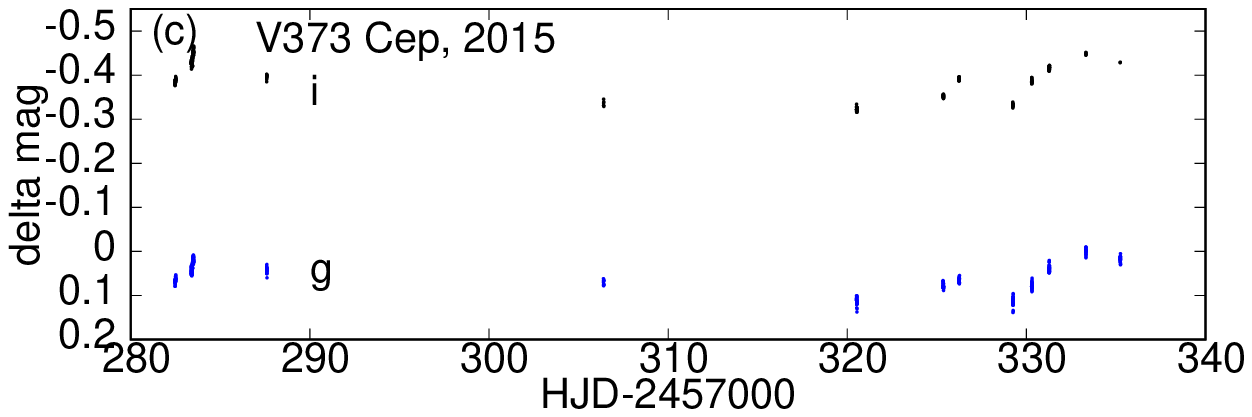}
\includegraphics[width=0.5\linewidth]{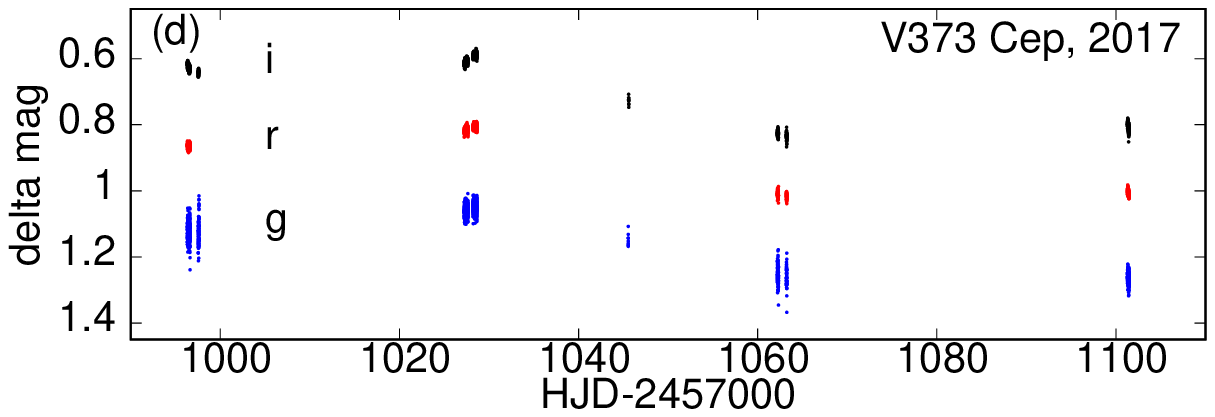}\\
\includegraphics[width=0.5\linewidth]{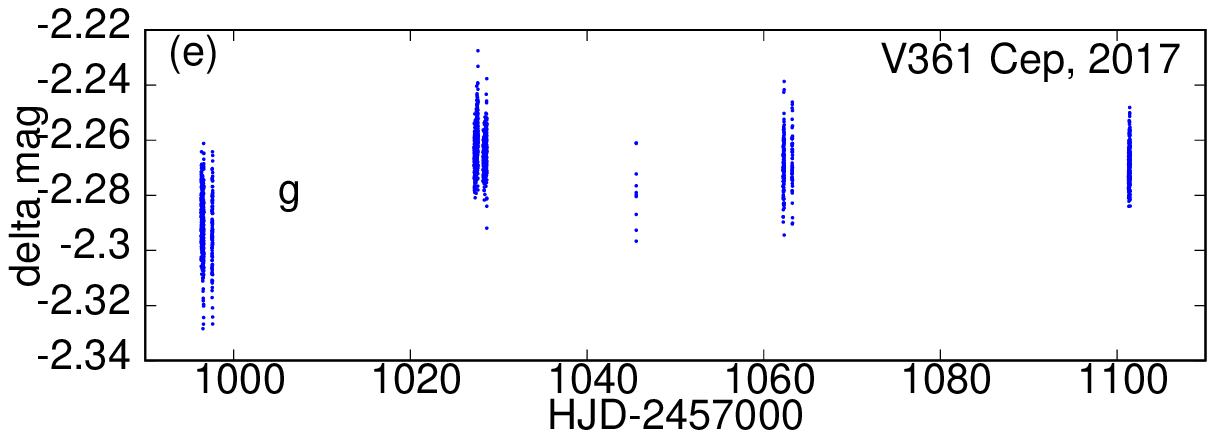}
\includegraphics[width=0.5\linewidth]{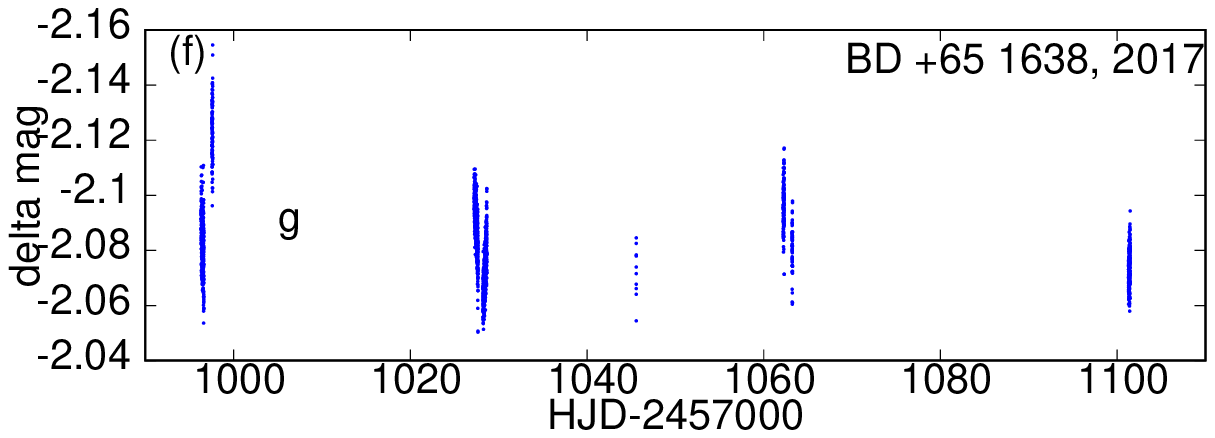}\\
\includegraphics[width=0.5\linewidth]{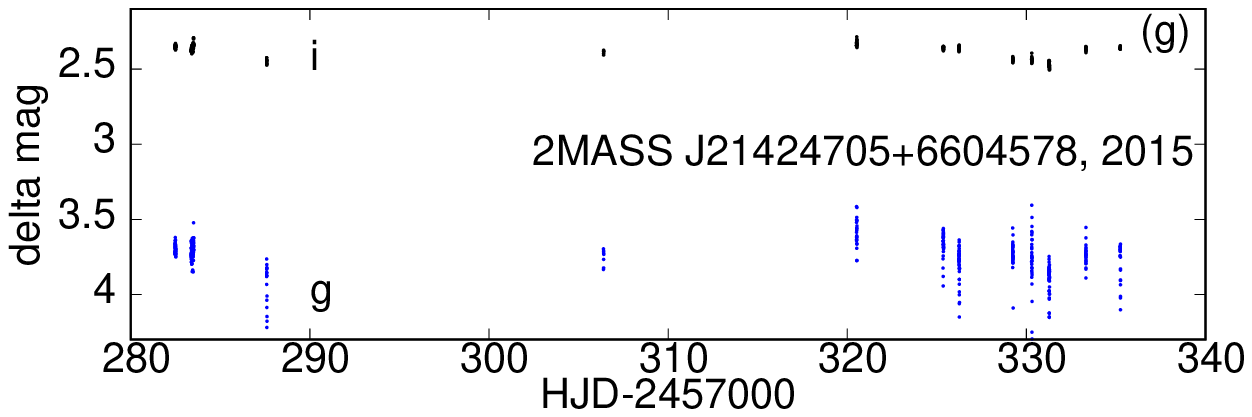}
\includegraphics[width=0.5\linewidth]{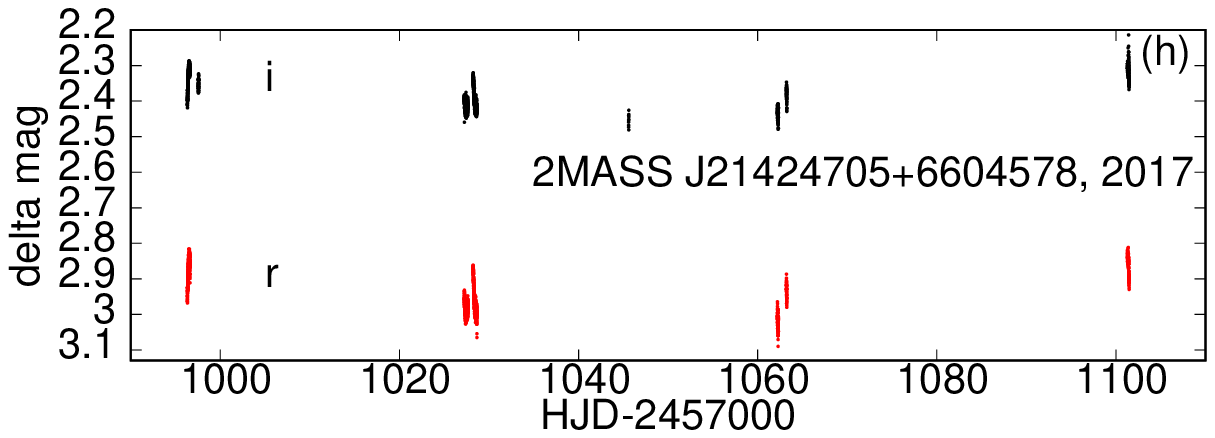}\\
\includegraphics[width=0.5\linewidth]{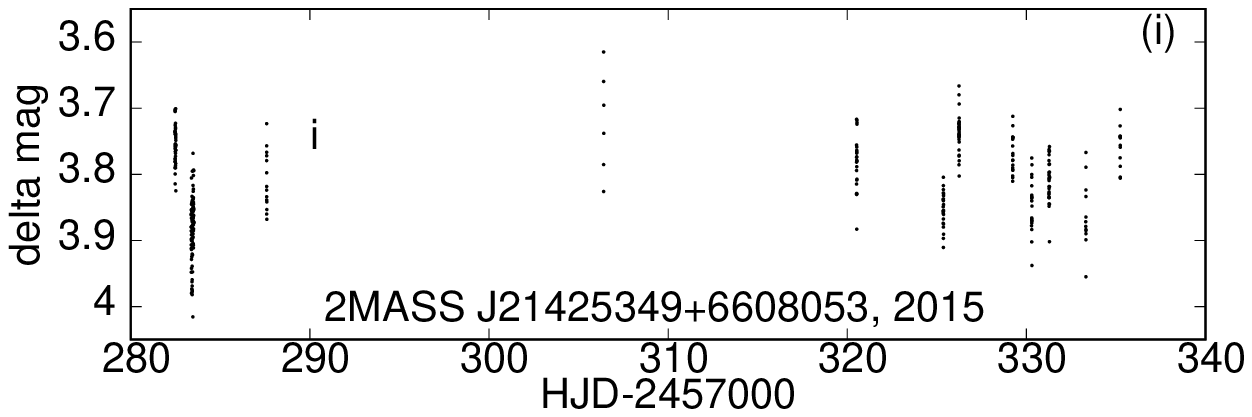}
\includegraphics[width=0.5\linewidth]{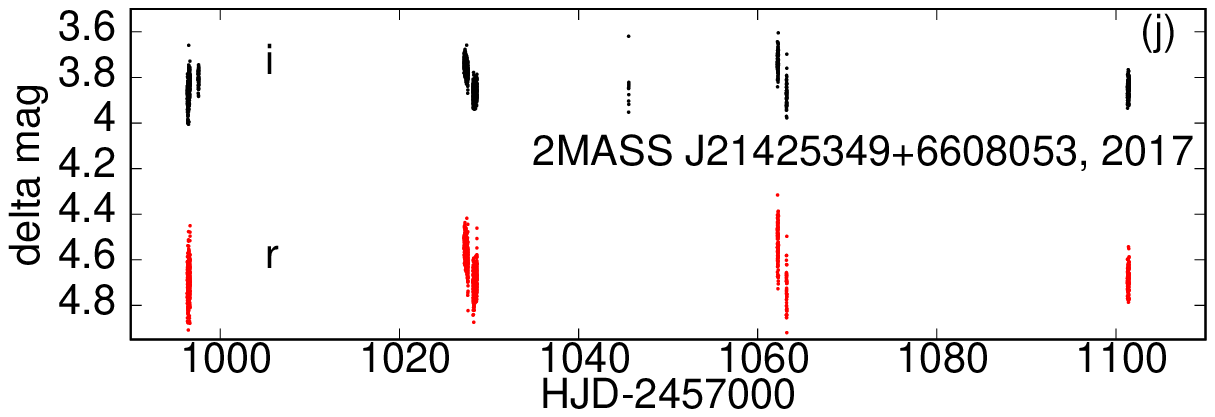}\\
\includegraphics[width=0.5\linewidth]{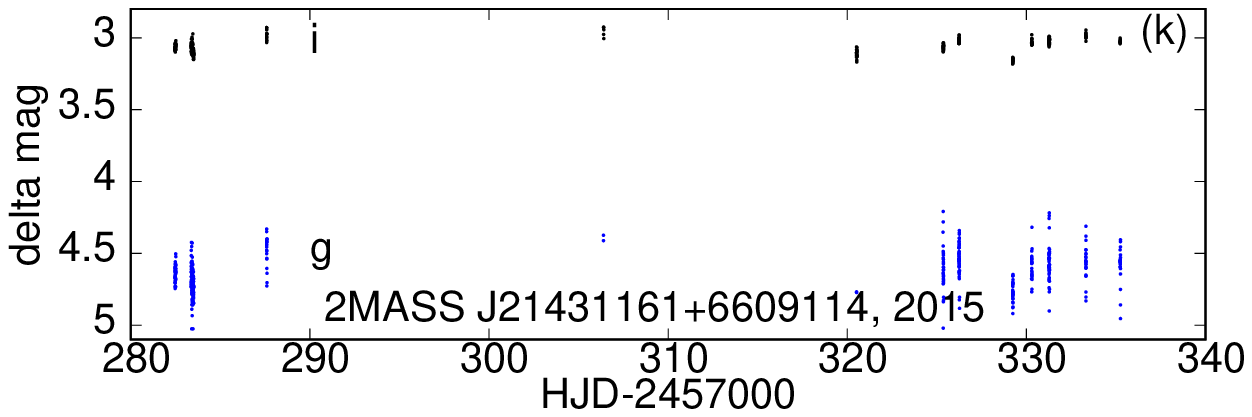}
\includegraphics[width=0.5\linewidth]{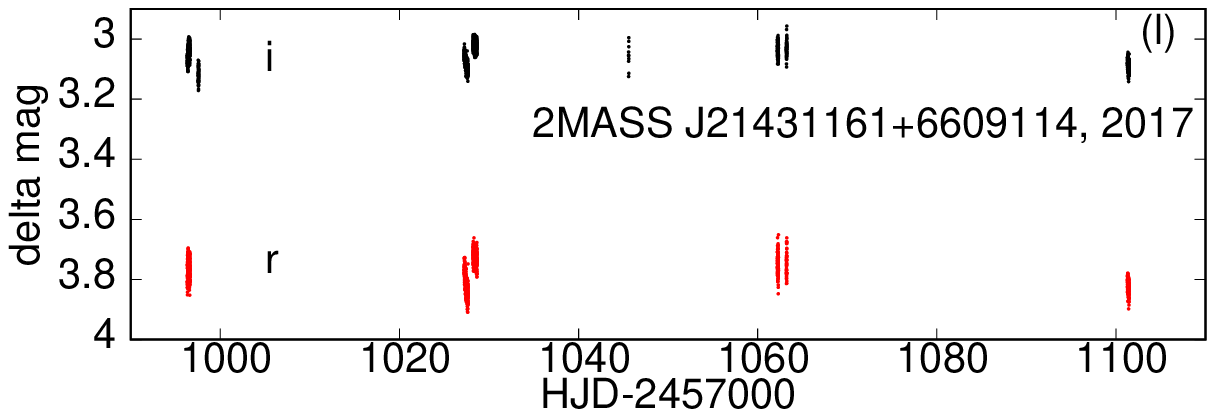}\\
\includegraphics[width=0.5\linewidth]{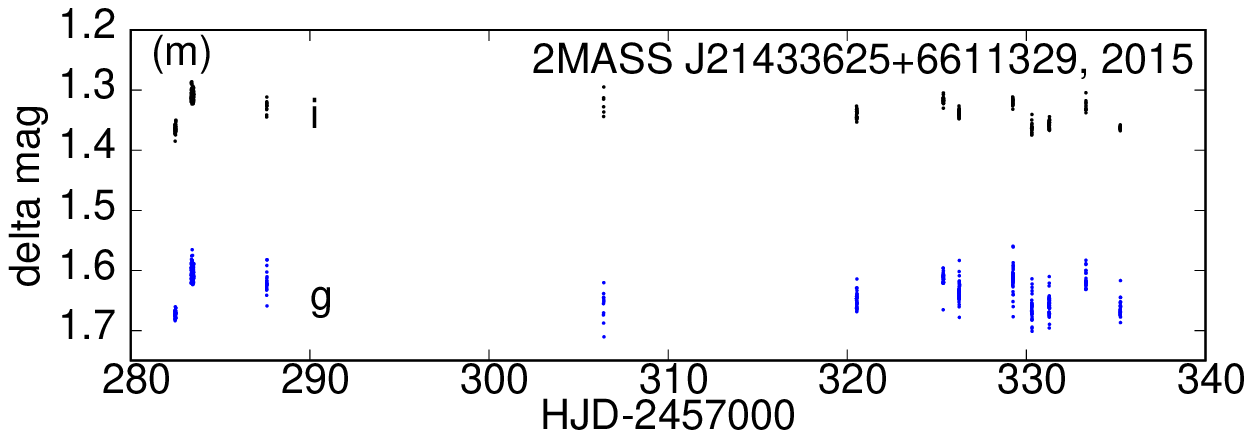}
\includegraphics[width=0.5\linewidth]{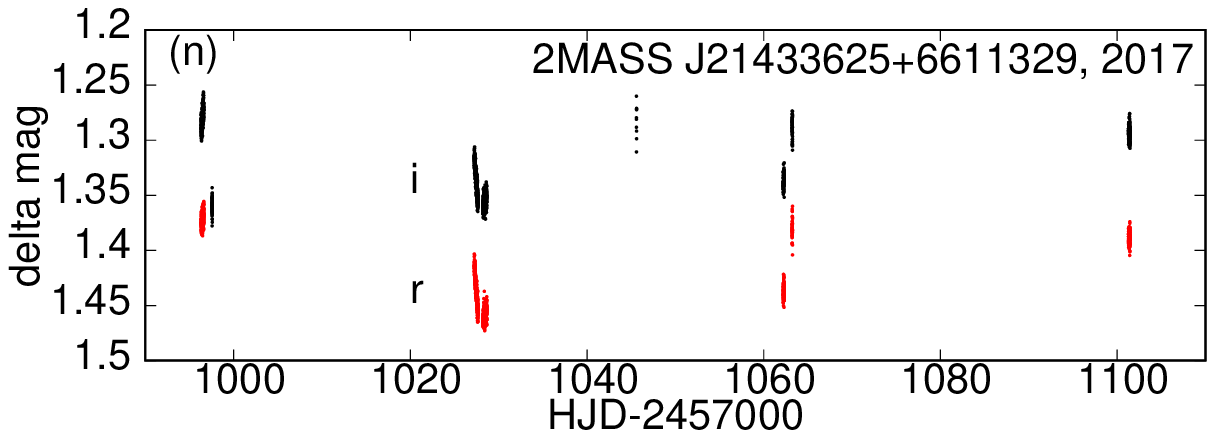}\\
\includegraphics[width=0.5\linewidth]{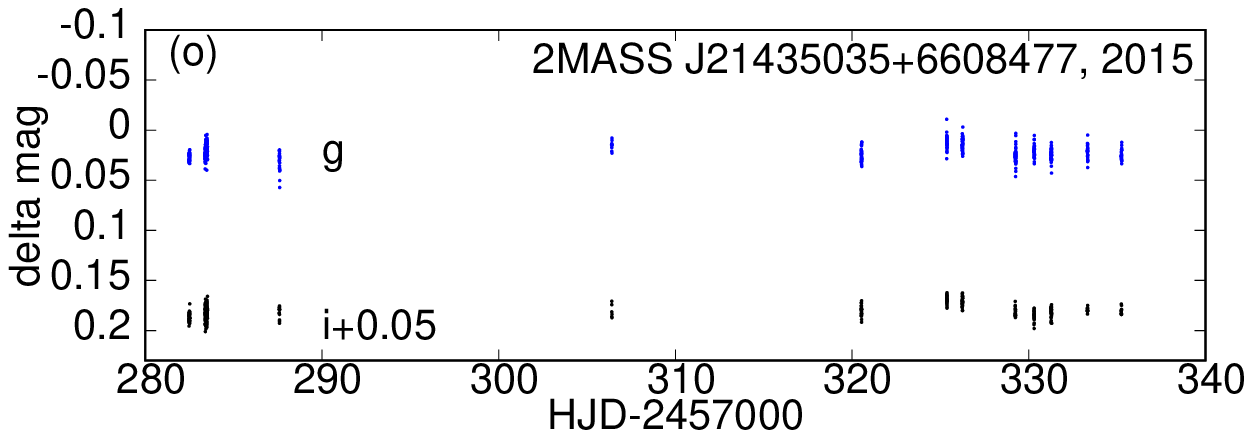}
\includegraphics[width=0.5\linewidth]{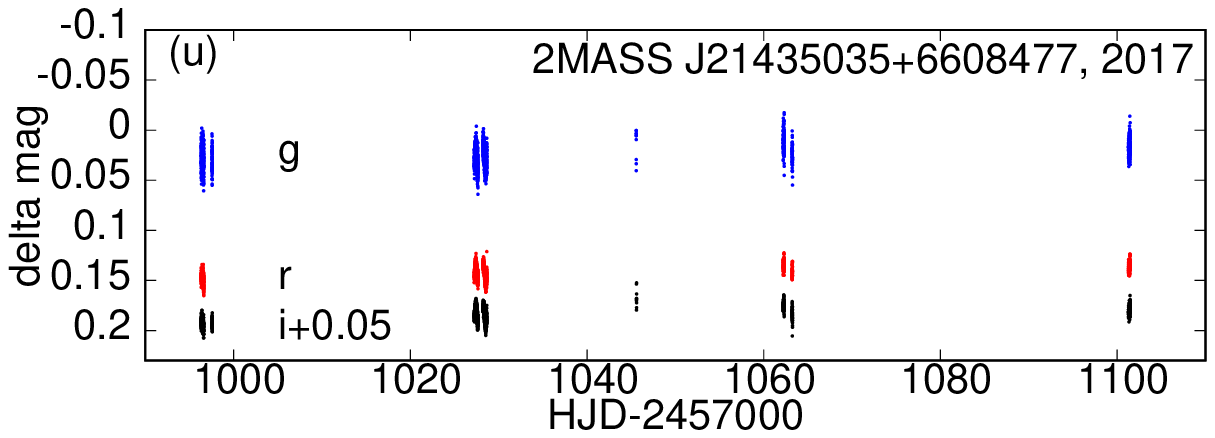}
\end{tabular}}
\FigCap{Results of 2015 and 2017 observations of young variable stars from field \#11 (NGC~7129).}
\end{figure}

\begin{figure}[]
\centerline{%
\begin{tabular}{l@{\hspace{0.1pc}}}
\includegraphics[width=.5\linewidth]{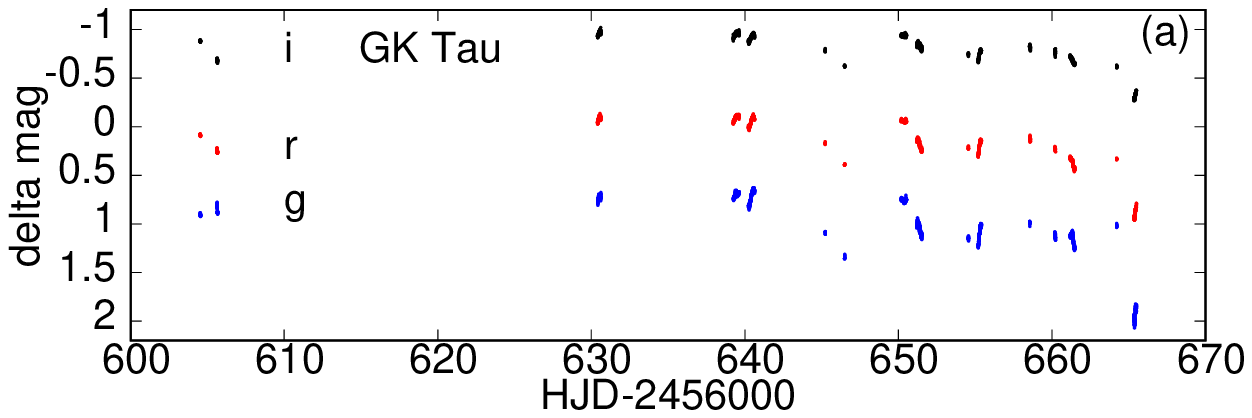} 
\includegraphics[width=.5\linewidth]{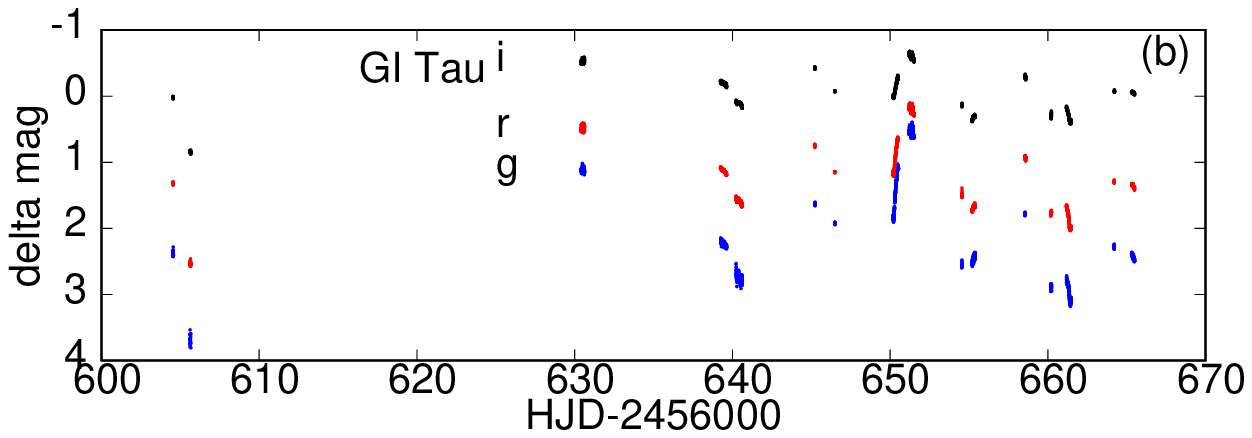}\\
\includegraphics[width=.5\linewidth]{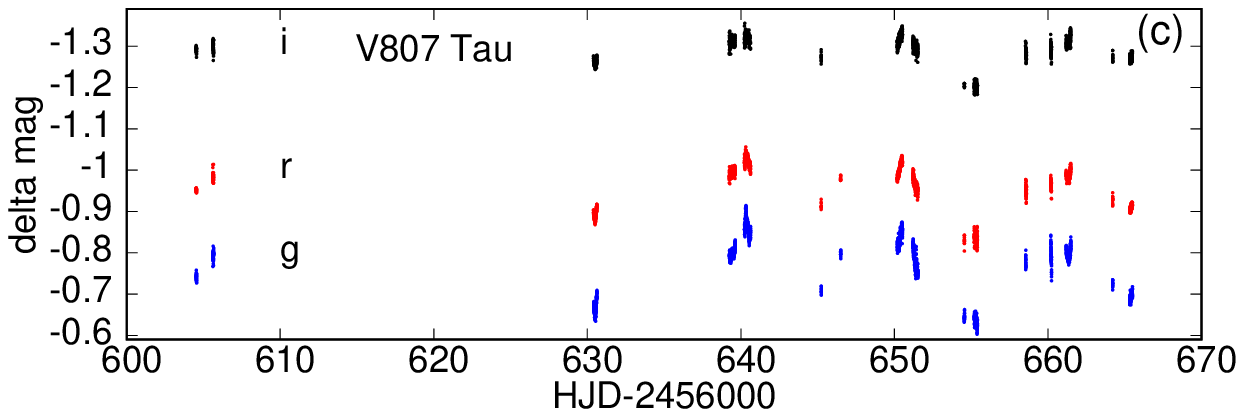}
\includegraphics[width=.5\linewidth]{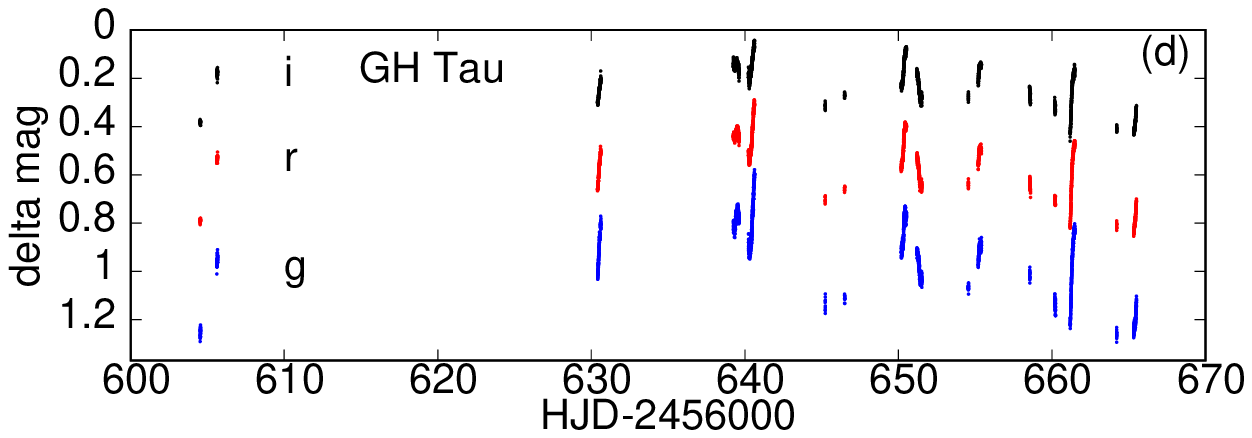}\\
\includegraphics[width=.5\linewidth]{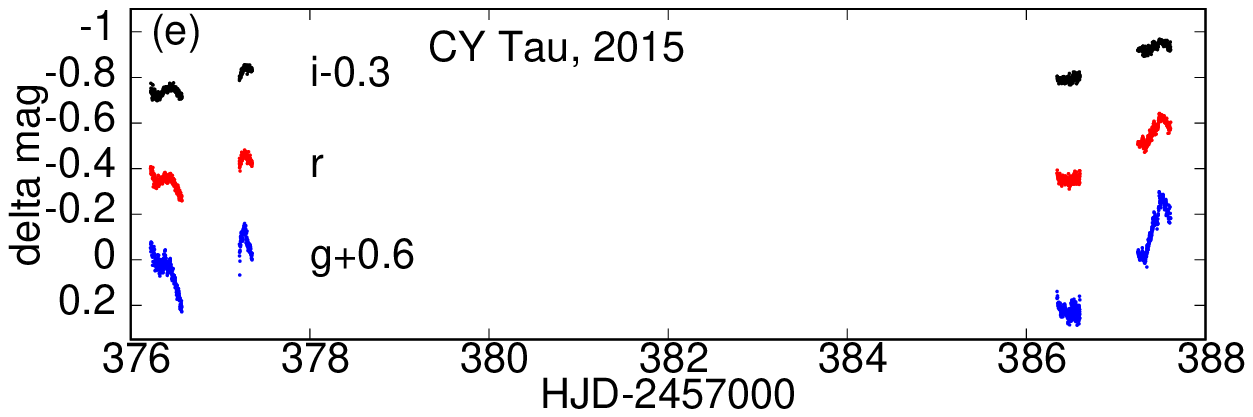}
\includegraphics[width=.5\linewidth]{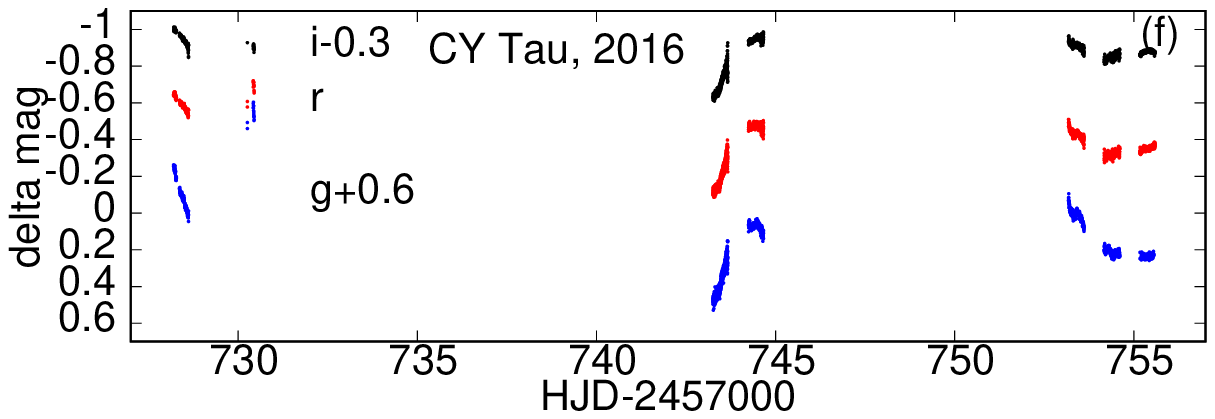}\\  
\includegraphics[width=.5\linewidth]{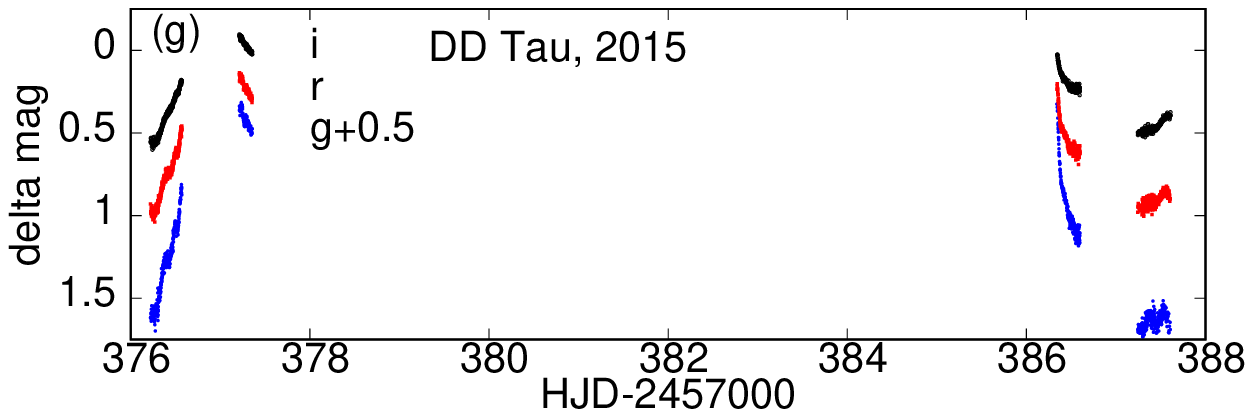}
\includegraphics[width=.5\linewidth]{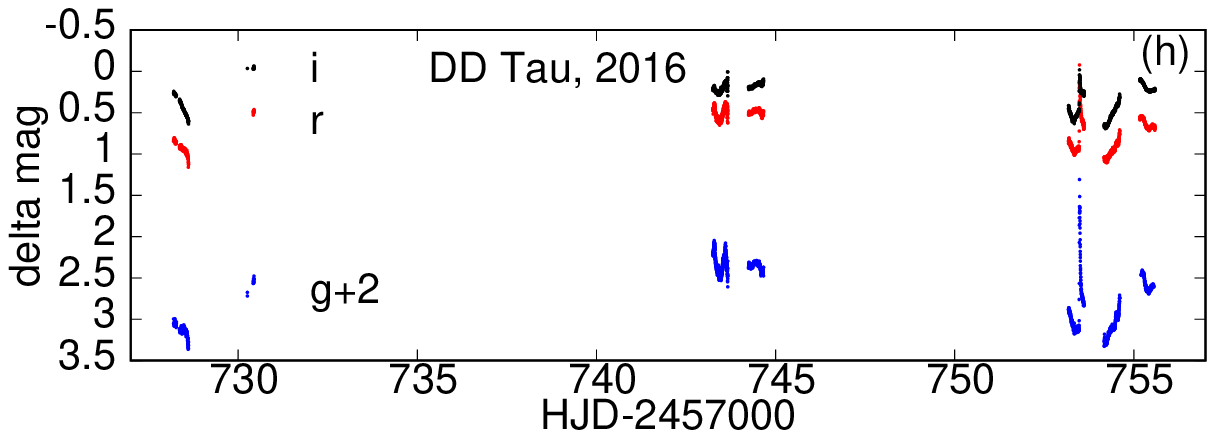}\\
\includegraphics[width=.5\linewidth]{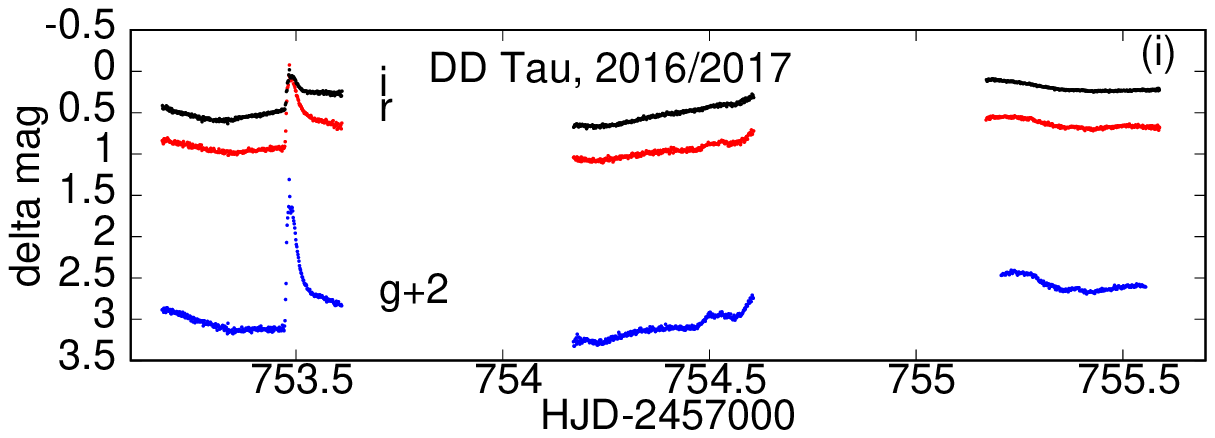}
\includegraphics[width=.5\linewidth]{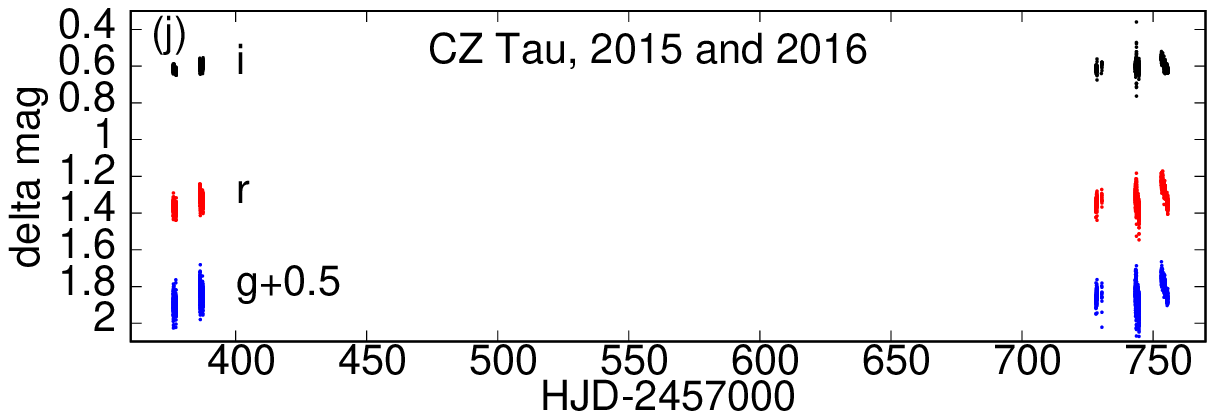}\\
\includegraphics[width=.5\linewidth]{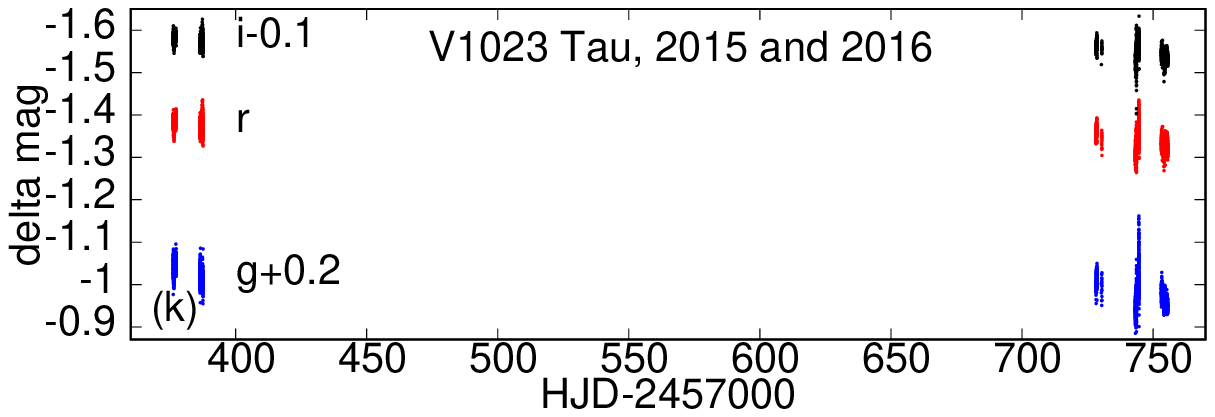}
\includegraphics[width=.5\linewidth]{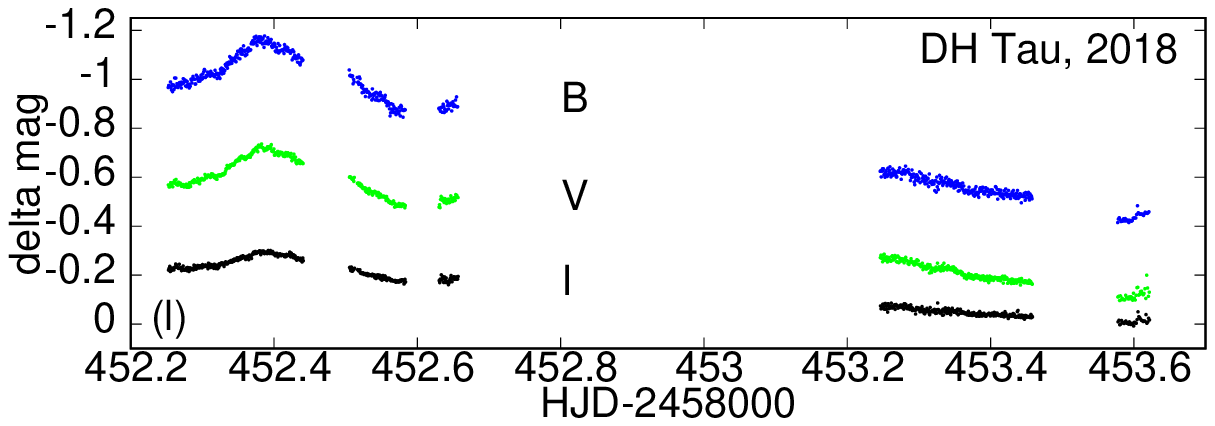}\\
\includegraphics[width=.5\linewidth]{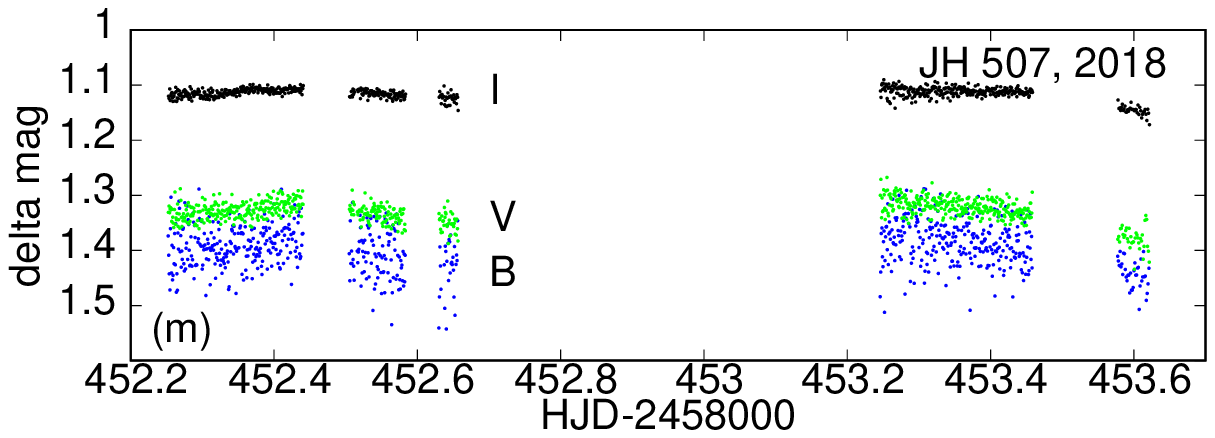}
\end{tabular}}
\FigCap{Results for young variable stars from fields \#4, \#8, and \#14 in ''Taurus-Auriga'' SFR.}
\end{figure}

\begin{figure}[]
\centerline{%
\begin{tabular}{l@{\hspace{0.1pc}}}
\includegraphics[width=.5\linewidth]{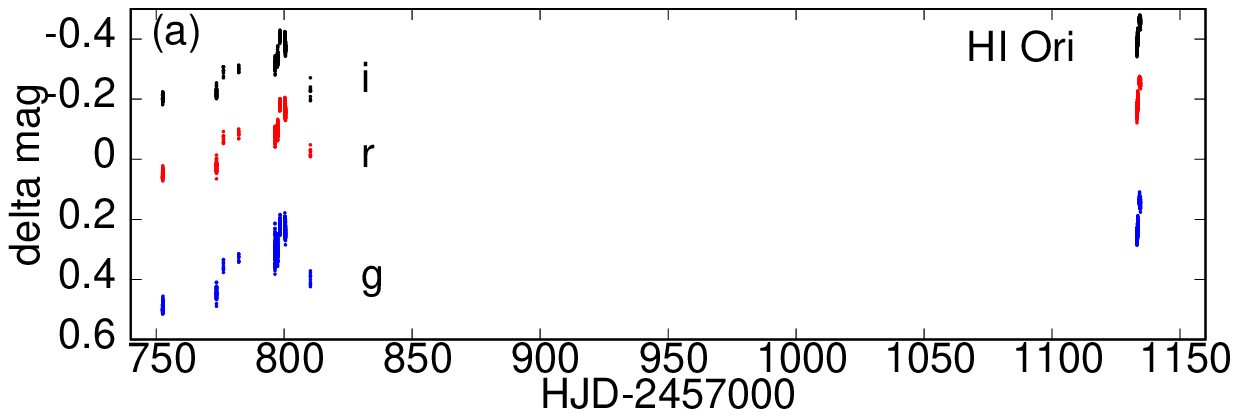} 
\includegraphics[width=.5\linewidth]{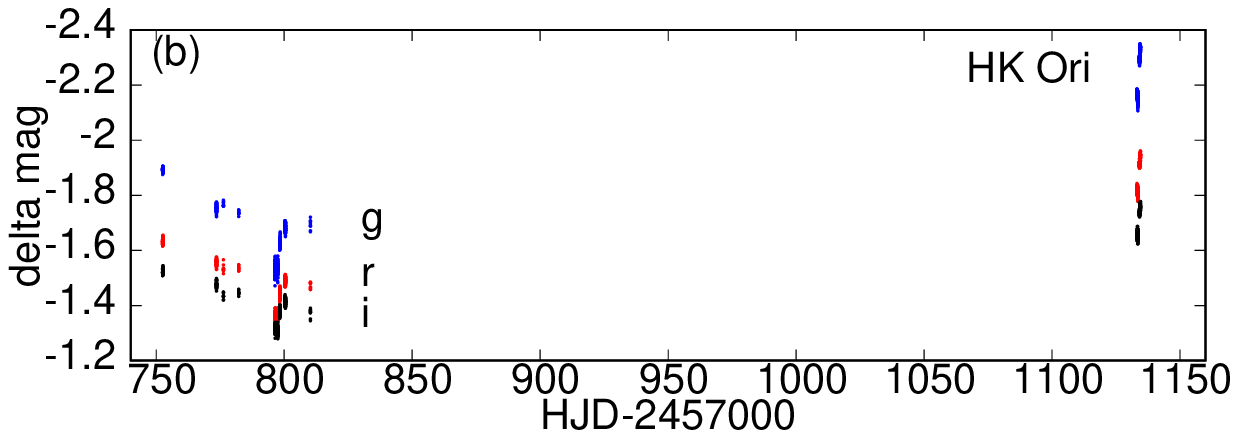}\\
\includegraphics[width=.5\linewidth]{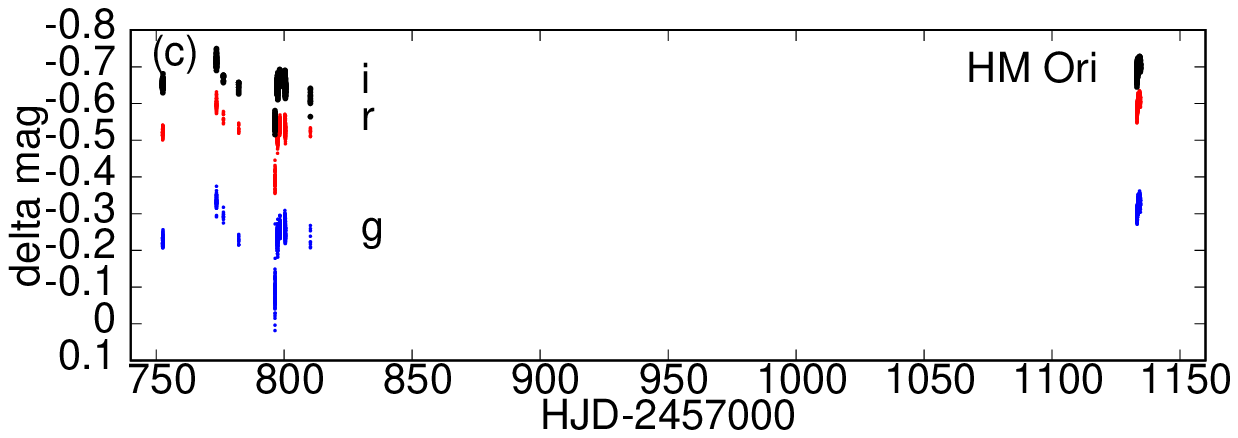} 
\includegraphics[width=.5\linewidth]{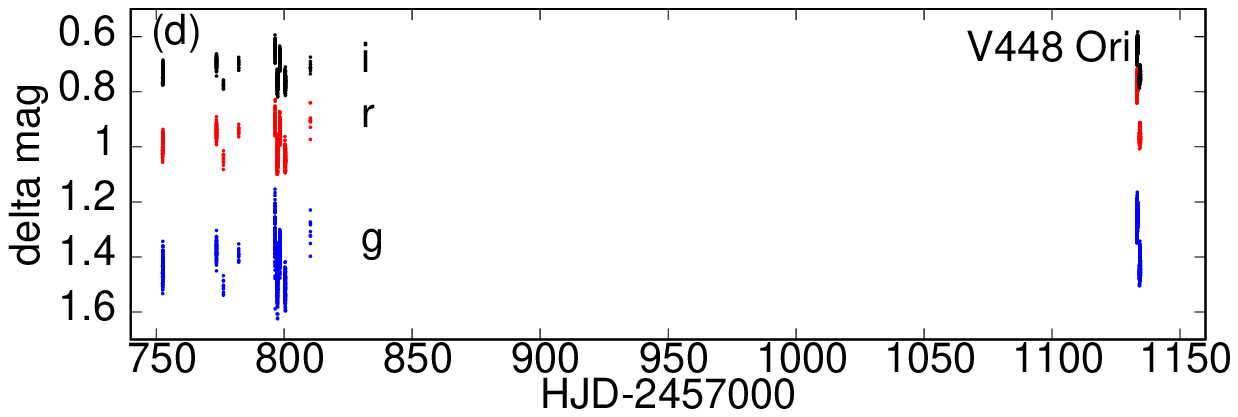}\\
\includegraphics[width=.5\linewidth]{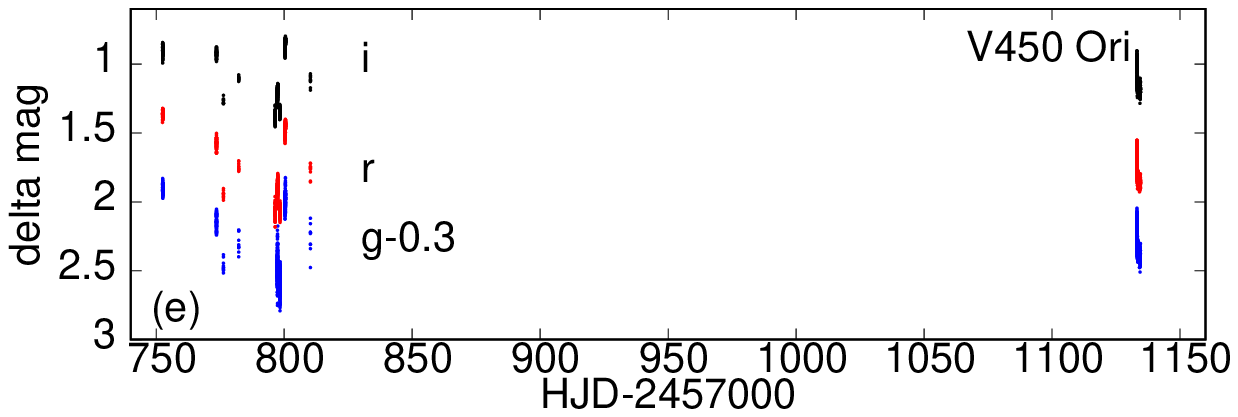}
\includegraphics[width=.5\linewidth]{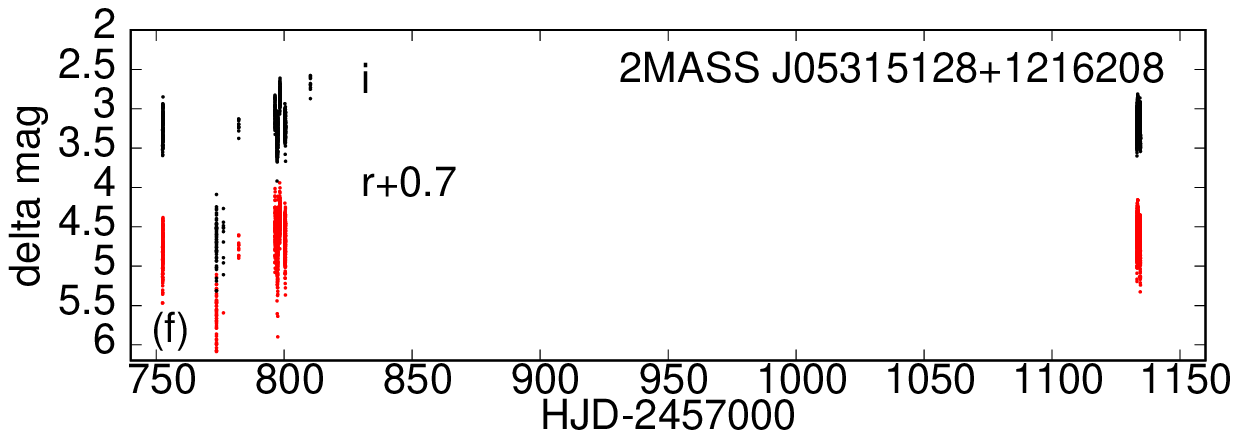}\\
\includegraphics[width=.5\linewidth]{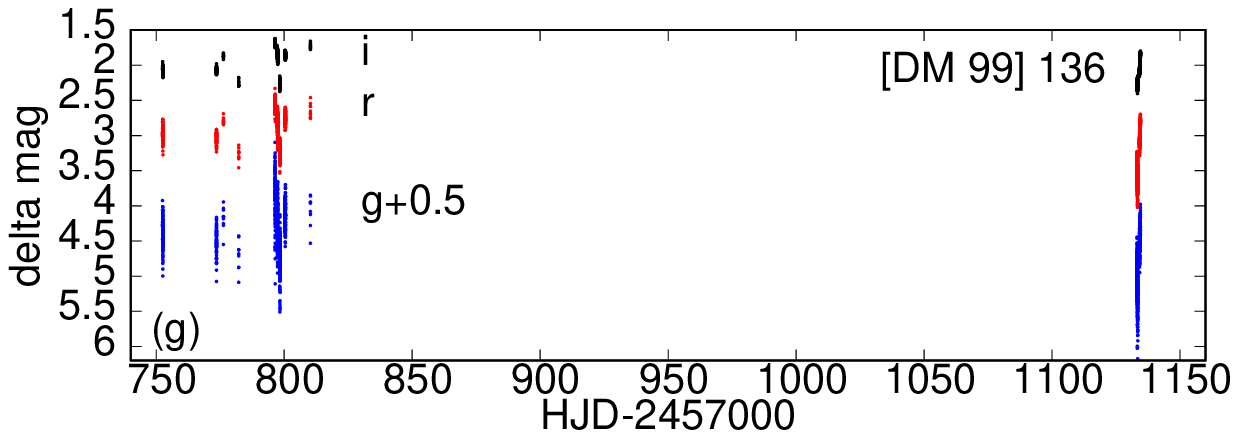}
\includegraphics[width=.5\linewidth]{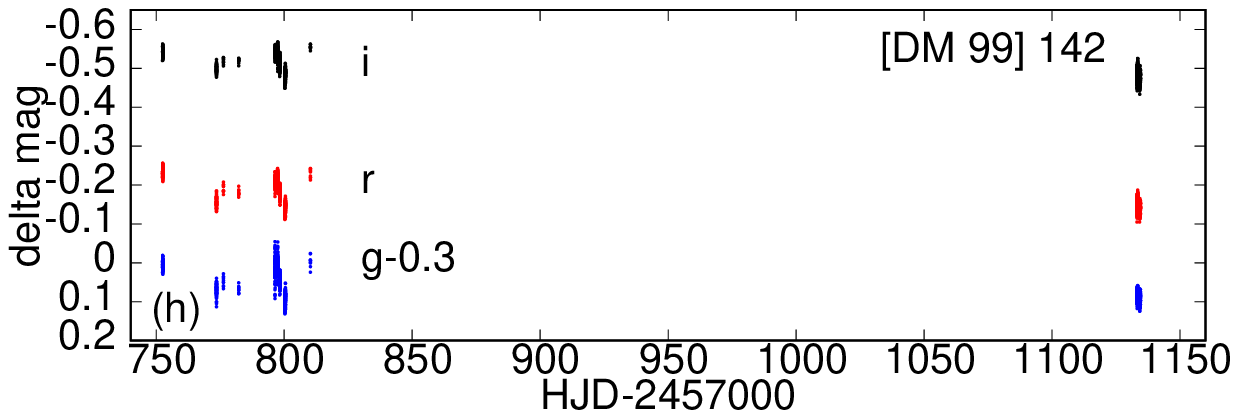}\\
\includegraphics[width=.5\linewidth]{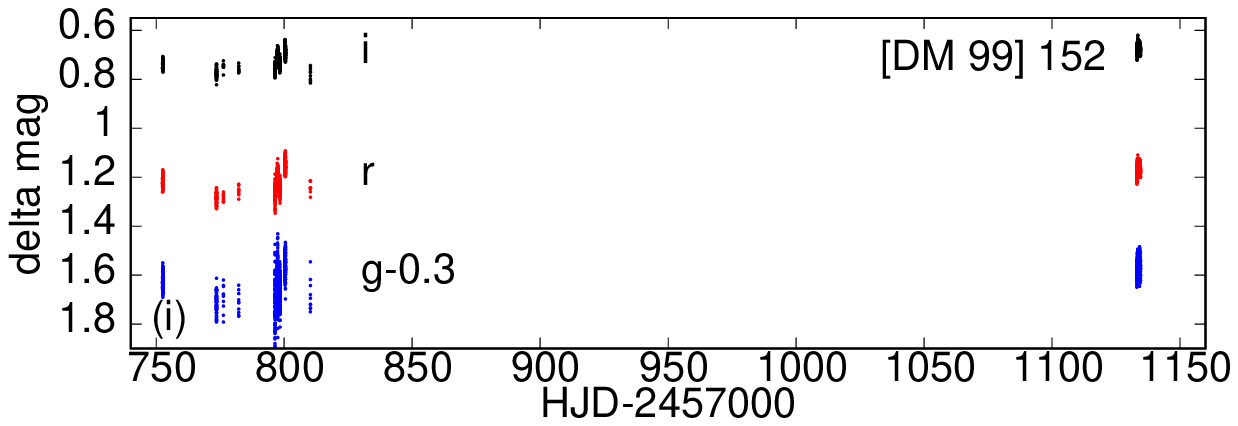}
\includegraphics[width=.5\linewidth]{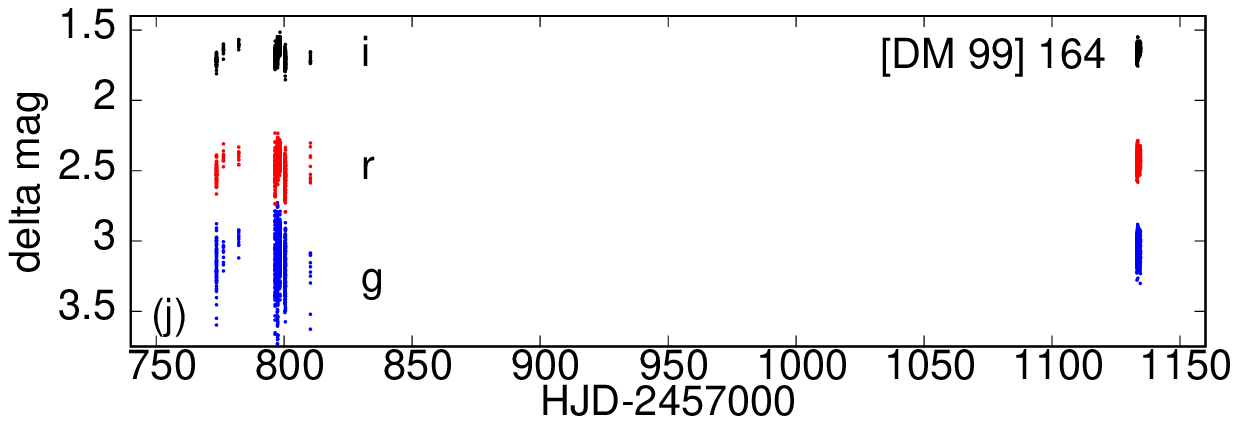}\\
\includegraphics[width=.5\linewidth]{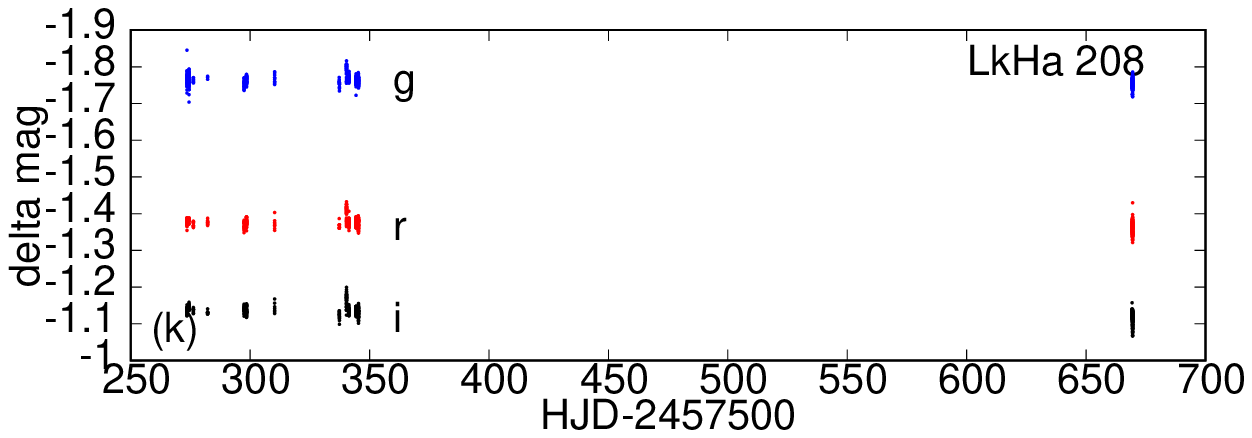}
\includegraphics[width=.5\linewidth]{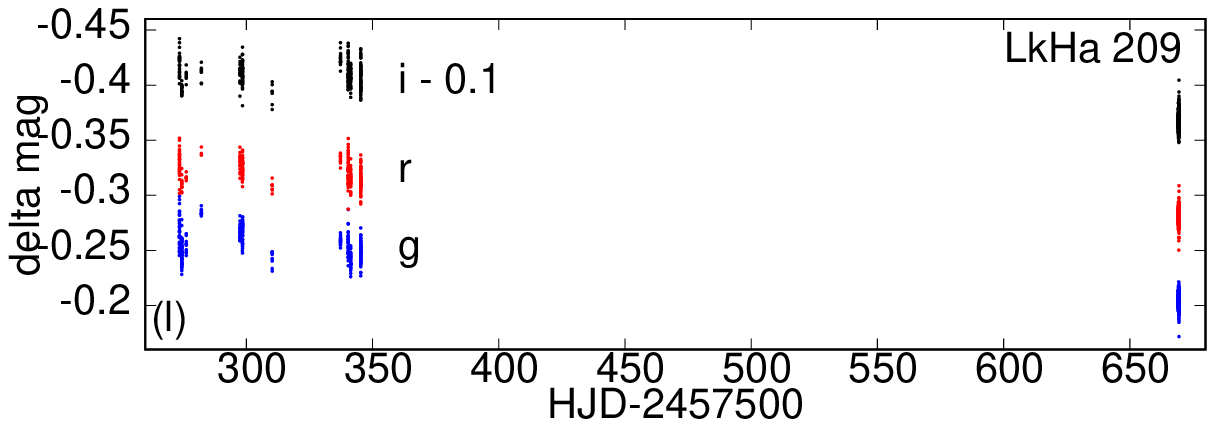}
\end{tabular}}
\FigCap{Results for young variable stars from field \#10 in ''Taurus-Auriga'' SFR.
The last two panels show results for variables from field \#13.}
\end{figure}

\begin{figure}[]
\centerline{%
\begin{tabular}{l@{\hspace{0.1pc}}}
\includegraphics[width=.5\linewidth]{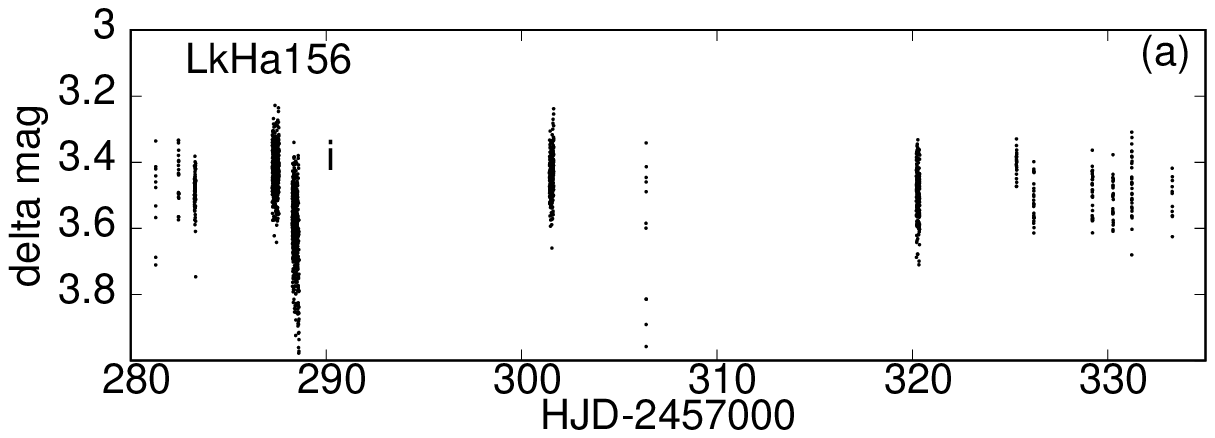} 
\includegraphics[width=.5\linewidth]{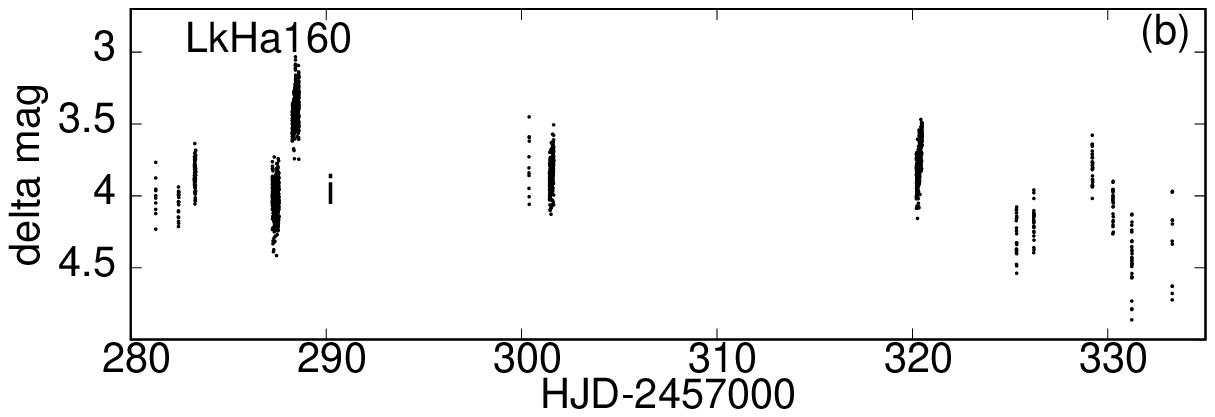}\\
\includegraphics[width=.5\linewidth]{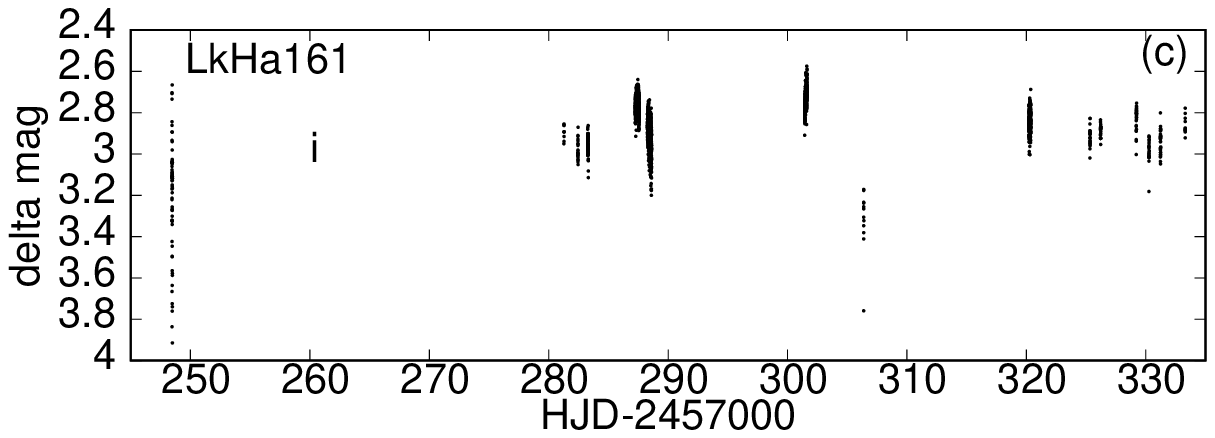}
\includegraphics[width=.5\linewidth]{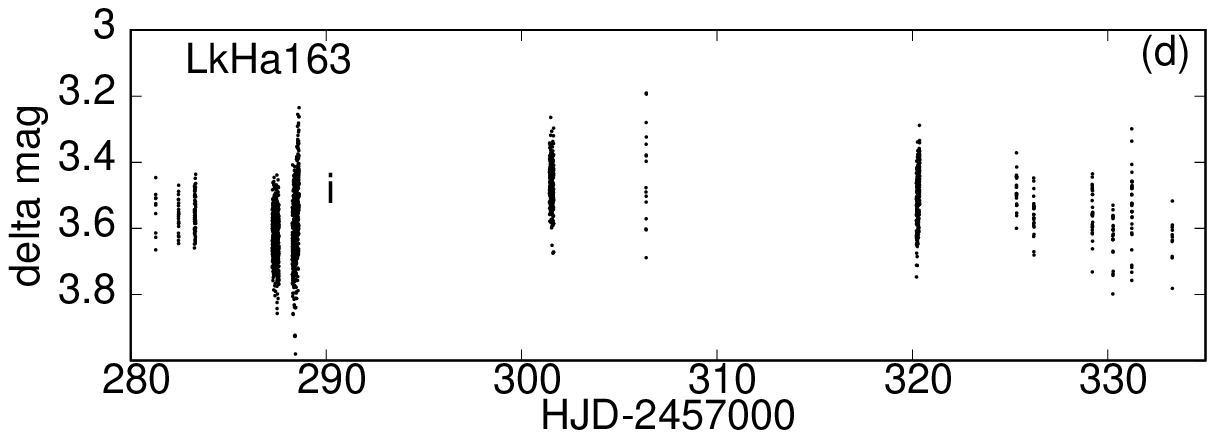}\\
\includegraphics[width=.5\linewidth]{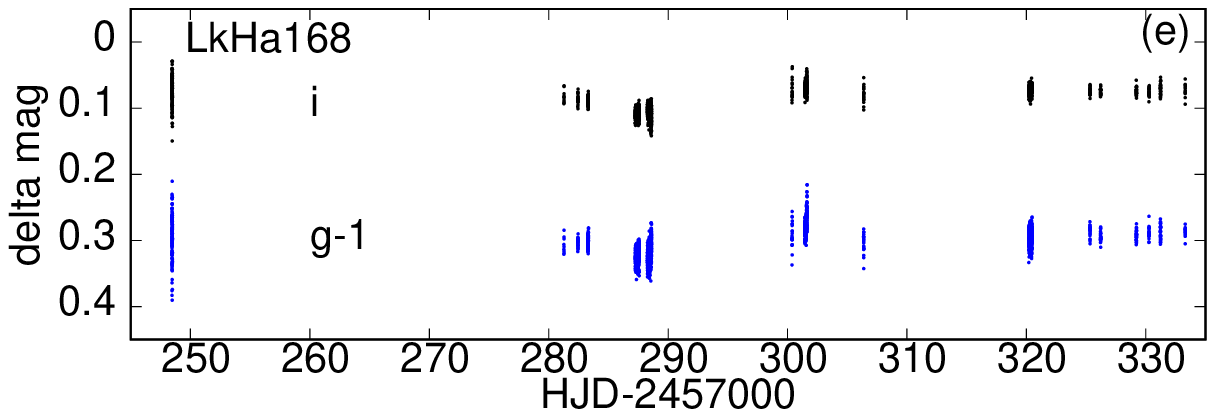}
\includegraphics[width=.5\linewidth]{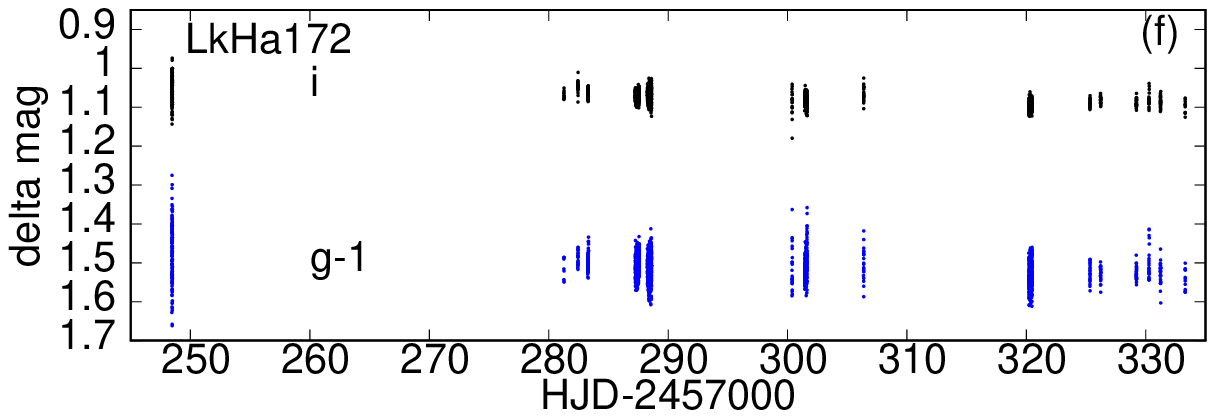}\\  
\includegraphics[width=.5\linewidth]{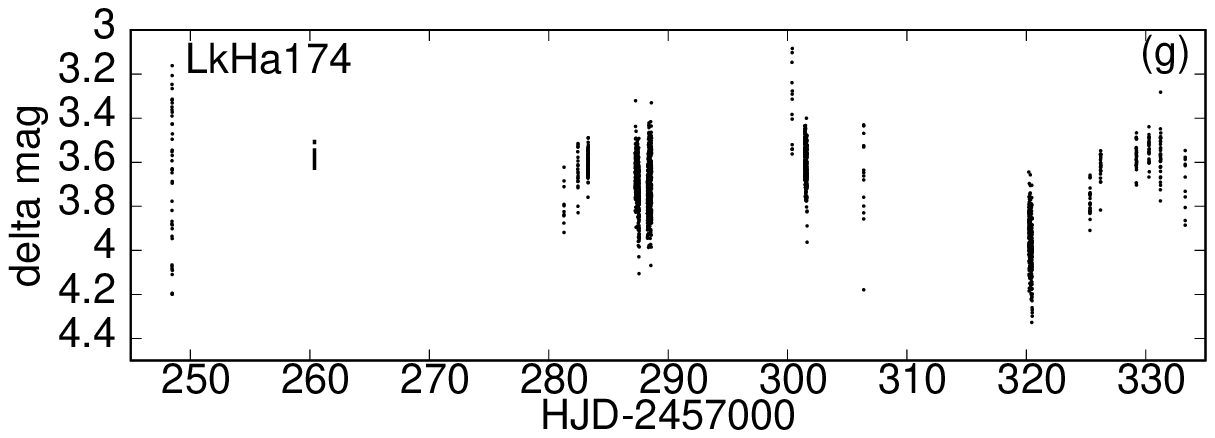}
\includegraphics[width=.5\linewidth]{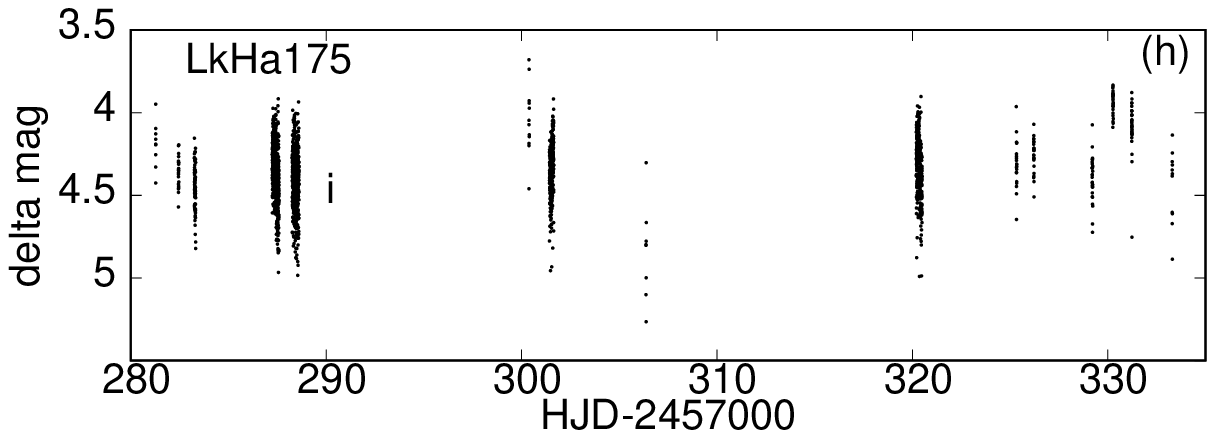}\\
\includegraphics[width=.5\linewidth]{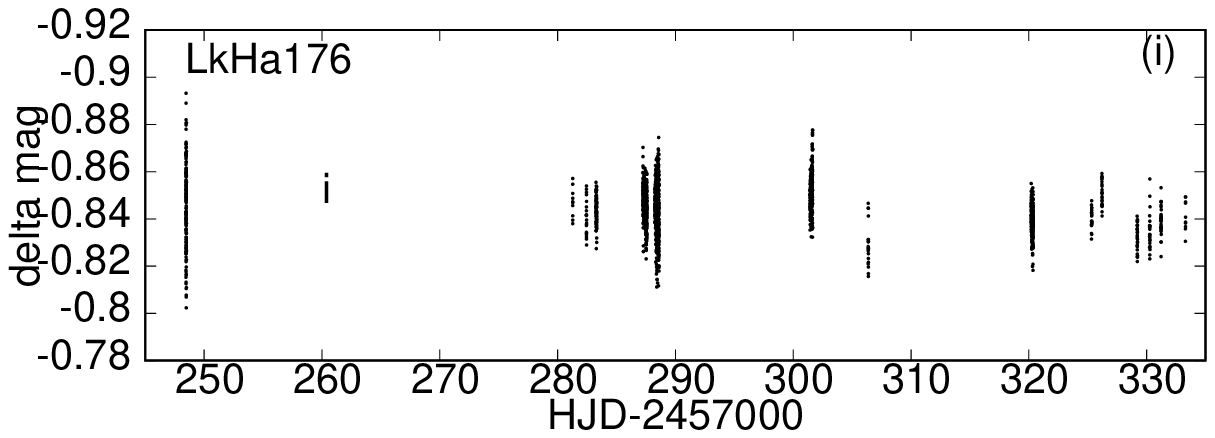}
\includegraphics[width=.5\linewidth]{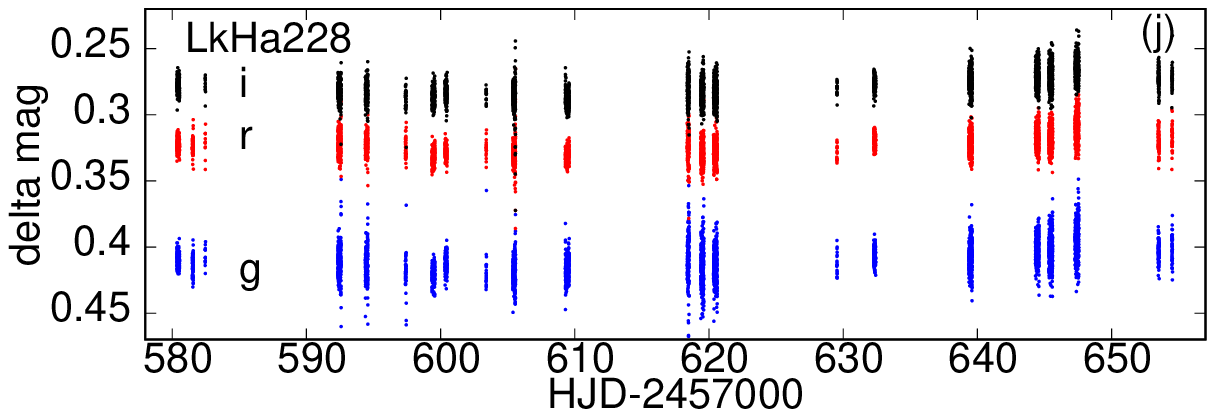}\\
\includegraphics[width=.5\linewidth]{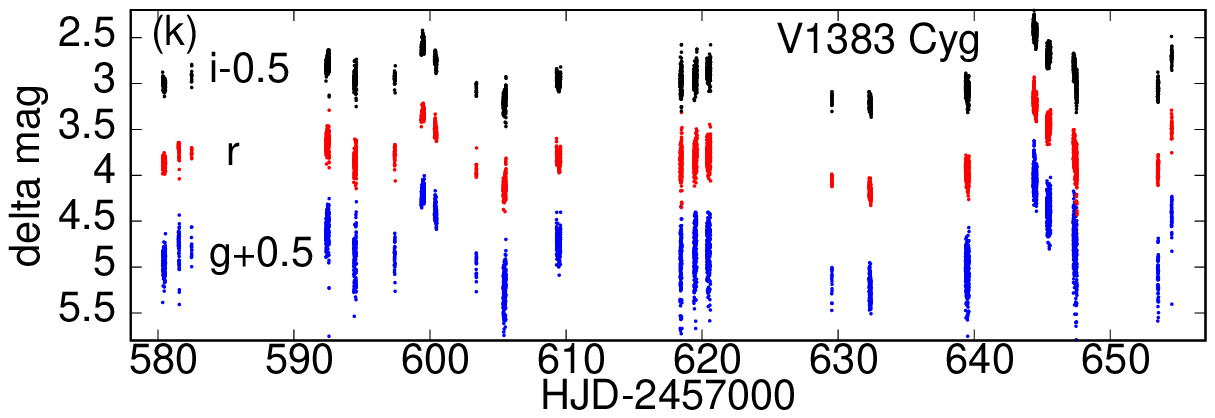}
\includegraphics[width=.5\linewidth]{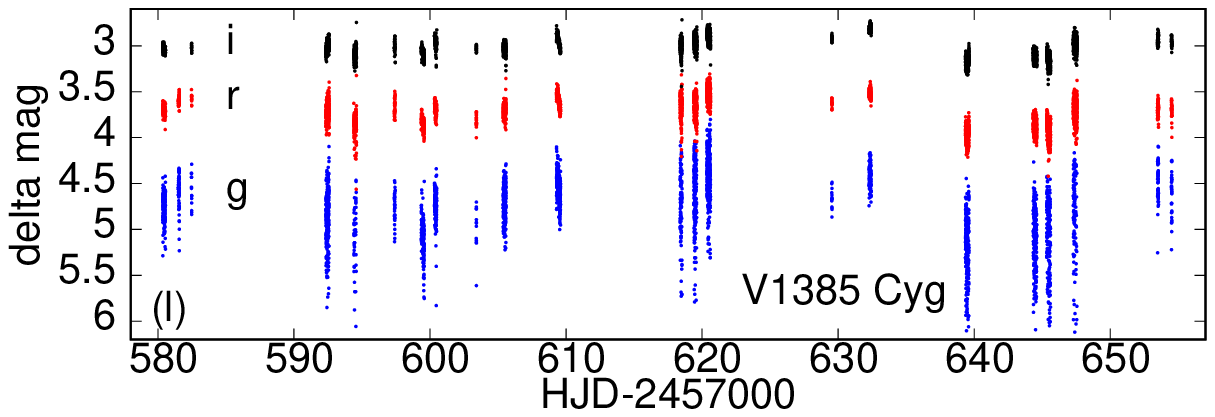}\\
\includegraphics[width=.5\linewidth]{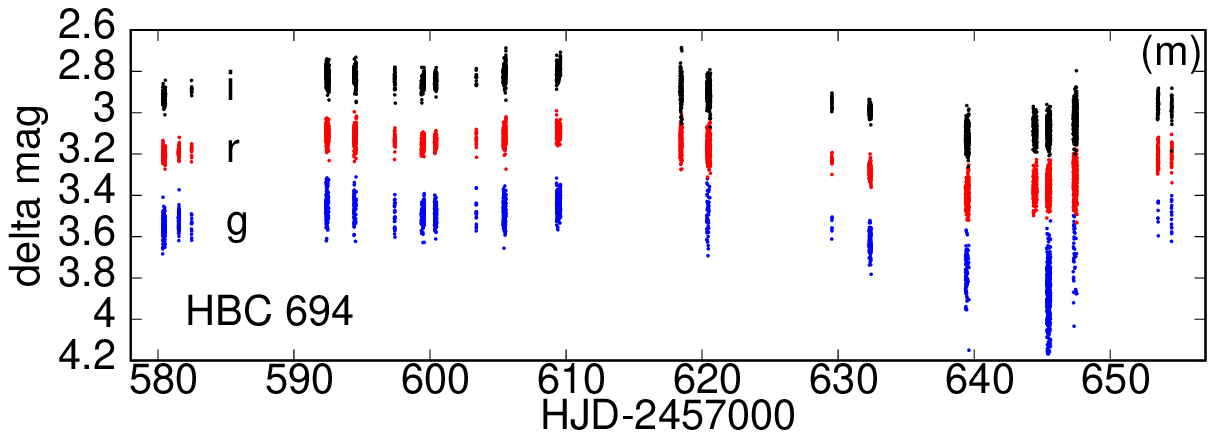}
\includegraphics[width=.5\linewidth]{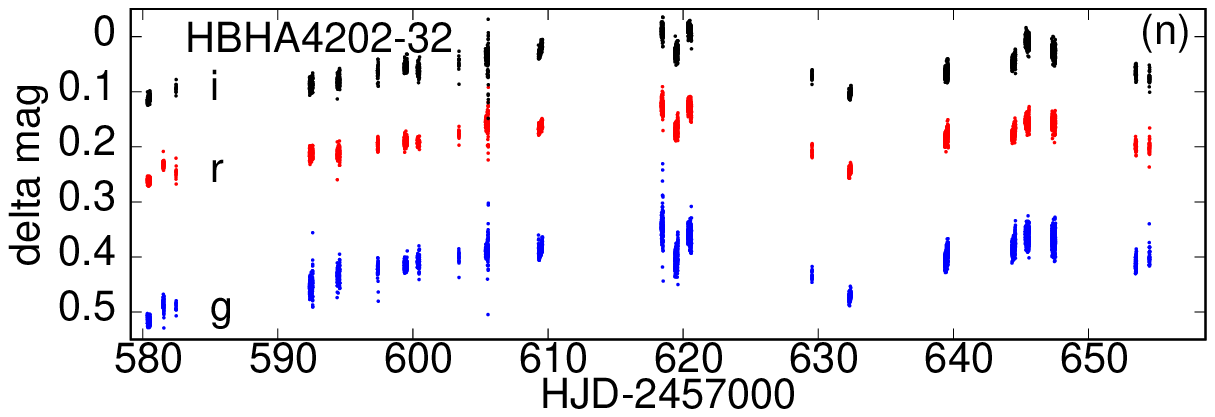}\\
\includegraphics[width=.5\linewidth]{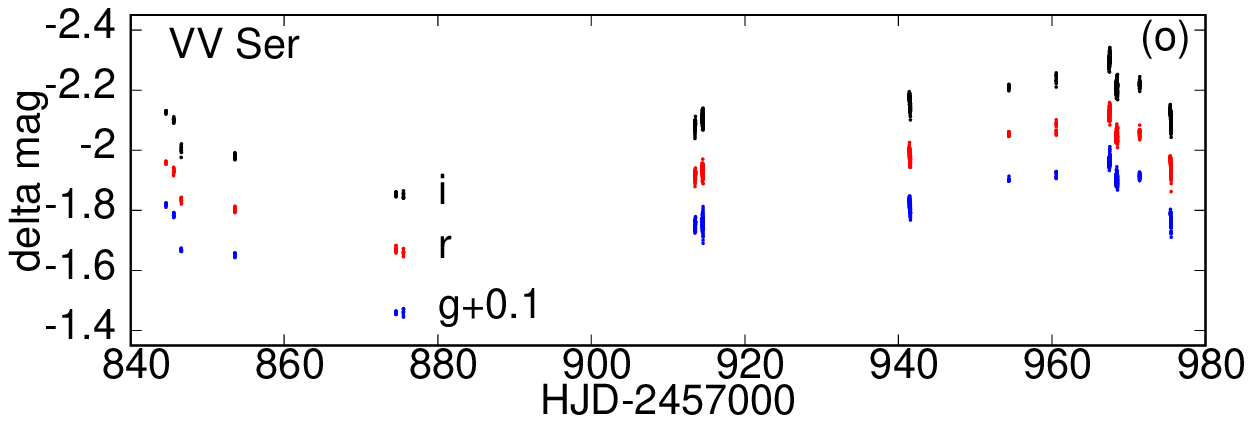}
\includegraphics[width=.5\linewidth]{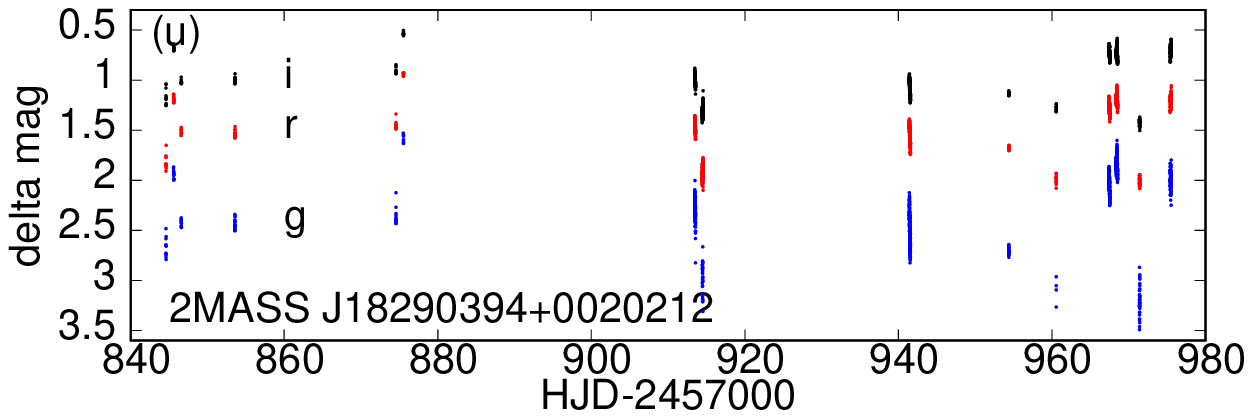}
\end{tabular}}
\FigCap{Results for young variable stars from fields \#7 and \#9 in Cygnus 
and for the field \#12.}
\end{figure}

\Acknow{The authors acknowledge Polish National Science Centre for two grants supporting 
this research: 2012/05/E/ST9/03915 (MS, MD, KG, WO, MW) and 2011/03/D/ST9/01808 (GS). 
This research has made use of the {\sc SIMBAD} database, operated at CDS, Strasbourg, France. 
Special thanks are due prof. Stanislaw Zola and dr. Pawel Zielinski for taking short series 
of the data on a few occasions.}

\end{document}